\documentclass[a4paper,11pt]{article}
\usepackage{jheppub}
\usepackage[T1]{fontenc}
\usepackage{setspace}
\usepackage{amsmath,amssymb,mathtools}
\usepackage{physics}
\usepackage{float}
\usepackage{graphicx}
\usepackage{subcaption}
\usepackage{xcolor}
\usepackage{soul}

\newcommand{\diffnote}[1]{}
\newcommand{\errnote}[1]{}
\newcommand{\bluecheck}{}
\newcommand{\reasoncheck}{}
\newcommand{\calccheck}{}

\newcommand{\cH}{\mathcal{H}}
\newcommand{\cM}{\mathcal{M}}
\newcommand{\cP}{\mathcal{P}}
\newcommand{\cB}{\mathcal{B}}
\newcommand{\cO}{\mathcal{O}}
\newcommand{\cU}{\mathcal{U}}
\newcommand{\cV}{\mathcal{V}}
\newcommand{\cW}{\mathcal{W}}
\newcommand{\cK}{\mathcal{K}}
\newcommand{\cI}{\mathcal{I}}
\newcommand{\cT}{\mathcal{T}}
\newcommand{\wh}{\widehat}
\newcommand{\eps}{\epsilon}

\title{Closed String Field Theory in 25.99 Dimensions}
\author[a]{Amr Ahmadain}
\author[b]{Alexander Frenkel}
\author[c]{Xi Yin}
\affiliation[a]{Department of Physics, Swansea University, Swansea, SA2 8PP, UK}
\affiliation[b]{Simons Center for Geometry and Physics, Stony Brook University, Stony Brook, NY 11794, USA}
\affiliation[c]{Jefferson Physical Laboratory, Harvard University, Cambridge, MA 02138, USA}
\emailAdd{amrahmadain@gmail.com}
\emailAdd{frenkelalexander1@gmail.com}
\emailAdd{xiyin@fas.harvard.edu}

\abstract{
We return to and refine Zwiebach's formulation of closed string field theory (CSFT) built around non-critical backgrounds \cite{Zwiebach:1996jc,Zwiebach:1996ph}, restricting our attention to genus zero. The  structure involves a special string state $F$ that  encodes the failure of worldsheet BRST invariance, and a metric-dependent  descent operator $\mathcal{B}$ adapted to the  Weyl frame. We construct the mixed  moduli spaces needed for the classical  BV  action,  prove their existence, and extend the Sen-Zwiebach  background independence argument to first order off of the conformal locus. We apply the formalism to the mildest deviation away from criticality -- worldsheet CFTs with nonzero central charge: we consider both $D = 26-\epsilon$ dimensional flat space and linear dilaton profiles in bosonic string theory, focusing for simplicity on building solutions that depend on only one of the $D$ dimensions.}
\date{\today}

\begin{document}
\maketitle

\section{Introduction}

Closed string field theory (CSFT) is built on a chosen worldsheet conformal field theory (CFT), and in that sense it knows about one background at a time. Within the conformal locus the situation is well controlled: Sen showed that infinitesimal changes of background are generated by specific worldsheet insertions and analyzed the resulting on-shell amplitudes and field redefinitions in a sequence of papers \cite{Sen:1990hh,Sen:1990na,Sen:1992pw}, and Sen and Zwiebach proved that the resulting string field theories are related by antibracket-preserving field redefinitions \cite{Sen:1993mh,Sen:1993kb}. The difficulty begins when one tries to leave the conformal locus, because the BRST charge that organizes the entire gauge structure ceases to be conserved and nilpotent as soon as conformal invariance is lost.

Zwiebach addressed this problem in two papers in 1996 \cite{Zwiebach:1996jc,Zwiebach:1996ph}. His key insight was that the full geometric machinery of closed string field theory---string vertices as chains in the bundle of punctured spheres, the BV master equation as a boundary condition on those chains---can be preserved even when the BRST charge fails. The price is that in addition to the ordinary punctures carrying the dynamical string field $\Psi$, one must introduce \emph{special punctures} carrying a fixed Grassmann-odd state $F$ of ghost number 3 that encodes the BRST defect. The corresponding vertices live on punctured spheres equipped with Hermitian metric data, so that the local geometry at each special puncture is determined by the metric rather than by an independent choice of local coordinate.

Zwiebach's original construction is powerful but compressed, and some of its details have aged unevenly against the language now standard in discussions of background independence, BV geometry, homotopy-algebraic structure, and conformal perturbation theory \cite{Ghoshal:1991pu,Hata:1993gf,Sen:2014dqa,Sen:2016qap,Hohm:2017pnh,Pius:2014gza,Pius:2014iaa,Sen:2015offshell,Sen:2024nfd}. Meanwhile closed SFT has emerged as a natural framework for organizing conformal perturbation theory and regulating worldsheet short-distance singularities \cite{Sen:2019jpm,Mazel:2024alu,Frenkel:2025wko,Kim:2026stringloops}, with the choice of string vertices playing the role of renormalization data and the identification of spacetime fields depending on the string-field frame \cite{Mazel:2024alu,Mazel:2025diffeo}. A formulation that works slightly off the conformal locus is therefore needed both for background independence and practical calculations. Our aim is to reorganize Zwiebach's construction into a logically sharper account with three concrete outputs: recursion relations for the mixed vertices, an adaptation of the Sen-Zwiebach proof of background independence to first order in the off-shell deformation, and an explicit central-charge deformation example.

The distinction between the two kinds of punctures is central to everything that follows. At an ordinary puncture one chooses a local coordinate exactly as in critical CSFT. At a special puncture the Hermitian metric determines a preferred local coordinate up to phase through the Bergman--Zwiebach normalization reviewed in section \ref{sec:offcritical} and worked out explicitly in Appendix \ref{app:antighost-descent}. The genus-zero mixed vertices still satisfy \(\dim\Gamma_{0,n;m}=2n+3m-6\), but the extra \(m\) directions are interpolation directions in this enlarged metric bundle, not additional complex-structure moduli of the punctured sphere.

A proposal for the structure of background independent string field theory has been given in \cite{Grigoriev:2021bes}. It remains unclear exactly how the construction in this paper relates to this work -- this relationship is an interesting avenue for further investigation.

\subsection{The Off-Critical Architecture}

In the critical theory the basic data are punctured Riemann surfaces with chosen local coordinates at each puncture; off criticality, that is no longer sufficient. Correlators depend on a Hermitian metric, and the special punctures that carry the BRST-defect state must be treated differently from the ordinary punctures that carry the dynamical string field.

The dynamical field \(\Psi\) still sits at ordinary punctures and is inserted exactly as in critical CSFT. The new ingredient is a fixed Grassmann-odd state \(F\) placed at special punctures. We emphasize that $F$ is \textit{not} a dynamical string field -- it is a fixed set of parameters that enter the target space action. Its relationship to the broken BRST structure is given in section \ref{sec:offcritical}. The genus-zero action is built from mixed vertices \(\{\Psi^n;F^m\}\), or equivalently from chains \(\Gamma_{0,n;m}\), whose role is to absorb the BV boundary terms produced by the failure of BRST conservation and nilpotency. At $m=1$ these are in fact the same interpolating spaces appearing in the proof of background independence \cite{Sen:1993mh}. These mixed vertices are the minimal geometric data needed to make the BV equation close in the presence of the off-critical defect, and the full metric-dependent BRST identity simultaneously keeps track of the trace/Weyl response of the correlator. The dimension of $\Gamma_{0,n;m}$ differs from the usual moduli space dimension of $n+m$ punctured spheres: $\dim \Gamma_{0,n;m}=2n+3m-6$. The extra $m$ directions come from the metric dependence of the special local coordinates and are the geometric data needed to absorb the off-critical boundary terms.

At genus zero the relevant geometric space is the enlarged bundle \(\wh{\cP}^{\,\omega}_{0,n+m}\) of punctured spheres equipped with a Hermitian metric, ordinary local coordinates, and `metric-adapted' special local coordinates; section \ref{sec:offcritical} gives the precise definition. At an ordinary puncture the local coordinate map $f_i(w_i)$ is chosen independently, exactly as in critical CSFT. At a special puncture the Hermitian metric determines a preferred local coordinate map $g_a(w_a)$ up to a constant phase by the Bergman--Zwiebach \cite{Bergman:1994qq} normalization reviewed in subsection \ref{subsec:states-weyl}: concretely, if the metric is written locally as \(ds=\rho^{w_a}(w_a,\bar w_a)|dw_a|\), then \(\rho^{w_a}(0)=1\) and all pure holomorphic and antiholomorphic derivatives of \(\rho^{w_a}\) vanish at the origin.

The identification of the local coordinate bundle with data from the Weyl frame bundle around special punctures is essential for defining new string vertices that interpolate between Weyl frames. Because the off-critical correlator depends on the global Hermitian metric, we construct an operator-valued one-form \(\cB\) \eqref{eq:descent}, that we call the descent operator for reasons that will be explained in Section \ref{sec:offcritical}, that acquires new contour terms built from the metric-determined maps \(g_a[h]\) at the special punctures; the precise formula is \eqref{eq:Boffcritical}. The mechanism is worth emphasizing: \(\cB\) sees the metric dependence of the special punctures not through an additional independent Weyl modulus, but because the special local coordinates themselves are fixed by the metric. Under a metric variation the induced change of the special local coordinates enters through the contour term built from \(\delta g_a[h]\), while the correlator varies because it is defined globally with respect to \(h\). 


The failure of BRST symmetry is encoded by a family of worldsheet forms related by the action of the descent operator $\mathcal{B}$ \eqref{eq:descent}. The essential point is that the same state \(F\) that measures the BRST defect also enters the geometry of the action through the mixed vertices. Section \ref{sec:offcritical} states precisely how the descendants are transgressed to bundle forms and how the Weyl-response completion enters. By `transgress' we mean the natural map of a form from the worldsheet cotangent space spanned by $dz, d\bar{z}$ to the fiber bundle cotangent subspace spanned by $\delta z, \delta \bar{z}$ (given by replacing the $d$'s with $\delta$'s).

The paper then proceeds in three steps. We first rewrite Zwiebach's genus-zero construction with special punctures defined in terms of metric-adapted local geometry. We then derive the corrected mixed-vertex recursion relations and prove the existence of the interpolation spaces. Finally, we test the formalism in the central-charge deformation, where the special state is the antighost dilaton and the descendant computation can be carried out explicitly.

\subsection{Summary of Results}

The key question is: what replaces the conserved, nilpotent BRST charge once conformal invariance is lost? Zwiebach's answer is that the failures of conservation and nilpotency are not eliminated but absorbed---encoded by a single fixed Grassmann-odd state carried at special punctures. In the formulation adopted here those punctures are tied to metric-adapted local coordinates, making the geometric origin of the descendants explicit. The paper's main conclusions are as follows.
\begin{enumerate}
\item At genus zero, the mixed vertices with $n$ ordinary punctures and $m$ special punctures, denoted by $\Gamma_{0,n;m}$ in section \ref{sec:offcritical}, satisfy recursion relations obtained by replacing exact BRST nilpotency with the defect terms carried by the special punctures. In particular, the sewing terms split ordinary punctures against ordinary punctures and special punctures against special punctures; this is the corrected index assignment used throughout the paper.
\item The first-order background-independence argument in \ref{sec:bgindep} extends the proof of Sen--Zwiebach \cite{Sen:1990hh,Sen:1993mh} off of the conformal locus once one identifies a ghost-number-two local worldsheet insertion $\cO_x$ associated with an infinitesimal deformation of the reference worldsheet theory $x$, together with the induced special state obtained from its BRST variation.
\item A possible all-orders extension is discussed in section \ref{sec:higher-order-bgi}, but it remains conditional on controlling the deformation $\cO_x$ in the presence of special punctures, the short-distance definition of its BRST variation, and possible extra contact terms.
\item For a matter CFT with central charge $c_m=26+\Delta c_m$, where $\Delta c_m$ measures the deviation from the critical value, the special state is the antighost dilaton. In this example the formalism is explicit enough to show that flat space has no linearized solution and that nearby linear-dilaton backgrounds obey the exact bookkeeping relation $\Delta c_m=12\beta\eps-12\eps^2$.
\end{enumerate}

\paragraph{Standing assumptions.}
The general construction assumes a contour-dependent odd operator \(Q_B[\gamma]\) attached to every oriented contour \(\gamma\) on the punctured worldsheet, together with a fixed Grassmann-odd local insertion \(F^{[0]}\) and local worldsheet $\mathcal{B}$-descendants \(F^{[1]}\), \(F^{[2]}\) measuring the failures of nilpotency and contour independence. In the abstract formalism \(Q_B[\gamma]\) is part of the off-critical worldsheet data; the paper does not claim a first-principles construction for an arbitrary non-conformal worldsheet theory, but instead takes the descendant family stated in section \ref{sec:offcritical} as given. In the off-critical central-charge example of section \ref{sec:deltac}, by contrast, \(Q_B[\gamma]\) is written explicitly as a contour integral of off-critical BRST currents, and the formulas \eqref{eq:jBheuristic}--\eqref{eq:jBheuristicDiv} serve as a working hypothesis for the current-level realization. The main subtlety is the short-distance regularization needed to interpret \(Q_B[\gamma]^2\), which we do not develop in full generality. The formalism treats \(Q_B[\gamma]\) and the special string field state $F$ satisfying \eqref{eq:descent} as input data.

The same operator-valued one-form \(\cB\) that enters the vertex measure also maps the local descendants \(F^{[1]}\), \(F^{[2]}\) to bundle forms along the motion of the special puncture, while the full bundle-level BRST identity further involves the local trace insertion \(\Theta\) governing infinitesimal Weyl variation. Precise statements of the transgression relation, the Weyl-response completion, and the bundle/worldsheet grading conventions are given in section \ref{sec:offcritical}.

For the background-independence analysis we further assume the existence of ghost-number-two local worldsheet insertions \(\cO_x\), where \(x\) labels the reference worldsheet theory, together with their local descendants and the associated deformation insertion \(\phi_x\). The precise deformation is formulated in section \ref{sec:bgindep}. Section \ref{sec:deltac-linear} presents a linear dilaton example in which these assumptions can be examined concretely, but they remain genuine assumptions in the general off-shell worldsheet theory.

\paragraph{Notation.}
Throughout the paper, $\Psi$ denotes the dynamical string field and $F$ the fixed off-critical state. When the same symbol is repeated, as in $\{\Psi^n;F^m\}$, it means repeated insertion of the same state at all punctures of that type. The integers $n$ and $m$ count ordinary and special punctures, so $\Gamma_{0,n;m}$ is a genus-zero chain with $n$ ordinary punctures and $m$ special punctures, and the semicolon in $\{\Psi^n;F^m\}$ records that the two types of punctures are not treated symmetrically. A vertical bar denotes one distinguished special puncture inserted asymmetrically: in the unprimed chains $\Gamma_{0,n|1;m}$ this distinguished puncture is obtained by promoting one of the ordinary punctures, while in the primed chains $\Gamma'_{0,n|1;m}$ it is created independently in the complement of the ordinary coordinate disks. The symbol $x$ labels a reference worldsheet theory, $\cO_x$ the local insertion representing an infinitesimal deformation of that theory, and $\phi_x$ the corresponding two-dimensional deformation insertion. Symbols such as \(\delta m^\mu\), \(\delta\tau^k\), \(\delta z_i\), and \(\delta\bar z_i\) denote bundle one-forms, while unadorned \(dz\) and \(d\bar z\) denote worldsheet differentials.

\paragraph{Organization.}
The rest of the paper is organized as follows. Section \ref{sec:critical} recalls the genus-zero geometry of critical CSFT in the minimum form needed later. Section \ref{sec:offcritical} introduces the metric-adapted special punctures and derives the corrected recursion relations for the mixed vertices. Section \ref{sec:bgindep} reformulates local background independence in a way that interfaces directly with the off-critical construction, but only at first order in the deformation parameter. Section \ref{sec:deltac} treats the clean conformal example with nonzero central-charge defect, and section \ref{sec:deltac-linear} then applies the same formalism to the nearby nearly marginal linear-dilaton sector. Section \ref{sec:discussion} returns to the broader interpretation, including the heuristic relation to Weyl-orbit averaging and a tentative higher-order extrapolation.

\section{Critical Closed String Field Theory in Brief}
\label{sec:critical}

\subsection{Forms on Families of Punctured Spheres}

Let $\cH$ denote the BRST state space of a reference conformal background, restricted by $b_0^- \Psi = L_0^- \Psi = 0$.
Here $b_0^-:=b_0-\tilde b_0$ and $L_0^-:=L_0-\tilde L_0$ are the usual left-right antisymmetric zero modes. The basic geometric space is the bundle $\wh{\cP}_{0,n}\to \cM_{0,n}$ whose base is the moduli space of $n$-punctured spheres and whose fiber records a choice of local coordinate disk at each puncture.

For a family of punctured spheres equipped with local coordinates $z=f_i(w_i)$ at the punctures, and with real parameters $\tau^k$ on the family, define the operator-valued differential form
\begin{equation}
\cB
=
\sum_k \delta\tau^k\,\cB_{\tau^k}
-\sum_i \oint_{\partial D_i}
\left(
\frac{dz}{2\pi i}\,\delta f_i(w_i)\,b(z)
-\frac{d\bar z}{2\pi i}\,\delta \bar f_i(\bar w_i)\,\tilde b(\bar z)
\right). \bluecheck
\label{eq:Bdef}
\end{equation}
where $D_i$ is the unit disk in the local coordinate $w_i$, $\delta f_i(w_i):=\sum_k \delta\tau^k\,\partial_{\tau^k}f_i(w_i)$ is the infinitesimal change of the local coordinate map along the family, and $\cB_{\tau^k}$ is the usual $b$-ghost insertion paired with the Beltrami differential for the variation along the family coordinate $\tau^k$. Throughout the paper $\cB$ denotes the operator-valued one-form itself. It has form degree one but is Grassmann even: the $b$-ghost contour is Grassmann odd, and the accompanying differential one-form is also odd. For a tangent vector $v$ on the parameter space, we write $\cB(v):=\iota_v\cB$ for the contracted contour operator. Unlike $\cB$ itself, the contracted operator $\cB(v)$ is Grassmann odd, because the one-form factor has been removed. If $\cO$ is a local worldsheet insertion at a puncture, then the descendant one-form $\cB\cO$ is defined by $(\cB\cO)(v):=\cB(v)\cO$. Thus $\cB\cO$ is a local-insertion-valued one-form, and $\cB^2\cO$ is a local-insertion-valued two-form.

When $\cB$ acts on products of insertion-valued differential forms, the sign comes from two sources: the contracted operator $\cB(v)$ is Grassmann odd, and the resulting one-form must be moved past any pre-existing form degree. Equivalently, if $A$ is an insertion-valued $p$-form of Grassmann parity $|A|_{\mathrm{Gr}}$, then
\begin{equation}
\cB(AB)=(\cB A)B+(-1)^{|A|_{\mathrm{Gr}}+p}A\,\cB B. \bluecheck
\label{eq:Bleibniz}
\end{equation}
This is the sign convention used throughout the paper. For a matter insertion $\cO_{\mathrm m}$ on a locally flat surface, one finds for example
\begin{equation}
\cB(c\tilde c\,\cO_{\mathrm m})
=
\tilde c\,\cO_{\mathrm m}\,\delta z_i
-c\,\cO_{\mathrm m}\,\delta\bar z_i,
\qquad
\cB^2(c\tilde c\,\cO_{\mathrm m})
=
2c\tilde c\,\cO_{\mathrm m}\,\delta z_i\wedge \delta\bar z_i. \bluecheck
\label{eq:Bexample}
\end{equation}

Given $n$ string-field insertions, write $\underline\Psi = \Psi_1 \otimes \cdots \otimes \Psi_n$ for the ordered tensor product of all external states. The standard CSFT differential form\footnote{We follow the notational convention of Xi Yin's string book, publicly available \href{https://github.com/xiyin137/stringbook}{here.}} on $\wh{\cP}_{0,n}$ is
\begin{equation}
\Omega[\underline\Psi]
=
\frac{1}{(-2\pi i)^{n-3}}
\left\langle
e^{\cB}
\prod_{i=1}^n [\Psi_i(0)]^{f_i}
\right\rangle. \bluecheck
\label{eq:Omega}
\end{equation}
Here $[\Psi_i(0)]^{f_i}$ means that the local worldsheet insertion corresponding to the state $\Psi_i$ is placed at the origin of the local coordinate $w_i$ and then mapped to the punctured sphere by $z=f_i(w_i)$. The angle brackets denote the worldsheet correlator in the reference CFT on the underlying punctured sphere. The factor $(-2\pi i)^{-(n-3)}$ is the standard genus-zero normalization for a sphere with $n$ punctures.

Because the background is conformal, the BRST charge is conserved and nilpotent, and one has the fundamental identity
\begin{equation}
\Omega[Q_B\underline\Psi] = -\delta\,\Omega[\underline\Psi]. \bluecheck
\label{eq:QBcritical}
\end{equation}

\subsection{String Vertices and the Genus-Zero Master Equation}

Let $\Gamma_{0,n}\subset \wh{\cP}_{0,n}$ be a genus-zero string vertex, meaning an oriented real $(2n-6)$-dimensional chain with $n$ ordinary punctures. The associated multilinear functional is
\begin{equation}
\{\Psi_1,\ldots,\Psi_n\}
:=
\int_{\Gamma_{0,n}}\Omega[\underline\Psi]. \bluecheck
\label{eq:vertexcritical}
\end{equation}
We use the standard notation
\begin{equation}
\{\Psi^n\}:=\{\Psi,\ldots,\Psi\}.
\end{equation}
At low points,
\begin{equation}
\{\Psi\}=0,
\qquad
\{\Psi^2\}=\bra{\Psi}c_0^-Q_B\ket{\Psi}. \bluecheck
\label{eq:lowcriticalvertices}
\end{equation}
Then the classical action is
\begin{equation}
S_0^{\mathrm{crit}}[\Psi]
=
\sum_{n\ge 2}\frac{1}{n!}\{\Psi^n\}
=
\frac{1}{2}\bra{\Psi}c_0^-Q_B\ket{\Psi}
+\sum_{n\ge 3}\frac{1}{n!}\{\Psi^n\}. \bluecheck
\label{eq:criticalaction}
\end{equation}
Geometrically, gauge invariance is equivalent to the statement that the vertices satisfy
\begin{equation}
\partial \Gamma_0 + \frac{1}{2}\{\Gamma_0,\Gamma_0\}=0,
\qquad
\Gamma_0 := \sum_{n\ge 3}\Gamma_{0,n}. \bluecheck
\label{eq:mastercritical}
\end{equation}
where $\{\cdot,\cdot\}$ denotes the usual twist-sewing antibracket on chains in the bundle of punctured spheres with local coordinates: one ordinary puncture from each chain is sewn and the relative twist angle is integrated over \cite{Zwiebach:1992ie}.

\diffnote{Unlike \texttt{main.tex}, this review does not use the displayed formulas around \texttt{eqn:local-coordinate-boundary} (2.8), \texttt{eqn:twist-sewing-commutator} (2.10), and \texttt{eqn:vanilla-main-identity} (2.24), because those formulas were truncated or had incorrect dimension/sign factors. The argument here starts directly from the standard genus-zero master equation \eqref{eq:mastercritical}.}

\section{Genus-Zero Off-Critical SFT}
\label{sec:offcritical}

The off-critical genus-zero construction has four logically distinct layers: the geometric bundle on which the forms live (punctured spheres equipped with Hermitian metric and metric-adapted special coordinates); the BRST-defect data attached to that geometry (the contour-dependent charge \(Q_B[\gamma]\), the local worldsheet descendants \(F^{[1]},F^{[2]}\), and the Weyl-response completion); the mixed vertex forms and the action built from them; and finally the corrected bundle identity and recursion relations obtained by pushing BRST contours through those correlators. We explain these layers in order.

\subsection{States, Operators, and Metric-Adapted Special Coordinates}
\label{subsec:states-weyl}

\begin{figure}[ht]
\centering
\includegraphics[width=0.35\textwidth]{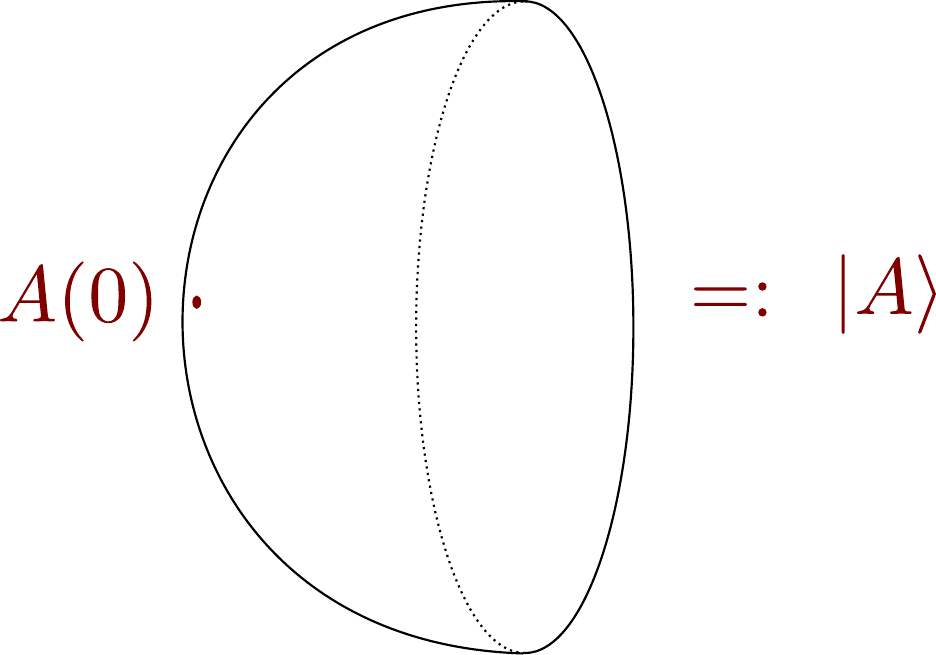}
\caption{A local worldsheet insertion $A$ prepares a state $\ket{A}$ only after one chooses the local metric data needed to identify a neighborhood of the puncture with a cylindrical end carrying the standard closed-string boundary circle. Changes of this auxiliary choice are absorbed into the geometric data of the string vertex.}
\label{fig:stateop}
\end{figure}

Away from a conformal background two issues invisible in the critical theory must be confronted. First, correlators depend on a choice of Hermitian metric, not just on the complex structure. Second, a puncture carries a Hilbert-space state only after one specifies the local geometry interpolating between the puncture and a cylindrical end ---the data needed to match the puncture to the boundary circle on which the closed-string Hilbert space is defined. In particular, the Kontsevich--Segal axioms \cite{Kontsevich:2021dmb} only allow states to be inserted on boundaries that have some neighborhood whose metric is that of a finite-length cylinder. Throughout the paper we use ``state'' for an element of the closed-string Hilbert space and ``local insertion'' for a local worldsheet field; the two are related by Euclidean evolution over a hemisphere in a chosen Hermitian metric, in the spirit of the Kontsevich--Segal axioms. Figure \ref{fig:stateop} illustrates this state-preparation picture.

\begin{figure}[ht]
    \centering
    \begin{subfigure}[b]{0.48\textwidth}
        \centering
        \includegraphics[width=0.82\textwidth]{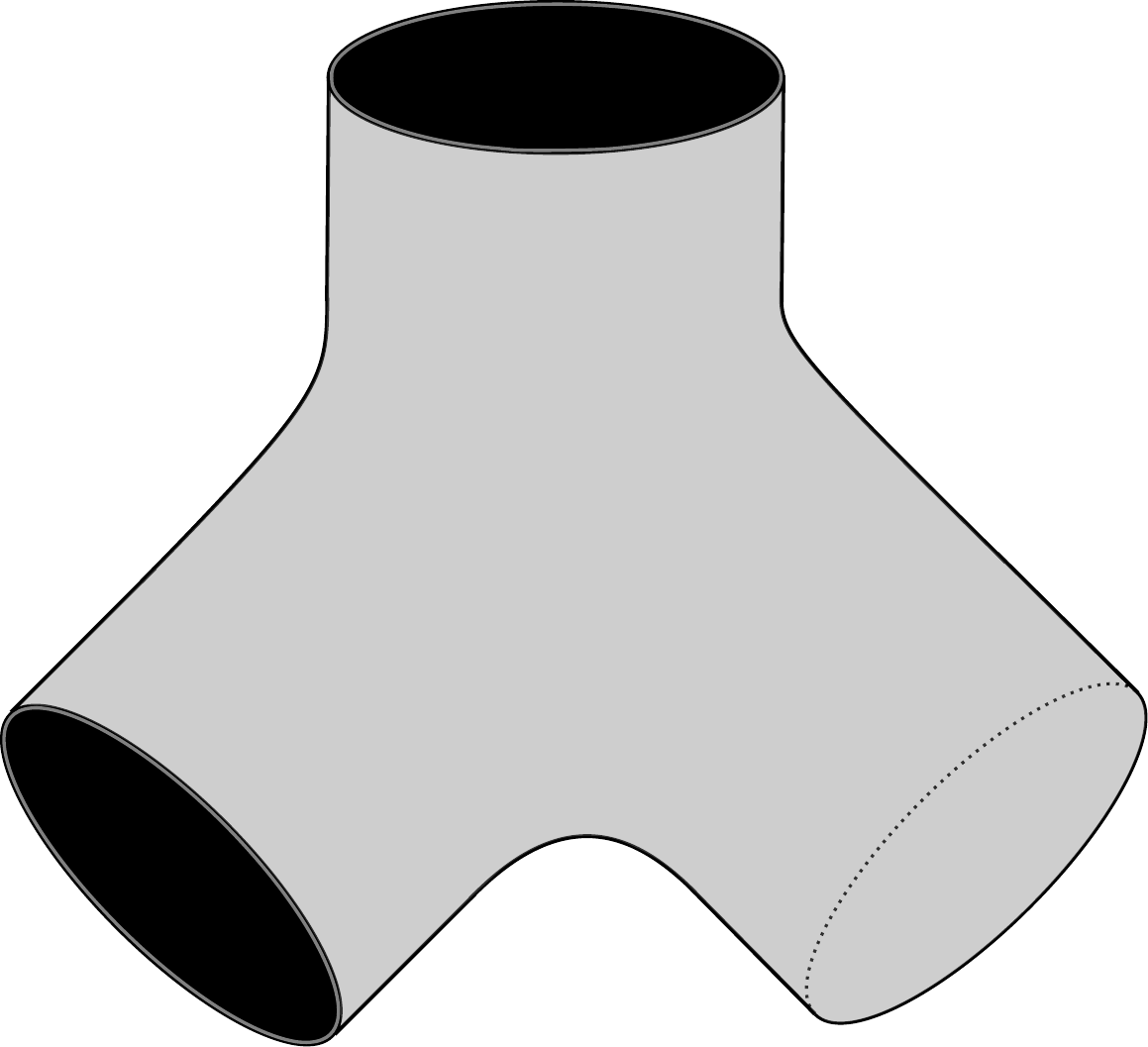}
        \caption{Three-string vertex with Weyl-framed ordinary punctures.}
    \end{subfigure}
    \hfill
    \begin{subfigure}[b]{0.48\textwidth}
        \centering
        \includegraphics[width=0.82\textwidth]{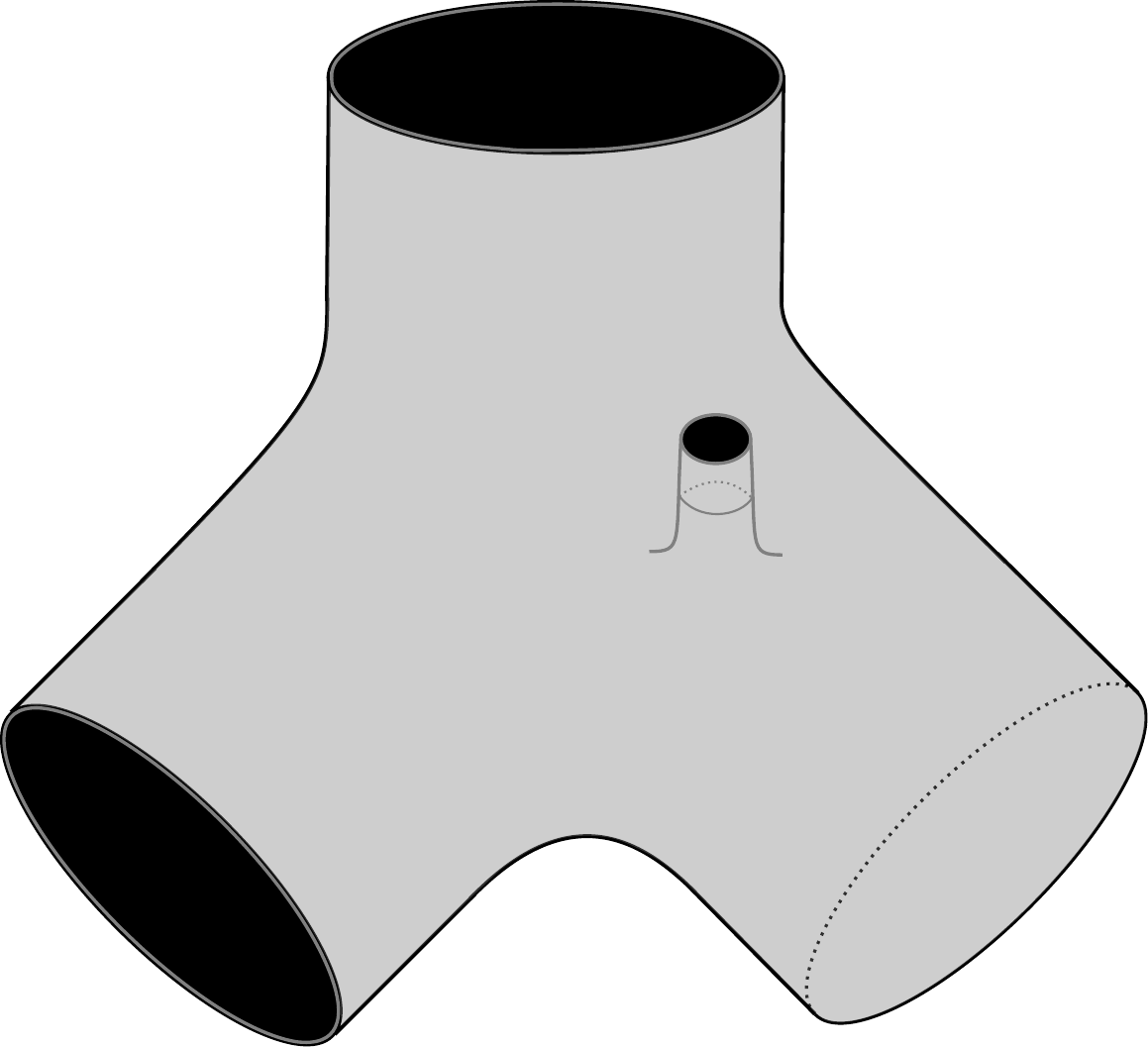}
        \caption{Four-string vertex with a movable ordinary puncture.}
    \end{subfigure}
    \caption{Weyl-framed string vertices. At each ordinary puncture the local geometry must interpolate to a cylindrical end so that the puncture carries the standard closed-string boundary circle. This is the geometric content of the Kontsevich--Segal requirement that state-carrying boundaries admit cylindrical neighborhoods.}
    \label{fig:weylvertex}
\end{figure}

\begin{figure}[ht]
\centering
\includegraphics[width=0.6\textwidth]{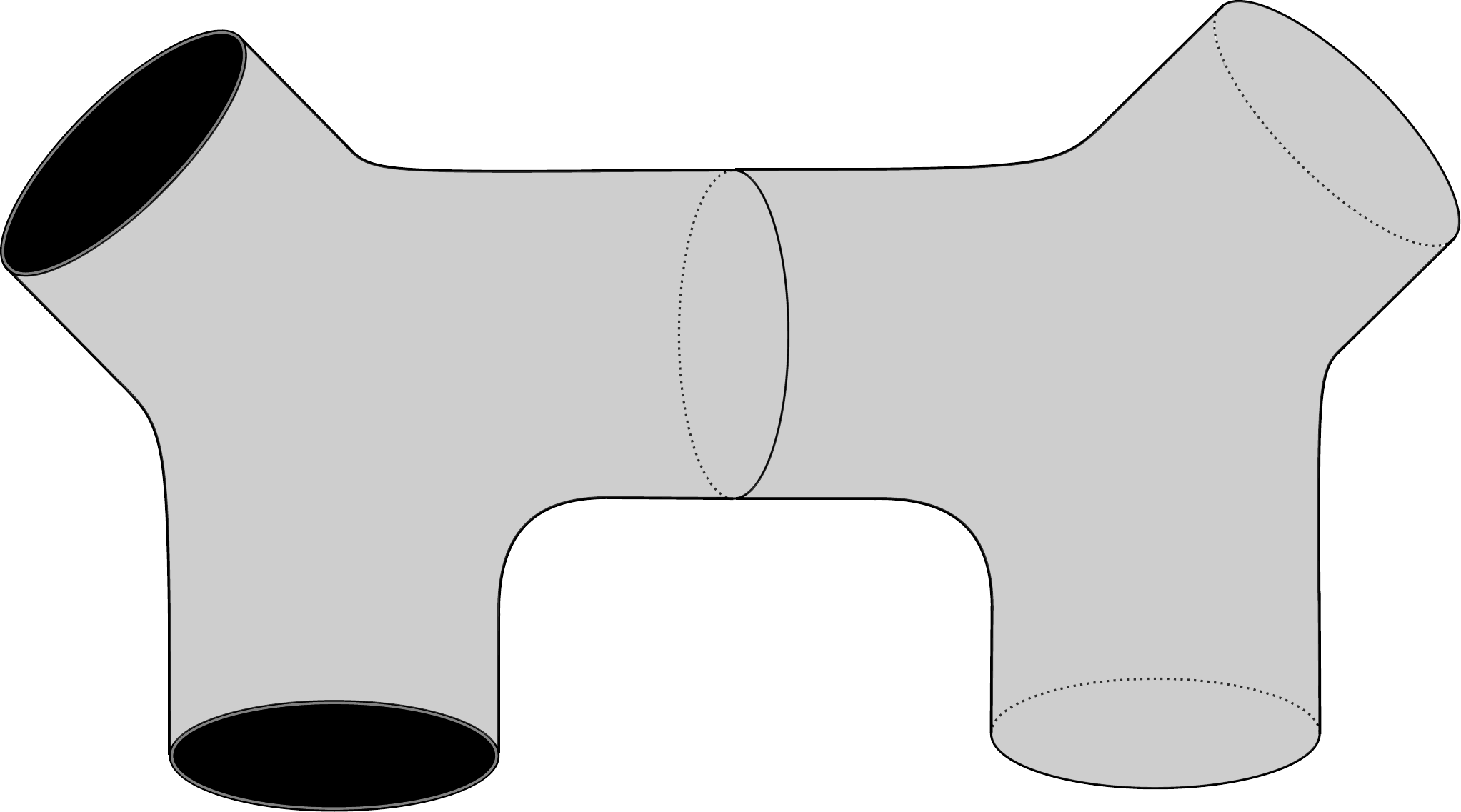}
\caption{Twist-sewn three-point vertices with their Weyl frame. This sewn worldsheet is part of the boundary data for the four-string vertex in figure \ref{fig:weylvertex}(b): near a sewing degeneration, the Weyl-framed four-string vertex must approach this geometry.}
\label{fig:sewnweylvertex}
\end{figure}

$\{\Psi^n;F^m\}$ therefore live in an enlarged bundle \(\wh{\cP}^{\,\omega}_{0,n+m}\) whose points are punctured spheres equipped with a Hermitian metric. $\omega$ in ${\cP}^{\,\omega}_{0,n+m}$ is the Weyl frame defined by metric element $ds = e^{\omega(z,\bar{z})}$. Thus an ordinary puncture consists not just of a marked point and a local-coordinate disk \(z=f_i(w_i)\), but also carries a local Weyl frame that extends that disk to a cylindrical end. Figure \ref{fig:weylvertex} shows this for the three-string and four-string vertices. In particular, the twist-sewn three-point worldsheet of figure \ref{fig:sewnweylvertex} is part of the boundary data that the four-string vertex must approach near a sewing degeneration, so the Weyl frame of the four-string vertex is constrained by the same boundary condition. These cylindrical-end conditions are imposed only on ordinary punctures, since they are the punctures that carry dynamical string states and participate in twist sewing. A special puncture is different. If \(p_a\) is a special puncture and the metric is written locally as \(ds=\rho^{w_a}(w_a,\bar w_a)|dw_a|\), then the Bergman--Zwiebach prescription \cite{Bergman:1994qq} fixes the local coordinate \(w_a\) up to constant phase by
\begin{equation}
ds=\rho^{w_a}(w_a,\bar w_a)|dw_a|,
\qquad
\rho^{w_a}(0)=1,
\qquad
\partial_{w_a}^n\rho^{w_a}|_{w_a=0}
=
\partial_{\bar w_a}^n\rho^{w_a}|_{w_a=0}
=0
\quad (n\ge 1). \bluecheck
\label{eq:specialmetriccoord}
\end{equation}
Equivalently, the metric determines a normalized local-coordinate map \(z=g_a(w_a)\) at each special puncture. Appendix \ref{app:antighost-descent} explicitly shows how these relations input the worldsheet geometry into the action of $\mathcal{B}$.

More precisely, a point of \(\wh{\cP}^{\,\omega}_{0,n+m}\) is an isomorphism class of data
\begin{equation}
(\Sigma;\,p_1,\dots,p_n;\,q_1,\dots,q_m;\,h;\,f_1,\dots,f_n),
\end{equation}
where \(\Sigma\) is a genus-zero Riemann surface, \(p_i\) are the ordinary punctures, \(q_a\) are the special punctures, \(h\) is a Hermitian metric on \(\Sigma\), and each \(f_i\) is a holomorphic embedding of the unit disk with \(f_i(0)=p_i\). Two such collections are identified if they are related by a biholomorphism of \(\Sigma\) that preserves the punctures, pulls back one Hermitian metric to the other, and matches the local-coordinate germs modulo independent constant phase rotations of the local disks. At each special puncture \(q_a\), the metric \(h\) determines a local coordinate \(w_a\) and hence a holomorphic embedding \(g_a[h]\) by \eqref{eq:specialmetriccoord}, unique up to constant phase rotation. As in ordinary CSFT, the hat indicates the quotient by these phase rotations. The projection \(\wh{\cP}^{\,\omega}_{0,n+m}\to\cM_{0,n+m}\) forgets the metric and all local-coordinate data, retaining only the underlying punctured sphere.

This is also how \(\cB\) knows about Weyl structure. On a family in \(\wh{\cP}^{\,\omega}_{0,n+m}\), with ordinary local-coordinate maps \(f_i\) and metric-adapted special maps \(g_a\), the operator-valued one-form takes the explicit off-critical form
\begin{equation}
\begin{split}
\cB
=&\,
\sum_\mu \delta m^\mu\,\cB_{m^\mu}\\
&-\sum_{i=1}^n \oint_{\partial D_i}
\left(
\frac{dz}{2\pi i}\,\delta f_i(w_i)\,b(z)
-\frac{d\bar z}{2\pi i}\,\delta \bar f_i(\bar w_i)\,\tilde b(\bar z)
\right)\\
&-\sum_{a=1}^m \oint_{\partial D_a}
\left(
\frac{dz}{2\pi i}\,\delta g_a[h](w_a)\,b(z)
-\frac{d\bar z}{2\pi i}\,\delta \bar g_a[h](\bar w_a)\,\tilde b(\bar z)
\right),
\end{split}
\label{eq:Boffcritical}
\end{equation}
where \(m^\mu\) are coordinates on the moduli space of the underlying punctured sphere. Along a pure metric variation the Beltrami term vanishes, but the contour term built from \(\delta g_a[h]\) still contributes to \(\cB\). See appendix \ref{app:antighost-descent} for an example of the technology that converts the metric-adapted local coordinates to explicit differentials along the Weyl fibers. When an additional BRST-anomaly insertion is included due to shrinking a $Q_B$ contour to zero (see for example the definition of \(\mathfrak{k}\) in \eqref{eq:kOmega} below), the same metric-adapted contour term also generates the bundle-degree completion required by the global Weyl response of the correlator. Special punctures therefore do not carry an independent extra Weyl modulus. Rather, the metric itself determines their local-coordinate data. The twist-sewing antibracket still acts only on ordinary punctures.

The same geometric distinction also admits a more heuristic interpretation given in \ref{sec:discussion}, relating the off-shell vertices to explicit averages of Weyl gauge orbits.

\subsection{Broken BRST Structure}

\paragraph{Formal input.}
With the geometric setting in place we can state the algebraic structure that replaces BRST nilpotency. Because the BRST current is no longer conserved, the charge becomes contour-dependent: one must write \(Q_B[\gamma]\) rather than \(Q_B\). In the general formalism \(Q_B[\gamma]\) is assumed abstractly as part of the worldsheet data; in the central-charge deformation of section \ref{sec:deltac} it is realized explicitly as a contour integral of off-critical BRST currents. Zwiebach's construction then posits a fixed Grassmann-odd state \(F \equiv F^{[0]}\) together with descendants $F^{[1]}$ and $F^{[2]}$ satisfying
\begin{equation}
\lim_{\gamma\to 0} Q_B[\gamma]\,F^{[0]}(0)=0,
\qquad
Q_B[\gamma]^2 = \frac{1}{2\pi i}\oint_\gamma F^{[1]},
\qquad
Q_B[\gamma_2]-Q_B[\gamma_1]
=
-\frac{1}{2\pi i}\int_{M}F^{[2]},
\bluecheck
\label{eq:descent}
\end{equation}
Here \(M\) denotes the oriented two-dimensional region swept between the contours, with $\partial M=\gamma_2-\gamma_1$.
\eqref{eq:descent} does not by itself completely define the data attached to the special state \(F\): the Weyl-response completion \eqref{eq:F2Weyl} introduced below is also part of the off-critical input. In the central-charge deformation studied in \S\ref{sec:deltac}, our working hypothesis is that the BRST current formulas \eqref{eq:jBheuristic} and \eqref{eq:jBheuristicDiv} furnish the BRST current-level realization of the descendant relations \eqref{eq:descent} and the Weyl-response completion \eqref{eq:F2Weyl}. The main subtlety is the treatment of the short-distance terms implicit in \(Q_B[\gamma]^2\): products of contour-integrated currents can generate power-divergent contact terms, and a complete derivation of \eqref{eq:descent}--\eqref{eq:F2Weyl} would require a consistent regularization of those divergences. We believe such a regularization exists, but we do not attempt to formulate it in full generality in this paper. Accordingly,  the family \eqref{eq:descent}--\eqref{eq:F2Weyl} is taken as part of the off-critical input.

\paragraph{Local descendants and their transgression.}
To state the remaining off-critical input precisely, let
\begin{equation}
\pi:\cU^{\,\omega}_{0,n;m}\to \wh{\cP}^{\,\omega}_{0,n+m}
\end{equation}
denote the universal punctured surface whose fiber over a point
\begin{equation}
(\Sigma;\,p_1,\dots,p_n;\,q_1,\dots,q_m;\,h;\,f_1,\dots,f_n)
\end{equation}
is the punctured worldsheet \(\Sigma\setminus\{p_i,q_a\}\). Then \(F^{[1]}\) is a local-insertion-valued worldsheet one-form on \(\cU^{\,\omega}_{0,n;m}\), and \(F^{[2]}\) is a local-insertion-valued worldsheet two-form on \(\cU^{\,\omega}_{0,n;m}\). They are not bundle forms. If a family carries one distinguished special puncture \(q\), then every tangent vector \(v\) to the bundle induces an infinitesimal displacement \(\dot q(v)\in T_q\Sigma\). The bundle descendants obtained from the local descendants are the transgressed forms
\begin{equation}
(\cB F^{[0]})(v)=\iota_{\dot q(v)}F^{[1]},
\qquad
\left(\frac12\cB^2 F^{[0]}\right)(v_1,v_2)=\iota_{\dot q(v_2)}\iota_{\dot q(v_1)}F^{[2]}. \bluecheck
\label{eq:Bdescent}
\end{equation}
The superscript $[p]$ records worldsheet form degree: $F^{[0]}$ is a local insertion placed at a special puncture, $F^{[1]}$ is its local worldsheet one-form descendant integrated along contours, and $F^{[2]}$ is its local worldsheet two-form descendant integrated over two-dimensional regions. Thus \(\cB\) that supplies measure factors in \eqref{eq:Omega} (and \eqref{eq:OmegaF} below) also converts the local descendant data into bundle forms. On the enlarged bundle \(\wh{\cP}^{\,\omega}_{0,n+m}\), its new components come from the metric dependence of the special local coordinates described in \eqref{eq:specialmetriccoord}.

\paragraph{Weyl-response completion.}
To make the full bundle-level BRST identity literal, one needs one more piece of descendant data. Let \(v_\omega\) be a tangent vector to \(\wh{\cP}^{\,\omega}_{0,n+m}\) whose infinitesimal deformation is a pure Weyl variation \(\delta\omega_v\) at fixed puncture position. We then assume that when the distinguished extra special puncture carrying \(F^{[2]}\) is inserted as in \eqref{eq:kOmega}, the same metric-adapted one-form \(\cB\) produces the local Weyl-response insertion:
\begin{equation}
(\cB F^{[2]})(v_\omega)=\Xi[v_\omega]. \bluecheck
\label{eq:F2Weyl}
\end{equation}
Here \(\Xi[v_\omega]\) is the local worldsheet two-form inserted at the distinguished extra puncture. Because \(\mathfrak{k}\Omega\) carries the standard extra-puncture factor together with the additional contour-orientation minus sign explained below, \(\Xi[v_\omega]\) differs by a net factor of \(+2\pi i\) from the bare Weyl-anomaly insertion that differentiates \(\Omega[\underline\Psi;F^{\otimes m}]\) itself. The universal Weyl-response insertion is therefore
\begin{equation}
\Xi[v_\omega]
:=
(2\pi i)\left(-\frac{1}{2\pi}\,\delta\omega_v\,T^a{}_a\,d^2\sigma\,\sqrt g\right)
=
-i\,\delta\omega_v\,T^a{}_a\,d^2\sigma\,\sqrt g
=
-4i\,\delta\omega_v\,\Theta\,d^2\sigma\,\sqrt g. \reasoncheck
\label{eq:F2WeylDeltac}
\end{equation}
The first equality in \eqref{eq:F2WeylDeltac} is the standard Weyl-anomaly insertion multiplied by the net factor \(+2\pi i\) required by the normalization of \(\mathfrak{k}\Omega\)\footnote{Our sign conventions differ from Zwiebach's. The genus-zero form is normalized by \((-2\pi i)^{-(n-3)}\), each additional puncture contributes one further factor of \((-2\pi i)^{-1}\), the odd ghost \(b\) anticommutes with the odd one-forms \(dz,d\bar z\), and the BRST contour that produces \(\mathfrak{k}\) is shrunk on the outside of the ordinary local-coordinate disks. The last point contributes an additional minus sign relative to the naive extra-puncture normalization. With these conventions, the standard Weyl variation \(\delta\log Z=-(2\pi)^{-1}\int d^2\sigma\,\sqrt g\,\delta\omega\,T^a{}_a\) gives the sign in \eqref{eq:F2WeylDeltac}. Readers comparing signs with Zwiebach should translate for his different bookkeeping of \(b\) and \(dz\).}; the later equalities use \(T^a{}_a=4\Theta\). The Weyl-response insertion is written in the invariant measure convention to be used in section \ref{sec:deltac-linear}, while the local contour descendants \(F^{[1]}\) and \(F^{[2]}\) continue to be written in the complex-form basis \(dz\wedge d\bar z\). Section \ref{sec:deltac} later exhibits the local ghost-to-bundle exchange mechanism in the central-charge deformation case and compares it with this universal normalization.

Both \eqref{eq:Bdescent} and \eqref{eq:F2Weyl} are part of the defining off-critical input and necessary for the consistency of the formalism. The first passes from local descendants on the universal punctured surface to the bundle forms that enter the vertex calculus; the second is needed for the technical modification of the identity \eqref{eq:QBcritical} in the off-critical case (which we will state in \eqref{eq:QBplusk}).

\paragraph{BRST current-level realization.}
In the central-charge deformation, our working hypothesis is that the off-critical BRST currents are
\begin{equation}
j_B = cT^{\mathrm m}+bc\partial c+\frac{3}{2}\partial^2 c+\tilde c\,\Theta,
\qquad
\bar j_B = \tilde c\,\bar T^{\mathrm m}+\tilde b\tilde c\bar\partial\tilde c+\frac{3}{2}\bar\partial^2\tilde c+c\,\Theta,
\label{eq:jBheuristic}
\end{equation}
for which
\begin{equation}
\bar\partial j_B+\partial\bar j_B
=
(\partial c+\bar\partial\tilde c)\,\Theta.
\label{eq:jBheuristicDiv}
\end{equation}
\eqref{eq:jBheuristicDiv} explains why the same local trace insertion \(\Theta\) appears both in the contour-defect descendant \(F^{[2]}\) and in the Weyl-response term produced by \(\cB\) on the distinguished extra special puncture.

\begin{figure}[ht]
\centering
\includegraphics[width=\textwidth]{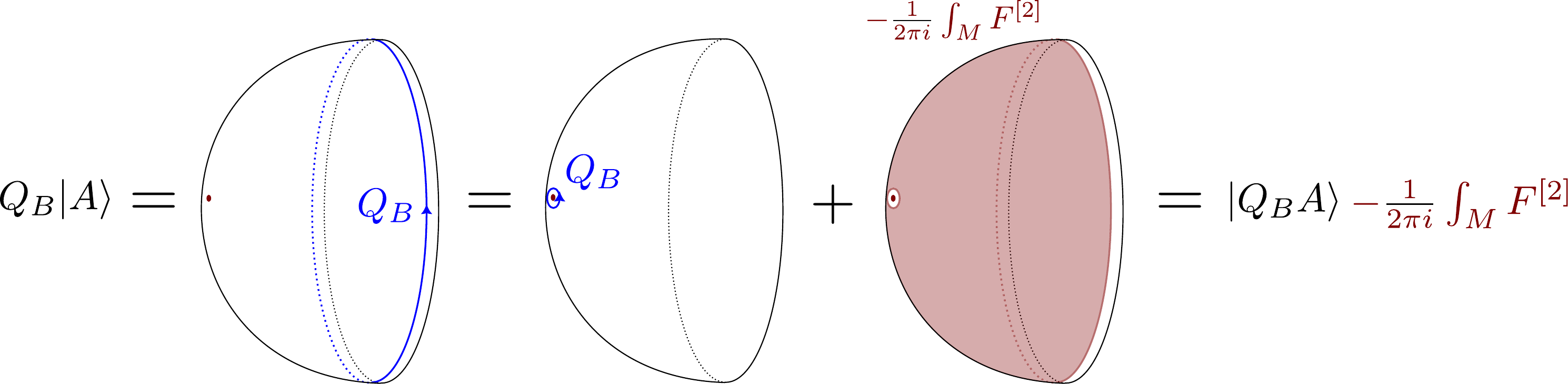}
\caption{Two notions of BRST action away from a conformal background. On the left, $Q_B$ acts on a local insertion by a contour that shrinks to the puncture. On the right, $Q_B$ acts on the corresponding state by a contour on the boundary of the hemispherical state-preparation region. The two actions differ by the integrated two-form descendant over the shaded region between the two contours.}
\label{fig:QB-action}
\end{figure}

\subsection{Mixed Vertices with Special Punctures}

\paragraph{Definition of the mixed worldsheet form.}
Given $n$ ordinary punctures carrying string fields, write
\begin{equation}
\underline\Psi:=\Psi_1\otimes\cdots\otimes\Psi_n.
\end{equation}
For $m$ special punctures carrying the same fixed state $F$, define the differential form
\begin{equation}
\Omega[\underline\Psi;F^{\otimes m}]
:=
\frac{1}{(-2\pi i)^{n+m-3}}
\left\langle
e^{\cB}
\prod_{i=1}^n [\Psi_i(0)]^{f_i}
\prod_{a=1}^m [F(0)]^{g_a[h]}
\right\rangle_{(\Sigma,h)}. \bluecheck
\label{eq:OmegaF}
\end{equation}
The subscript on the correlator emphasizes that the off-critical worldsheet expectation value is defined on the punctured sphere \((\Sigma,h)\) with its chosen global Hermitian metric. The prefactor $(-2\pi i)^{-(n+m-3)}$ is again the standard genus-zero normalization, now with the total puncture count \(n+m\). Thus \(\Omega[\underline\Psi;F^{\otimes m}]\) is a differential form on \(\wh{\cP}^{\,\omega}_{0,n+m}\): the coefficient functions vary with the global metric \(h\), and the induced variation of the special maps \(g_a[h]\) enters through the same contour term in \(\cB\) that already treats ordinary local-coordinate variations. Ordinary punctures are symmetrized, while special punctures are antisymmetrized. The corresponding vertices are chains
\begin{equation}
\Gamma_{0,n;m}\subset \wh{\cP}^{\,\omega}_{0,n+m},
\qquad
\dim \Gamma_{0,n;m} = 2n+3m-6. \bluecheck
\label{eq:dimmixed}
\end{equation}
\(\Gamma_{0,n;m}\) is an \(m\)-parameter family inside the enlarged bundle of punctured spheres equipped with a Hermitian metric and with metric-adapted special coordinates. Equivalently, compared with the ordinary \(2(n+m)-6\) puncture-position moduli, one has one additional interpolation direction for each special puncture coming from the metric dependence of its normalized local coordinate. This is the geometric content of \eqref{eq:dimmixed} used by Zwiebach \cite{Zwiebach:1996jc,Zwiebach:1996ph}.

\paragraph{Definition of the mixed vertices and action.}
We define the mixed multilinear functionals
\begin{equation}
\{\Psi_1,\ldots,\Psi_n;F^m\}
:=
\int_{\Gamma_{0,n;m}}\Omega[\underline\Psi;F^{\otimes m}]. \bluecheck
\label{eq:mixedvertex}
\end{equation}
when all fields are the same ($\Psi_1 = \ldots = \Psi_n = \Psi$), we write
\begin{equation}
\{\Psi^n;F^m\}:=\{\Psi, \ldots, \Psi ;F^m\}.
\end{equation}
The low-point conventions (we use `low-point' to refer to spheres with three or less total punctures) are
\begin{equation}
\{\Psi^2\}=\bra{\Psi}c_0^-Q_B[\gamma]\ket{\Psi},
\qquad
\{\Psi;F\}=\bra{F}c_0^-\ket{\Psi}. \bluecheck
\label{eq:lowpointdefs}
\end{equation}
The contour label on $Q_B[\gamma]$ is essential: away from conformality, moving the contour changes the BRST charge by an integrated descendant term. The second equation in \eqref{eq:lowpointdefs} simply says that with one ordinary puncture and one special puncture the vertex reduces to the BPZ pairing between the fixed special state $F$ and the dynamical string field $\Psi$.
The tree-level off-critical action is then
\begin{equation}
S_0[\Psi;F]
=
\sum_{n\ge 2}\frac{1}{n!}\{\Psi^n\}
+\sum_{m\ge 1,\, n\ge 1}\frac{1}{m!\,n!}\{\Psi^n;F^m\}.
\bluecheck
\label{eq:offaction}
\end{equation}

The presence of the linear term $\{\Psi;F\}$ means that $\Psi=0$ is not a solution when the reference worldsheet theory is genuinely off-critical: the special state $F$ records the defect directly in the string-field equation of motion.

\subsection{The Off-Critical Replacement for \texorpdfstring{$\Omega[Q_B\underline\Psi]=-\delta\Omega[\underline\Psi]$}{Omega[QBPsi]=-deltaOmega}}

\paragraph{Definition of the anomaly operator.}
The identity \eqref{eq:QBcritical} must be modified. Dragging the BRST contour across an off-critical correlator still produces the usual boundary terms, but now also a bulk anomaly. This anomaly is encoded by an operator $\mathfrak{k}$ on the differential forms $\Omega[\underline\Psi;F^{\otimes m}]$: if $D_{\mathrm{tot}}\subset\Sigma$ denotes the union of the ordinary local-coordinate disks \(D_i\), then $\mathfrak{k}\Omega[\underline\Psi;F^{\otimes m}]$ is the form obtained by inserting one additional distinguished special puncture carrying the local two-form descendant $F^{[2]}(w,\bar w)$ and integrating its worldsheet position over $\Sigma\setminus D_{\mathrm{tot}}$
\begin{equation}
\mathfrak{k}\Omega[\underline\Psi;F^{\otimes m}]
:=
-\int_{\Sigma\setminus D_{\mathrm{tot}}}
\frac{1}{(-2\pi i)^{n+m-2}}
\left\langle
e^{\cB}
\prod_{i=1}^n [\Psi_i(0)]^{f_i}
\prod_{a=1}^m [F(0)]^{g_a[h]}
F^{[2]}(w,\bar w)
\right\rangle_{(\Sigma,h)}. \bluecheck
\label{eq:kOmega}
\end{equation}
The integral in this definition is over the worldsheet position $(w,\bar w)$ of the additional special puncture insertion, not over moduli space. Equivalently, \(\mathfrak{k}\Omega[\underline\Psi;F^{\otimes m}]\) is the asymmetric mixed correlator with one additional special puncture, distinguished from the \(m\) fully antisymmetrized special punctures already present. The factor $(-2\pi i)^{-(n+m-2)}$ is the standard genus-zero normalization for the total puncture count \(n+m+1\), but the overall minus sign in \eqref{eq:kOmega} is equally important: when the BRST contour acting on an ordinary puncture is moved to the outside of the ordinary local-coordinate disks, it becomes the negatively oriented boundary of \(\Sigma\setminus D_{\mathrm{tot}}\), so the descent relation \eqref{eq:descent} produces \(+(2\pi i)^{-1}\int_{\Sigma\setminus D_{\mathrm{tot}}}F^{[2]}\) rather than \(-(2\pi i)^{-1}\int_{\Sigma\setminus D_{\mathrm{tot}}}F^{[2]}\). Because that extra puncture is metric adapted just like the others, the same factor \(e^{\cB}\) acts on its descendant insertion as well. By \eqref{eq:F2Weyl}, this produces not only the familiar vertical worldsheet two-form contribution but also the bundle-degree completion required when the global Hermitian metric varies. More explicitly, if \(v_\omega\) is a pure Weyl tangent vector, then the bundle-degree-one contribution coming from the distinguished puncture is
\begin{equation}
\bigl(\iota_{v_\omega}\mathfrak{k}\Omega\bigr)_{\rm dist}
=
-\int_{\Sigma\setminus D_{\mathrm{tot}}}
\frac{1}{(-2\pi i)^{n+m-2}}
\left\langle
e^{\cB}
\prod_{i=1}^n [\Psi_i(0)]^{f_i}
\prod_{a=1}^m [F(0)]^{g_a[h]}
\,(\cB F^{[2]})(v_\omega)
\right\rangle_{(\Sigma,h)}. \reasoncheck
\label{eq:kOmegaWeyl}
\end{equation}
Using \eqref{eq:F2Weyl}, this becomes the integrated local Weyl-response insertion
\begin{equation}
-\int_{\Sigma\setminus D_{\mathrm{tot}}}
\frac{1}{(-2\pi i)^{n+m-2}}
\left\langle
e^{\cB}
\prod_{i=1}^n [\Psi_i(0)]^{f_i}
\prod_{a=1}^m [F(0)]^{g_a[h]}
\,\Xi[v_\omega](w,\bar w)
\right\rangle_{(\Sigma,h)}.
\end{equation}
By \eqref{eq:F2WeylDeltac}, the distinguished contribution is the integrated local \(\Theta\)-insertion. Thus the \(\Theta\)-integral arises precisely from \(\cB\) acting on the distinguished \(F^{[2]}\) insertion.

\paragraph{Resulting bundle identity.}
With the definitions above, the proof of the BRST identity runs parallel to the critical case: one acts with \(Q_B[\gamma]\) on an ordinary external state, pushes the contour across the correlator, uses the standard commutator with the \(b\)-ghost measure to obtain the \(-\delta\) term, and applies Stokes' theorem to the nonconservation of the BRST current on the complement of the local-coordinate disks. The resulting integrated anomaly insertion is precisely the distinguished-extra-special-puncture contribution \(\mathfrak{k}\Omega[\underline\Psi;F^{\otimes m}]\). The local trace insertion \(\Theta\) governing Weyl response is therefore part of the same completed anomaly descendant rather than a separate correction. With this interpretation, the BRST identity becomes
\begin{equation}
\Omega[Q_B\underline\Psi;F^{\otimes m}]
=
\left(-\delta+\mathfrak{k}\right)\Omega[\underline\Psi;F^{\otimes m}]. \reasoncheck
\label{eq:QBplusk}
\end{equation} \eqref{eq:QBplusk} still relies on \eqref{eq:F2Weyl}. In the central-charge deformation example analyzed later, the current-side descendant \eqref{eq:F2current}, the local bundle-descent formula of Appendix~\ref{app:antighost-descent}, and the anomaly-normalized Weyl-response insertion agree once the contour-orientation sign in \eqref{eq:kOmega} is included. A fully general derivation of \eqref{eq:F2Weyl}, starting from the BRST current, for an arbitrary non-conformal worldsheet theory remains open.

\paragraph{Order counting.}
On a correlator that already contains $m$ copies of $F$, the operation \(\mathfrak{k}\) inserts one additional off-critical datum. We therefore assign \(\mathfrak{k}\) the same off-critical order as one extra special insertion. In the central-charge deformation of section \ref{sec:deltac}, this agrees with the fact that \(\mathfrak{k}\) is linear in the same defect parameter \(\Delta c_m\). Hence on a vertex computed only up to order $F^m$ one may still use
\begin{equation}
\int_{\Gamma_{0,n;m}}\Omega[Q_B\underline\Psi;F^{\otimes m}]
=
-\int_{\partial\Gamma_{0,n;m}}\Omega[\underline\Psi;F^{\otimes m}]
+ O(F^{m+1}). \reasoncheck
\label{eq:QBapprox}
\end{equation}

\begin{figure}[ht]
\centering
\includegraphics[width=\textwidth]{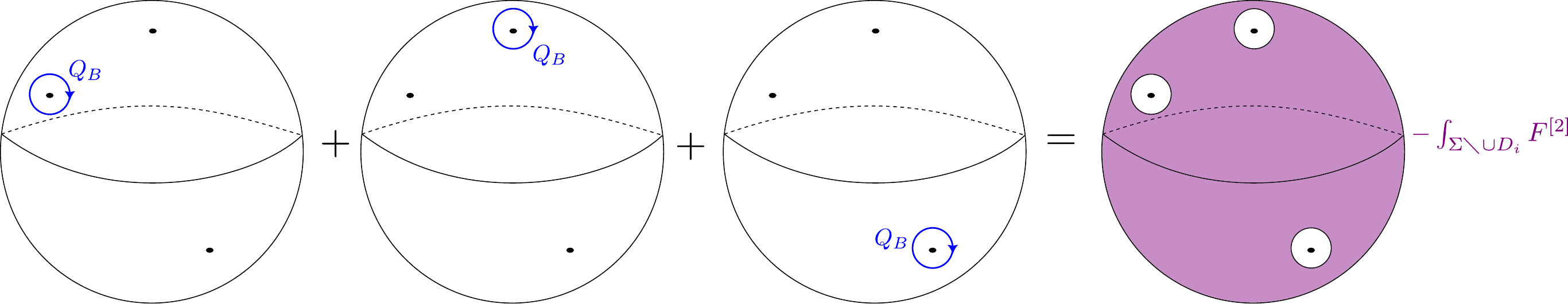}
\caption{The off-critical correction to the standard BRST-to-boundary argument. When the BRST contour is contracted, the failure of current conservation leaves behind an integrated $F^{[2]}$ insertion over the complement of the local-coordinate disks. Geometrically this behaves like an asymmetric vertex with one special puncture.}
\label{fig:qbf2}
\end{figure}

Figure \ref{fig:qbf2} illustrates the geometric meaning of the \(\mathfrak{k}\)-term. In effect, shrinking a BRST contour no longer annihilates the correlator; instead it leaves behind an integrated $F^{[2]}$ insertion on the complement of the local-coordinate disks. Geometrically this is the asymmetric insertion of one additional special puncture, and the figure shows its worldsheet projection. The same metric-adapted \(e^{\cB}\) acting on that extra puncture is what generates the bundle-degree completion needed on the full bundle \(\wh{\cP}^{\,\omega}_{0,n+m}\).

\subsection{First Order in the Special State}\label{subsec:first-order-action}

\begin{figure}[ht]
\centering
\includegraphics[width=\textwidth]{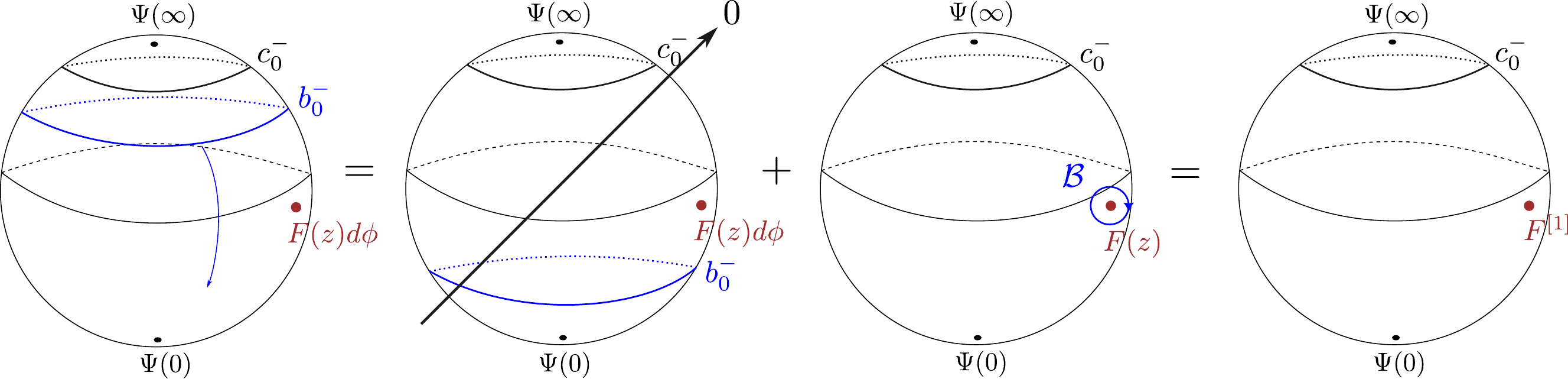}
\caption{The basic order-$F$ identity. The contour insertion of $F^{[1]}$ produced by $Q_B^2$ can be traded for an asymmetric three-string vertex with one special puncture. This is the first step in the order-by-order construction of the off-critical action.}
\label{fig:zw316}
\end{figure}

The first nontrivial consistency condition comes at order $F$. The failure of BRST nilpotency in the kinetic term produces a distinguished copy of $F$ inserted asymmetrically over the complement of the ordinary local-coordinate disks; this defines the \emph{primed} chain $\Gamma'_{0,n|1}$. The corresponding \emph{unprimed} chain $\Gamma_{0,n|1}$ arises instead by promoting one of the ordinary punctures to the distinguished special puncture and summing over all choices. Concretely, the kinetic term gives
\begin{equation}
(S_K,S_K)
=
-\bra{\Psi}c_0^-Q_B^2\ket{\Psi}
=
-\frac{1}{2\pi i}\bra{\Psi}c_0^-\oint_\gamma F^{[1]}\ket{\Psi}.
\label{eq:SKSK}
\end{equation}
For arbitrary external states \(\Psi_1,\Psi_2\), the contour insertion of the one-form descendant on the rotationally invariant choice of Weyl frame defining the inner product is exactly the asymmetric three-point vertex with one distinguished special puncture:
\begin{equation}
\frac{1}{2\pi i}\bra{\Psi_1}c_0^-\Bigl(\oint_\gamma F^{[1]}\Bigr)\ket{\Psi_2}
=
\int_{\Gamma'_{0,2|1}}\Omega[\Psi_1,\Psi_2;F]. \reasoncheck
\label{eq:F1bridge}
\end{equation}
\eqref{eq:F1bridge} is a rewrite of Zwiebach's basic order-$F$ identity \cite[sec.~3]{Zwiebach:1996jc}: the contour of \(F^{[1]}\) along the central geodesic of the canonical cylinder is traded, with the same twist-sewing normalization that defines the BPZ form, for the one-dimensional asymmetric three-point chain of figure \ref{fig:zw316}. At higher order, \eqref{eq:kKbridge} below plays the analogous role for the two-form descendant \(F^{[2]}\), where the distinguished special puncture is swept through a two-dimensional region rather than along a one-dimensional contour. Therefore the quadratic kinetic term \emph{forces} a correction linear in \(\Psi\), namely the term \(\{\Psi;F\}\) in \eqref{eq:offaction}; the BV equation requires it as soon as the BRST charge ceases to be nilpotent.

Collecting the order-$F$ terms in $(S_0,S_0)$ gives, for arbitrary external states,
\begin{equation}
\int_{\partial \Gamma_{0,n;1}}\Omega[\underline\Psi;F]
-\int_{\Gamma'_{0,n|1}}\Omega[\underline\Psi;F]
+\int_{\Gamma_{0,n|1}}\Omega[\underline\Psi;F]
+\sum_{l=1}^{n-2}
\int_{\{\Gamma_{0,l+1;1},\Gamma_{0,n-l+1}\}}
\Omega[\underline\Psi;F]
=0. \reasoncheck
\label{eq:orderFbalance}
\end{equation}
Since the integrand is arbitrary, the chain-level statement is
\begin{equation}
\partial \Gamma_{0,n;1}
=
\Gamma'_{0,n|1}
- \Gamma_{0,n|1}
- \sum_{l=1}^{n-2}\{\Gamma_{0,l+1;1},\Gamma_{0,n-l+1}\}. \reasoncheck
\label{eq:gamma1}
\end{equation}
This is precisely the genus-zero interpolation-space equation that already appeared in the Sen--Zwiebach proof of local background independence \cite{Sen:1993mh}.

\begin{figure}[ht]
\centering
\includegraphics[width=\textwidth]{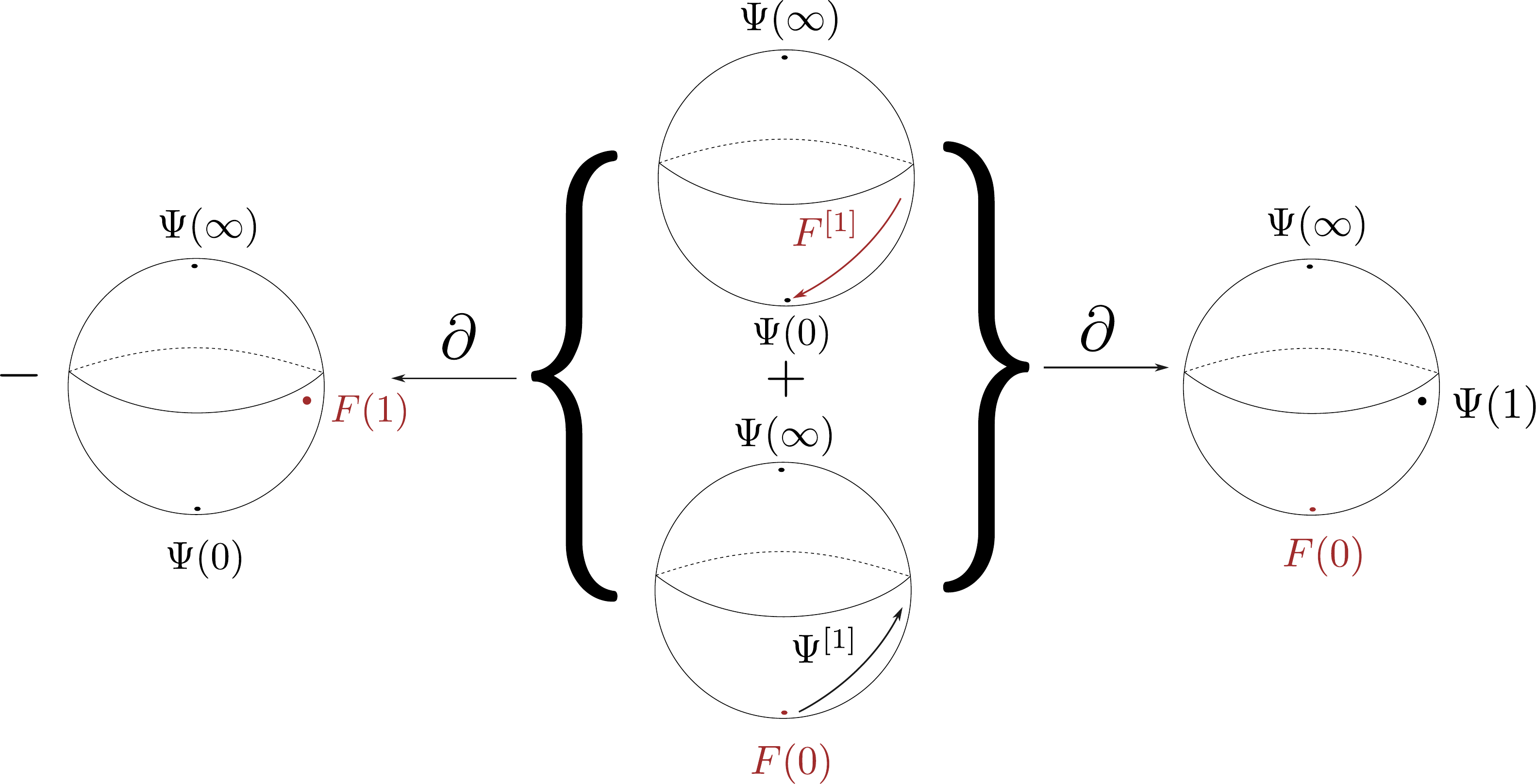}
\caption{We depict one component of the lowest-dimensional mixed interpolation space $\Gamma_{0,2;1}$, which interpolates between the primed asymmetric boundary component $\Gamma'_{0,2|1}$ and the unprimed asymmetric component $\Gamma_{0,2|1}$. In this one-dimensional example the interpolation literally exchanges the role of the ordinary puncture and the special puncture along a path on the sphere. The two other components exchange the $F$ puncture at $z=1$ and the $\Psi$ puncture at $z = \infty$, and move the $F$ puncture at $z=1$ to the choice of local coordinate frame in the three-vertex $\Gamma_{0,3}$. These three components must be avaraged, each carrying a factor of 1/3.}
\label{fig:Gamma-021-viz-streamlined}
\end{figure}

\diffnote{The dedicated \(\Gamma_{0,2;1}\) visualization from \texttt{main.tex} figure \texttt{fig:Gamma-021-viz} is included here because it lowers the reader's burden in the first nontrivial mixed-vertex example.}

\diffnote{Equation \eqref{eq:gamma1} is the clean first-order bridge between the 1996 off-critical construction and the 1993 background-independence papers. It corrects and streamlines the logic around \texttt{main.tex} formulas \texttt{eqn:SK-SK} (2.43), \texttt{eqn:Gamma-021-constraint} (2.45), and \texttt{eqn:Gamma-semicolon-1-space-boundaries} (2.59).}

\subsection{Quadratic Order and the General Recursion Relation}

At higher orders, one needs a precise dictionary between the form-level anomaly insertions and the asymmetric chains that enter the mixed-vertex recursion.
\subsubsection{From \texorpdfstring{$\mathfrak{k}$}{k} and the BV antibracket to \texorpdfstring{$\cK$}{K} and \texorpdfstring{$\cI$}{I}}
\label{subsec:KI-dictionary}


The insertion of a new special puncture carrying the special state $F$ as in $\mathfrak{k}$ \eqref{eq:kOmega} may be re-interpreted as a purely geometric operation on moduli spaces (following \cite{Sen:1993mh} and \cite{Zwiebach:1996jc}). Before stating the general recursion relations defining $\Gamma_{0,n;m}$ (and therefore all generalized vertices $\{\Psi^n;F^m\}$) we introduce geometric operations $\mathcal{K}$ and $\mathcal{I}$ to remove any reference to an explicit choice of special string state.

Write a point of \(\Gamma_{0,n;m}\) as \((\Sigma;D_1,\dots,D_n;q_1,\dots,q_m)\), where \(D_i=f_i(\{|w_i|<1\})\) are the ordinary local-coordinate disks and \(q_a\) are the special punctures. Then \(\cK\) adjoins one more special puncture \(q_*\) and integrates it over the complement of the ordinary disks while keeping it distinguished from the antisymmetrized set \(\{q_a\}\):
\begin{equation}
\Gamma'_{0,n|1;m}
=
\cK\Gamma_{0,n;m}
:=
\left\{
\begin{array}{c}
(\Sigma;D_1,\dots,D_n;q_1,\dots,q_m\mid q_*)
\\[2pt]
(\Sigma;D_1,\dots,D_n;q_1,\dots,q_m)\in\Gamma_{0,n;m},
\quad
q_*\in \Sigma\setminus \bigcup_{i=1}^n D_i
\end{array}
\right\}. \reasoncheck
\label{eq:Kdef}
\end{equation}
For the purpose of this geometric identification, the excluded disks are precisely the ordinary coordinate disks \(D_i\). The pre-existing special punctures are already encoded by the metric-adapted insertions \([F(0)]^{g_a[h]}\), so the distinguished puncture created by \(\cK\) is not accompanied by an additional independent family of special-coordinate disks. This is the same domain denoted by \(D_{\mathrm{tot}}\) in \eqref{eq:kOmega}.
Let \(\pi_{\cK}:\Gamma'_{0,n|1;m}\to\Gamma_{0,n;m}\) forget the distinguished puncture. Fiber integration along \(\pi_{\cK}\) uses the orientation induced by the contour-shrinking argument for \(\mathfrak{k}\): equivalently, on \(q_*\in\Sigma\setminus D_{\mathrm{tot}}\) it is the negative of the naive worldsheet orientation, so that the boundary of the complement reproduces the positively oriented contour around the ordinary local-coordinate disks. With this convention, the fiber integral matches the sign in \eqref{eq:kOmega}. On the total space of \(\Gamma'_{0,n|1;m}\), the factor attached to the distinguished puncture is \(e^{\cB}[F(0)]^{g_*[h]}\). Decompose \(\cB=\cB_{\mathrm{base}}+\cB_{q_*}\), where \(\cB_{q_*}\) is the contour one-form generated by moving \(q_*\) while keeping the other punctures fixed. Only the fiber-degree-two part survives the pushforward along \(\pi_{\cK}\), and by \eqref{eq:Bdescent} that part is
\begin{equation}
\frac{1}{2}\cB_{q_*}^2F=F^{[2]}.
\label{eq:Kfiberdesc}
\end{equation}
More precisely, because \(\cB_{\mathrm{base}}\) has zero \(q_*\)-fiber degree, the distinguished-puncture factor pushes forward as
\begin{equation}
(\pi_{\cK})_*\!\left(e^{\cB}[F(0)]^{g_*[h]}\right)
=
e^{\cB_{\mathrm{base}}}\!\left(\frac{1}{2}\cB_{q_*}^2F\right)
=
e^{\cB_{\mathrm{base}}}F^{[2]}.
\label{eq:kpushfactor}
\end{equation}
The pushforward of the \((m+1)\)-special-puncture form is therefore the anomaly operator of \eqref{eq:kOmega}:
\begin{equation}
(\pi_{\cK})_*\Omega[\underline\Psi;F^{\otimes(m+1)}]
=
\mathfrak{k}\,\Omega[\underline\Psi;F^{\otimes m}]. \reasoncheck
\label{eq:kKpush}
\end{equation}
Thus \eqref{eq:kKpush} is an identity of total forms on the base. Its vertical part is the integrated local insertion \(F^{[2]}\), while its bundle-degree-one Weyl component comes from \(\cB_{\mathrm{base}}\) acting on that same descendant. In particular, for a pure Weyl tangent vector \(v_\omega\), one uses \eqref{eq:F2Weyl} to replace \((\cB_{\mathrm{base}}F^{[2]})(v_\omega)\) by \(\Xi[v_\omega]\). There is therefore no additional term missing from \eqref{eq:kKpush}; the Weyl-response piece is the base-space-degree-one component of the same pushforward.

Integrating \eqref{eq:kKpush} over \(\Gamma_{0,n;m}\) gives
\begin{equation}
\int_{\Gamma_{0,n;m}}\mathfrak{k}\,\Omega[\underline\Psi;F^{\otimes m}]
=
\int_{\Gamma'_{0,n|1;m}}\Omega[\underline\Psi;F^{\otimes(m+1)}]
:=
\int_{\cK\Gamma_{0,n;m}}\Omega[\underline\Psi;F^{\otimes(m+1)}]. \reasoncheck
\label{eq:kKbridge}
\end{equation}
Integrating the form identity \eqref{eq:QBplusk} over \(\Gamma_{0,n;m}\) and using Stokes' theorem together with \eqref{eq:kKbridge} gives the off-critical replacement for \(\int_{\Gamma}\Omega[Q_B\underline\Psi]=-\int_{\partial\Gamma}\Omega[\underline\Psi]\):
\begin{equation}
\int_{\Gamma_{0,n;m}}\Omega[Q_B\underline\Psi;F^{\otimes m}]
=
-\int_{\partial\Gamma_{0,n;m}}\Omega[\underline\Psi;F^{\otimes m}]
+\int_{\Gamma'_{0,n|1;m}}\Omega[\underline\Psi;F^{\otimes(m+1)}]. \reasoncheck
\label{eq:QBchain}
\end{equation}
This identity is the analog of the geometric master equation to the BV master equation. In the BV expansion, the first term in \eqref{eq:QBchain} is the boundary contribution of the \(m\)-special-puncture vertex, while the second is the primed asymmetric term produced when the BRST contour leaves behind the distinguished \(F^{[2]}\) insertion.
The new puncture created by \(\cK\) carries the same odd state \(F\) as every special puncture, so it belongs to the special sector rather than the ordinary one. It is nevertheless kept \emph{distinguished}, rather than absorbed into the fully antisymmetrized set of special punctures, because \(\mathfrak{k}\) produces one specific extra puncture by integrating a local \(F^{[2]}\) insertion over \(\Sigma\setminus D_{\mathrm{tot}}\). If one were to forget this distinction and fully antisymmetrize over all \(m+1\) special punctures, the combinatorics of the bulk-anomaly term would be off by a factor of \(m+1\).
The vertical bar records exactly this distinction: \(q_*\) is a special puncture because it carries \(F\), but it is not included in the pre-existing antisymmetrization of the \(m\) punctures \(q_a\).

The second asymmetric operation comes from the BV antibracket term with one explicit \(F\)-insertion:
\begin{equation}
\bra{A}c_0^-\ket{[\Psi^n;F^m]}
:=
\int_{\Gamma_{0,n+1;m}}\Omega[A,\underline\Psi;F^{\otimes m}], \reasoncheck
\label{eq:offcriticalproduct}
\end{equation}
where \(A\) is inserted at an ordinary puncture, because the sewing antibracket still acts only on ordinary punctures. Setting \(A=F\) gives
\begin{equation}
\bra{F}c_0^-\ket{[\Psi^n;F^m]}
=
\int_{\Gamma_{0,n+1;m}}\Omega[F,\underline\Psi;F^{\otimes m}]. \reasoncheck
\label{eq:Ifromproduct}
\end{equation}
Because ordinary punctures are symmetrized in the form \(\Omega\), the integral on the right-hand side may be rewritten as a sum over the possible choices of which ordinary puncture carries the inserted state \(F\). Once that puncture carries the same odd state \(F\) as the special punctures it is no longer ordinary: it is itself reclassified as a special puncture.
\begin{equation}
\bra{F}c_0^-\ket{[\Psi^n;F^m]}
=
\int_{\Gamma_{0,n|1;m}}\Omega[\underline\Psi;F^{\otimes(m+1)}]
:=
\int_{\cI\Gamma_{0,n+1;m}}\Omega[\underline\Psi;F^{\otimes(m+1)}]. \reasoncheck
\label{eq:Ibridge}
\end{equation}
Together, \eqref{eq:QBchain} and \eqref{eq:Ibridge} are the two identities needed to pass from the geometric recursion on chains to the BV master equation for the action \eqref{eq:offaction}: \eqref{eq:QBchain} converts the BRST variation of an \(m\)-special-puncture vertex into its boundary plus the primed asymmetric chain, while \eqref{eq:Ibridge} converts the explicit special-state insertion from the antibracket into the unprimed asymmetric chain. In the latter case the new puncture is again kept distinguished, because the antibracket remembers which puncture was converted.
The operation \(\cI\) is therefore not merely a relabeling of punctures. Once the chosen puncture is reclassified from ordinary to distinguished special, its independent local coordinate \(f_j\) is replaced by the metric-adapted special coordinate determined by the Hermitian metric at the same center \(p_j\). This is exactly what allows the resulting insertion to lie in the same special-puncture defect sector as the pre-existing \(m\) insertions of \(F\).

If \(p_j=f_j(0)\) denotes the center of the \(j\)th ordinary coordinate disk, then \(\cI\) reclassifies it as a distinguished special puncture, and sums over the possible choices of \(j\):
\begin{equation}
\Gamma_{0,n|1;m}
=
\cI\Gamma_{0,n+1;m}
:=
\sum_{j=1}^{n+1}
\left\{
\begin{array}{c}
(\Sigma;D_1,\dots,\widehat{D_j},\dots,D_{n+1};q_1,\dots,q_m\mid p_j)
\\[2pt]
(\Sigma;D_1,\dots,D_{n+1};q_1,\dots,q_m)\in\Gamma_{0,n+1;m}
\end{array}
\right\}. \reasoncheck
\label{eq:Idef}
\end{equation}
The sum over \(j\) is the geometric expression of ordinary-puncture symmetrization, while the vertical bar indicates that the converted puncture is kept distinct from the pre-existing antisymmetrized special set. Thus \(\cK\) and \(\cI\) produce the same \emph{type} of object---a distinguished special puncture carrying \(F\)---but by two different mechanisms: \(\cK\) creates it from the integrated bulk descendant \(F^{[2]}\), whereas \(\cI\) creates it by reclassifying one of the already symmetrized ordinary punctures as special.
Consequently, in the order-\(F^{m+1}\) part of the BV master equation, every term of the form \(\int_{\Gamma_{0,n;m}}\Omega[Q_B\underline\Psi;F^{\otimes m}]\) is converted by \eqref{eq:QBchain} into the boundary term \(-\int_{\partial\Gamma_{0,n;m}}\Omega\) plus the primed asymmetric contribution \(\int_{\Gamma'_{0,n|1;m}}\Omega\), while every term with one explicit insertion of \(F\) is converted by \eqref{eq:Ibridge} into the unprimed asymmetric contribution \(\int_{\Gamma_{0,n|1;m}}\Omega\). Together with the usual sewing terms, this is the chain-level content of the recursion relations below.

\subsubsection{Quadratic Recursion}

We treat the quadratic recursion relation separately due to a couple of new subtleties that arise at this order. The first has to do with a correction to the cancellation of the term $\bra{F} c_0^{-} Q_B \ket{\Psi}$ in the BV antibracket. At this order we must keep track of the correction obtained from shrinking the $Q_B$ contour to zero around $F$ (see Fig. \ref{fig:QB-action})
\begin{equation}
\bra{\Psi}Q_B c_0^{-}\ket{F} = \frac{1}{2\pi i}\int_{D}\bra{\Psi}c_0^-F^{[2]}_{z\bar{z}}(z)\ket{F}d^2 z,
\end{equation}
where $D$ denotes the hemisphere surrounding $F$. Using the same contour deformation trick as shown in Fig. \ref{fig:zw316} allows us to use rotational invariance to exchange this two-form integral for the associated one-form $F^{1}$
\begin{equation}
\frac{1}{2\pi i}\int_D \bra{\Psi}c_0^- F^{[2]}_{z \bar{z}}(z)\ket{F}d^2z = \int_{\gamma_{\text{rad}}} \bra{\Psi}F^{[1]}\ket{F} =: \int_{\Gamma_{0,1|1;1}'}\Omega[\Psi; F^{\otimes 2}].
\end{equation}
where $\gamma_{\text{rad}}$ is a radial contour stretching from the boundary of $D$ to the $F$ insertion at its center. We denote this three-puncture moduli space as $\Gamma_{0,1|1;1}'$: an asymmetric variant of the interpolating space we encountered in \S\ref{subsec:first-order-action} (Fig. \ref{fig:Gamma-021-viz-streamlined} in particular). The interpolating space $\Gamma_{0,1;2}$ must now satisfy
\begin{equation}
\partial \Gamma_{0,1;2} = \Gamma_{0,1|1;1}' - \Gamma_{0,1|1;1}.
\end{equation}

We now have the ingredients necessary to state the quadratic order recursion relation. The boundary of \(\Gamma_{0,n;2}\) receives three distinct contributions: one from the primed asymmetric space, one from the unprimed asymmetric space, and one from sewings of lower-order mixed vertices. The result is
\begin{equation}
\partial \Gamma_{0,n;2}
=
\Gamma'_{0,n|1;1}
- \Gamma_{0,n|1;1}
- \frac{1}{2}\sum_{l+l'=n}\{\Gamma_{0,l+1;1},\Gamma_{0,l'+1;1}\}
- \sum_{l+l'=n}\{\Gamma_{0,l+1;2},\Gamma_{0,l'+1}\}. \reasoncheck
\label{eq:gamma2}
\end{equation}
Terms for which one of the sewn factors does not exist are understood to vanish.
\diffnote{Equation \eqref{eq:gamma2} differs from the unlabeled quadratic-order recursion displayed immediately after equation (2.63) of \texttt{main.tex}. Here both sewing terms carry minus signs, in agreement with Zwiebach's quadratic recursion and the all-orders formula \eqref{eq:gammageneral}; in \texttt{main.tex} those terms were written with plus signs, and the last sewing term also repeated the left index instead of using the primed one.}
For later comparison with Zwiebach's notation we record the translation in a footnote.\footnotemark
\footnotetext{In Zwiebach's notation the first two \(\cB^2\) equations are \(\partial \cB^2_1 = \cT^2_1 - \cI \cB^1_2\) and \(\partial \cB^2_2 = \cK \cB^1_2 - \cI \cB^1_3 - \frac{1}{2}\{\cB^1_2,\cB^1_2\} - \{\cV_3,\cB^2_1\}\). Here \(\cV_3\) is the ordinary genus-zero cubic vertex with three ordinary punctures; \(\cB^1_2\) and \(\cB^1_3\) are the first mixed interpolation spaces with one special puncture and, respectively, two and three ordinary punctures; \(\cB^2_1\) and \(\cB^2_2\) are the second-order interpolation spaces with two special punctures and, respectively, one and two ordinary punctures; \(\cT^2_1\) is the elementary asymmetric chain produced by the bulk-anomaly insertion; \(\cI\) is the operation that turns one ordinary puncture into a distinguished special puncture; and \(\{\cdot,\cdot\}\) is the usual sewing antibracket on chains. The superscripts count special punctures and the subscripts count ordinary punctures. These symbols should not be confused with the \(b\)-ghost differential-form operator \(\cB\) introduced in \eqref{eq:Bdef}. In the notation of this paper, the four terms in Zwiebach's quadratic equation are \(\Gamma'_{0,2|1;1}\), \(\Gamma_{0,2|1;1}\), \(\{\Gamma_{0,2;1},\Gamma_{0,2;1}\}\), and \(\{\Gamma_{0,3},\Gamma_{0,1;2}\}\), so \eqref{eq:gamma2} is exactly the same recursion rewritten in our \(\Gamma\)-notation.}

For $n \geq 2$ we absorb the proof of the consistency of this recursion relation into \S\ref{subsec:all-orders-action}. Here, we treat the case of $\Gamma_{0,1;2}$ separately due to its distinct nature. We must show that
\begin{equation}\label{eqn:Gamma-021-consistency}
\partial \Gamma_{0,1|1;1}' = \partial \Gamma_{0,1|1;1}.
\end{equation}
One boundary of both chains consists of the asymmetric three-punctured sphere with an ordinary puncture at one pole, and special punctures at the equator and the other pole. The other potential boundary component of both chains vanishes, as it involves inserting antisymmetrized special punctures at symmetric locations. In the case of $\Gamma_{0,1|1;1}'$ this involves inserting both special punctures at the same point at one pole, and in the case of $\Gamma_{0,1|1;1}$ this involves inserting special punctures at symmetric locations at either pole. This demonstrates \eqref{eqn:Gamma-021-consistency}.

\subsubsection{All-Orders Recursion and Closure}\label{subsec:all-orders-action}

In general the boundary of $\Gamma_{0,n;m}$ has three sources: a primed asymmetric piece, an unprimed asymmetric piece, and a sum of sewings of lower-order vertices:
\begin{equation}
\partial \Gamma_{0,n;m}
=
\Gamma'_{0,n|1;m-1}
- \Gamma_{0,n|1;m-1}
- \frac{1}{2}
\sum_{\substack{k+k'=m\\ l+l'=n}}
\{\Gamma_{0,l+1;k},\Gamma_{0,l'+1;k'}\},
\qquad m\ge 1. \reasoncheck
\label{eq:gammageneral}
\end{equation}
Here $k,k'$ count special punctures and $l,l'$ count ordinary punctures. The overall minus sign in front of the sewing term matches Zwiebach's sign convention for the extended sewing antibracket in the non-conformal recursion \cite[sec.~5]{Zwiebach:1996jc}.
Terms for which one of the sewn factors does not exist are understood to vanish.
This puncture assignment follows from the geometry of sewing: sewing removes one ordinary puncture from each factor, while the special punctures remain special after sewing, so the sewn term must split ordinary punctures against ordinary punctures and special punctures against special punctures. Equation \eqref{eq:gammageneral} is the off-critical correction to the genus-zero recursion. At quadratic order the closure statement can be written explicitly. The asymmetric operators satisfy
\begin{equation}
\cK^2=\cI^2=\cK\cI+\cI\cK=0. \bluecheck
\label{eq:KIprops}
\end{equation}
Their action on boundaries is
\begin{align}
\partial(\cK\Gamma_{0,n;m})&=\cK(\partial\Gamma_{0,n;m})+\{\Gamma'_{0,2|1},\Gamma_{0,n;m}\},\\
\partial(\cI\Gamma_{0,n;m})&=\cI(\partial\Gamma_{0,n;m}), \reasoncheck
\label{eq:KIboundary}
\end{align}
Here $\Gamma'_{0,2|1}$ is the elementary asymmetric chain that describes the collision of the distinguished special puncture with an ordinary puncture. Accordingly, dragging the boundary of a chain past a newly inserted distinguished puncture produces the extra sewing with $\Gamma'_{0,2|1}$ and the original chain, whereas turning an ordinary puncture into an antisymmetrized special puncture commutes with taking the boundary. Applying \eqref{eq:gamma1} then gives
\begin{align}
\partial\Gamma'_{0,n|1;1}
&=
-\cK\cI\Gamma_{0,n+1}
-\sum_{l=1}^{n-2}\cK\{\Gamma_{0,l+1;1},\Gamma_{0,n-l+1}\}
+\{\Gamma'_{0,2|1},\Gamma_{0,n;1}\},
\\
\partial\Gamma_{0,n|1;1}
&=
\cI\cK\Gamma_{0,n+1}
-\sum_{l=1}^{n-1}\cI\{\Gamma_{0,l+1;1},\Gamma_{0,n-l+2}\}. \reasoncheck
\label{eq:gamma21boundary}
\end{align}
Since $\cK$ and $\cI$ raise chain degree by two, they obey the ordinary Leibniz rule on sewn spaces,
\begin{align}
\cK\{A,B\}&=\{\cK A,B\}+\{A,\cK B\},
\\
\cI\{A,B\}&=\{\cI A,B\}+\{A,\cI B\}, \reasoncheck
\label{eq:KILeibniz}
\end{align}
for the chains that occur here. It follows that
\begin{equation}
\partial\Bigl(
\Gamma'_{0,n|1;1}
-\Gamma_{0,n|1;1}
-\frac12\sum_{l+l'=n}\{\Gamma_{0,l+1;1},\Gamma_{0,l'+1;1}\}
-\sum_{l+l'=n}\{\Gamma_{0,l+1;2},\Gamma_{0,l'+1}\}
\Bigr)=0. \reasoncheck
\label{eq:gamma2closure}
\end{equation}
The $\cK\cI$ and $\cI\cK$ terms cancel because of \eqref{eq:KIprops}. One then expands the differentiated sewings with \eqref{eq:KILeibniz} and substitutes \eqref{eq:gamma1}: the terms with one newly distinguished puncture reproduce the differentiated order-$F$ sewings, while the terms with one order-$F^2$ factor cancel against the differentiated mixed sewing in \eqref{eq:gamma2}.

For general \(m\ge 2\), the same mechanism gives an inductive closure proof for \eqref{eq:gammageneral}. Assume that the recursion relation is already known for all lower orders \(1\le r<m\). Applying \(\partial\) to the right-hand side of \eqref{eq:gammageneral} produces three kinds of terms. The derivatives of the primed and unprimed asymmetric pieces are rewritten with \eqref{eq:KIboundary}: one gets \(\cK(\partial\Gamma_{0,n;m-1})\), \(-\cI(\partial\Gamma_{0,n+1;m-1})\), and the distinguished-puncture collision term \(\{\Gamma'_{0,2|1},\Gamma_{0,n;m-1}\}\). The derivative of the sewing sum is then expanded using the standard boundary derivation property of the sewing antibracket. Substituting the lower-order recursion relations into \(\partial\Gamma_{0,n;m-1}\), \(\partial\Gamma_{0,n+1;m-1}\), and the differentiated sewn factors, the resulting contributions fall into three matching classes.

First, terms in which \(\cK\) or \(\cI\) acts on a lower-order asymmetric chain cancel pairwise because \(\cK^2=\cI^2=\cK\cI+\cI\cK=0\). Second, terms in which \(\cK\) or \(\cI\) acts on a lower-order sewing combine, via \eqref{eq:KILeibniz}, with the derivative of the explicit sewing sum in \eqref{eq:gammageneral}; these are exactly the differentiated sewings built from one lower-order asymmetric chain and one lower-order mixed vertex. Third, the extra boundary component term \(\{\Gamma'_{0,2|1},\Gamma_{0,n;m-1}\}\) is the missing contribution needed to close the Jacobi identity for the sewing antibracket among the sewings that carry the distinguished puncture. After summing over all splittings \(k+k'=m\) and \(l+l'=n\), every differentiated contribution appears twice with opposite sign, so the right-hand side of \eqref{eq:gammageneral} is closed. Equivalently, \(\partial^2\Gamma_{0,n;m}=0\) for all \(m\).

This is the higher-order cancellation in the BV master equation. For \(m\ge2\), the coefficient of order \(F^m\) in \((S_0,S_0)\) is the integral of \(\Omega[\underline\Psi;F^{\otimes m}]\) over
\begin{equation}
\partial\Gamma_{0,n;m}
-\Gamma'_{0,n|1;m-1}
+\Gamma_{0,n|1;m-1}
+\frac12
\sum_{\substack{k+k'=m\\ l+l'=n}}
\{\Gamma_{0,l+1;k},\Gamma_{0,l'+1;k'}\}.
\label{eq:BVorderm}
\end{equation}
The first term comes from the boundary part of \eqref{eq:QBchain} applied to the \(m\)-special-puncture vertex. The second term comes from the \(\mathfrak{k}\)-part of \eqref{eq:QBchain} applied to the \((m-1)\)-special-puncture vertex. The third term comes from \eqref{eq:Ibridge} applied to the BV antibracket term with one explicit insertion of \(F\) on a vertex with \(m-1\) special punctures. The last term is the ordinary sewing antibracket of lower-order mixed vertices whose special-puncture numbers add to \(m\). Requiring this order-\(F^m\) coefficient of \((S_0,S_0)\) to vanish for arbitrary external states is therefore equivalent to \eqref{eq:gammageneral}. The closure proof above shows that this chain has zero boundary, so the higher-order BV anomaly cancels.

\diffnote{The constructive quadratic-order consistency proof is written here in corrected notation corresponding to \texttt{main.tex} equations \texttt{eqn:KI-properties} (2.67), \texttt{eqn:Gamma-prime-boundary} (2.71), \texttt{eqn:KI-Leibniz} (2.72), and \texttt{eqn:Gamma-semicolon-2-consistency} (2.73).}

\diffnote{The ordinary/special puncture counting in \eqref{eq:gammageneral} is corrected relative to \texttt{main.tex} equation \texttt{eqn:general-Gamma-recursion} (2.75) and its repeated display later in the paper. There the sewing term inadvertently exchanged the roles of $n$ and $m$, so the all-orders consistency check written after \texttt{eqn:general-Gamma-recursion} (2.75) cannot simply be copied. In addition, the last sewing term in \texttt{main.tex} equation \texttt{eqn:Gamma-prime-boundary} (2.71) was not consistent with the general identity \eqref{eq:KIboundary}; the correct term is the sewing of \(\Gamma'_{0,2|1}\) with the original order-\(F\) chain, which is the geometric boundary described above.}

\subsection{Existence of the Mixed Vertices}\label{subsec:existence}

The recursion relation \eqref{eq:gammageneral} makes sense only if its right-hand side is a boundary in the enlarged bundle \(\wh{\cP}^{\,\omega}_{0,n+m}\). The existence argument has two steps: first one shows that the fibers of the bundle are contractible, then one uses this to reduce the existence question to a homological one on the base.

Fix a point of \(\cM_{0,n+m}\), so that the puncture locations \(z_1,\dots,z_{n+m}\) are held fixed, and choose two points in the corresponding fiber of \(\wh{\cP}^{\,\omega}_{0,n+m}\). These data consist of two Hermitian metrics, say \(h=e^{-2\omega(z,\bar z)}|dz|^2\) and \(\tilde h=e^{-2\tilde\omega(z,\bar z)}|dz|^2\), together with two choices of local coordinates \(f_i(w_i)\) and \(\tilde f_i(w_i)\) at the ordinary punctures. At the special punctures the local coordinates are not free: they are determined by the two metrics through \eqref{eq:specialmetriccoord}. After fixing the phase convention so that the linear coefficients \(f_i'(0)\) and \(\tilde f_i'(0)\) are positive real, each ordinary local coordinate may be written as
\begin{equation}
f_i(w_i)=z_i+w_i\,e^{\mathfrak f_i(w_i)},
\qquad
\tilde f_i(w_i)=z_i+w_i\,e^{\tilde{\mathfrak f}_i(w_i)}, \reasoncheck
\label{eq:fibercoords}
\end{equation}
where $\mathfrak f_i$ and $\tilde{\mathfrak f}_i$ are holomorphic on the unit disk. This isolates the universal simple zero at the puncture, so the interpolation acts only on the nonvanishing part of the coordinate map.

A concrete interpolation in the fiber is then
\begin{equation}
\begin{aligned}
\mathfrak f_i(w_i;\sigma)
&=
(1-\sigma)\mathfrak f_i(w_i)
\,+\,
\sigma \tilde{\mathfrak f}_i(w_i)
\,-\,
\log \lambda_i(\sigma),
\\
\omega(z;\sigma)
&=
(1-\sigma)\omega(z)+\sigma\tilde\omega(z),
\qquad 0\le \sigma\le1,
\end{aligned}
\reasoncheck
\label{eq:fiberinterpolation}
\end{equation}
with \(\lambda_i(\sigma)>0\) chosen continuously so that the linear coefficient of
\(f_i(w_i;\sigma):=z_i+w_i e^{\mathfrak f_i(w_i;\sigma)}\)
stays in the same phase convention and its Jacobian remains nondegenerate on the coordinate disk. The Hermitian metric interpolates linearly through \(\omega(z;\sigma)\), and at each special puncture the corresponding local coordinate \(g_a(w_a;\sigma)\) is then determined uniquely up to phase by \eqref{eq:specialmetriccoord}. Because that determination is local and algebraic in the jets of the metric, the phase convention may be fixed continuously along the interpolation. Since nondegeneracy is an open condition, this gives a path in the fiber from the first point to the second. Fixing one reference point and applying the same construction to every other point contracts the entire fiber. Thus the fibers of \(\wh{\cP}^{\,\omega}_{0,n+m}\) are contractible.

Projection to ordinary moduli space now does the rest. For $m\ge 2$, the chain on the right-hand side of \eqref{eq:gammageneral} has dimension $2n+3m-7$, whereas $\dim \cM_{0,n+m}=2(n+m)-6$, so the difference in degree is exactly $m-1>0$. Let $C$ denote that closed chain. Its projection $\pi_*C$ to $\cM_{0,n+m}$ has the same degree, but a manifold has no homology above its dimension, so $[\pi_*C]=0$. Because the fibers of $\wh{\cP}^{\,\omega}_{0,n+m}\to \cM_{0,n+m}$ are contractible, \(\pi_*\) is an isomorphism on homology. Hence \( [C]=0 \) upstairs as well, which is exactly the statement that \(C\) is a boundary. Therefore \(\Gamma_{0,n;m}\) exists with \(\partial\Gamma_{0,n;m}=C\) \cite{Zwiebach:1996ph}.

The same argument also explains the nonuniqueness of the mixed vertices. If two chains satisfy the same boundary condition \eqref{eq:gammageneral}, then their difference is a closed chain in $\wh{\cP}^{\,\omega}_{0,n+m}$ of the same degree. Its projection to $\cM_{0,n+m}$ again vanishes in homology for dimensional reasons, so the difference is itself a boundary upstairs. Different choices of interpolating spaces therefore differ only by homologically trivial deformations, which is the geometric origin of the usual field-redefinition ambiguity in the string vertices.

\section{Background Independence Off the Conformal Locus}
\label{sec:bgindep}

Throughout this section, \(x\) denotes a \emph{conformal reference worldsheet theory}. We compare it with a nearby theory \(y_\lambda\) whose worldsheet action is perturbed by
\begin{equation}
I_{\mathrm{ws}}(y_\lambda)=I_{\mathrm{ws}}(x)-\frac{\lambda}{2\pi}\int \phi_x+O(\lambda^2).
\label{eq:yperturbation}
\end{equation}
The nearby theory \(y_\lambda\) need not be conformal. In particular, we do \emph{not} assume that it carries a canonical BRST Hilbert space, a canonical state-operator map on arbitrary loops, or a canonical inner product on local insertions. All string-field states, BV brackets, and canonical-connection data remain attached to the reference space \(\cH_x\); the effect of moving toward \(y_\lambda\) is encoded first by distinguished local insertions on the \(x\)-worldsheet and only then, after choosing the same local-coordinate/cap geometry as in figures \ref{fig:stateop} and \ref{fig:QB-action}, by prepared special states such as \(Q_B\cO_x\).

This is more conservative than Zwiebach's 1996 proposal, which seeks to formulate the string action directly on the state space of a non-conformal two-dimensional theory. We do not attempt that stronger construction. Our assumption is only that the nearby family \(y_\lambda\) admits a tangent-level representation inside the reference conformal theory \(x\). The discussion below accordingly uses two distinct layers of data. The first layer is the critical BV geometry and canonical connection, living entirely on the reference theory \(x\). The second is the first-order insertion/state data encoding the nearby theory \(y_\lambda\), represented inside \(x\). Keeping these two layers separate is what allows the argument to stay within the part of the formalism that is under control.

\subsection{Reference Conformal Theory and Infinitesimal Background Changes}

Let \(x\) be the conformal reference theory introduced above, and let \(\cH_x\) denote its semi-relative closed-string state space, namely the BRST state space built from that conformal theory and restricted by the analogue of \(b_0^-\Psi=L_0^-\Psi=0\). A string field in that background may then be expanded in a local basis of \(\cH_x\), with coefficients denoted \(\psi_x^i\). For an infinitesimal change of \emph{conformal} background \(x^\mu\to x^\mu+\delta x^\mu\), Sen and Zwiebach write the field redefinition as
\begin{equation}
\psi_{x+\delta x}^i
=
\psi_x^i + \delta x^\mu f_\mu^i(\psi_x,x) + O(\delta x^2). \bluecheck
\label{eq:fdiffeo}
\end{equation}
At tree level the BV Laplacian and measure terms drop out, so the infinitesimal conditions of action equivalence and symplecticity reduce to
\begin{equation}
\partial_\mu \omega_{i'j'}
+
\frac{\partial_l f_\mu^i}{\partial \psi^{i'}}\,\omega_{ij'}
+
\omega_{i'j}\,\frac{\partial_r f_\mu^j}{\partial \psi^{j'}}
=0,
\qquad
\partial_\mu S + \frac{\partial_r S}{\partial \psi^i}f_\mu^i = 0. \reasoncheck
\label{eq:infbg}
\end{equation}
Here $\omega_{ij}$ is the odd symplectic form on string-field space, $\partial_l,\partial_r$ are the usual left and right BV derivatives, and $\partial_\mu$ differentiates with respect to the theory-space coordinate $x^\mu$. The primed indices $i',j'$ refer to the shifted background $x+\delta x$, while the unprimed indices refer to the original background $x$. The geometric meaning of the connection on the state-space bundle was clarified in \cite{Ranganathan:1992nb,Ranganathan:1993vj}.
Following \cite{Ranganathan:1992nb,Ranganathan:1993vj,Sen:1993mh}, decompose
\begin{equation}
f_\mu^i
=
-\Gamma_\mu^{~i}
-\wh\Gamma_{\mu j}^{~i}\,\psi^j
-h_\mu^i(\psi,x), \reasoncheck
\label{eq:fdecomp}
\end{equation}
where $\wh\Gamma$ is the canonical connection on the bundle of state spaces $\cH_x$ over theory space, $\Gamma_\mu^{~i}$ is the field-independent shift, and $h_\mu^i$ starts at quadratic order in $\psi$. When the symplectic form is covariantly constant under $\wh\Gamma$, the nonlinear part is Hamiltonian
\begin{equation}
h_\mu^i = \omega^{ij}\,\frac{\partial_l \cU_\mu}{\partial \psi^j}. \reasoncheck
\label{eq:ham}
\end{equation}
The tree-level background-independence equation is therefore
\begin{equation}
D_\mu(\wh\Gamma)\,S_0
- \frac{\partial_r S_0}{\partial \psi^i}\Gamma_\mu^{~i}
- \{S_0,\cU_\mu\}
=0.
\reasoncheck
\label{eq:classicalbi}
\end{equation}
Up to this point the discussion is still the ordinary critical Sen--Zwiebach framework: both the reference theory and the infinitesimally shifted theory are conformal, and the comparison is expressed entirely as a field redefinition on the corresponding BRST state spaces. In the non-conformal case considered below, we deliberately do \emph{not} replace \(\cH_x\) by some a priori Hilbert space attached to \(y_\lambda\). Rather, the nearby theory \(y_\lambda\) is represented perturbatively on the reference worldsheet of \(x\), first by local insertions and then by prepared special states obtained from those insertions using the state-preparation geometry of figure \ref{fig:QB-action}.
The modern flat-vertex analysis of weakly curved backgrounds teaches us that even when the physical content is fixed, the explicit identification of fields and gauge parameters with string-field components depends on the string-field frame, while only the field-redefinition class is invariant \cite{Mazel:2025diffeo}. The same lesson applies to the detailed representatives used for $\cO_x$, the mixed vertices, and the associated interpolation spaces.

\diffnote{The decomposition \eqref{eq:fdecomp} and the symplectic condition \eqref{eq:infbg} follow the index structure of Sen--Zwiebach. This corrects the contractions in \texttt{main.tex} equation \texttt{eqn:linear-classical-conditions} (3.11) and aligns the discussion with \texttt{eqn:f-U-expansion} (3.12).}
\diffnote{The present section also departs deliberately from the looser language of \texttt{main.tex}: we no longer speak as if a general nearby non-conformal theory came with its own canonical Hilbert space or state-operator map. All BV and connection data remain attached to the reference conformal theory \(x\), and the nearby theory \(y_\lambda\) is represented only through distinguished local insertions and the prepared states they define.}

\subsection{Tangent-Level Data for the Nearby Theory}

Fix the conformal reference theory \(x\) and the nearby theory \(y_\lambda\) introduced above, whose worldsheet action differs from that of \(x\) by the normalized integrated insertion \(-(2\pi)^{-1}\lambda\int\phi_x\) as in \eqref{eq:yperturbation}.
At first order in \(\lambda\) the nearby theory is represented inside the reference theory \(x\) by a ghost-number-two local insertion \(\cO_x\), its one-form descendant \(\cO_x^{[1]}\), and the worldsheet two-form perturbation \(\phi_x\). If \(r\) denotes the moving auxiliary insertion point in the asymmetric vertex, the corresponding transgressed bundle forms satisfy
\begin{equation}
(\cB \cO_x)(v)=\iota_{\dot r(v)}\cO_x^{[1]},
\qquad
\left(\frac12\cB^2 \cO_x\right)(v_1,v_2)=\iota_{\dot r(v_2)}\iota_{\dot r(v_1)}\phi_x. \bluecheck
\label{eq:Ox}
\end{equation}
This is the same starting point as in \cite{Sen:1990hh,Sen:1993mh}, except that the nearby theory \(y_\lambda\) is now allowed to be off critical. Here \(\cO_x\) is a local worldsheet insertion of the reference theory, not an operator on string-field space; \(\cO_x^{[1]}\) is its local one-form descendant and \(\phi_x\) the corresponding two-form. The worldsheet action uses the normalized integral \(-(2\pi)^{-1}\int\phi_x\), while the vertex calculus uses the transgressed bundle forms \(\cB\cO_x\) and \(\frac12\cB^2\cO_x\). Nothing at this stage is computed in an independently defined Hilbert space for \(y_\lambda\).
As a sanity check: in the ordinary conformal marginal case one takes \(\phi_x=V_{(1,1)}\,d^2z\) with \(V_{(1,1)}\) a matter primary and \(\cO_x=c\tilde c\,V_{(1,1)}\). Then \(Q_B\cO_x=0\) at linearized level, the special-state term in \eqref{eq:dQB} vanishes, and one recovers the standard Sen--Zwiebach setup.
The existence of the family \((\cO_x,\cO_x^{[1]},\phi_x,Q_B\cO_x)\), together with the contour-deformation statement \eqref{eq:dQB}, should therefore be read as an explicit first-order assumption---the non-conformal analogue, at the level of tangent vectors, of the canonical connection on the conformal manifold. We do not attempt a general construction of the spaces of states or local insertions for an arbitrary non-conformal worldsheet theory; for the present paper, the only assumption is that the nearby family \(y_\lambda\) admits a first-order representation inside the reference conformal theory \(x\) by this insertion. (A more general framework would presumably require the Kontsevich--Segal axioms, suitably extended to accommodate theories that are not fully UV complete.)

To represent the first-order variation of the string vertices, one inserts \(\phi_x\) asymmetrically over the complement of the local-coordinate disks around the ordinary punctures. This representative still lives entirely on the reference \(x\)-worldsheet. The BRST contour can therefore be deformed exactly as in the undeformed conformal theory, with the first-order effect encoded by the tangent deformation just introduced.

At first order, contour deformation of the BRST insertion across the integrated perturbation, still performed on the reference \(x\)-worldsheet with one distinguished movable insertion, gives the first-order variations
\begin{equation}
\delta_\lambda Q_B[\gamma]
=
\lambda\oint_\gamma \cO_x^{[1]}+O(\lambda^2),
\qquad
\delta_\lambda F
=
\lambda\,Q_B\cO_x+O(\lambda^2). \reasoncheck
\label{eq:dQB}
\end{equation}
Here \(\delta_\lambda\) denotes the first-order variation at \(\lambda=0\), represented inside the reference theory \(x\). The coefficient \(-(2\pi)^{-1}\) in the action deformation \eqref{eq:yperturbation} fixes the normalization of the perturbing two-form \(\phi_x\) and is what makes the trace variation come out as \eqref{eq:dThetaflat}. By contrast, \eqref{eq:dQB} is written in the standard descent normalization for the tangent deformation \((\cO_x,\cO_x^{[1]},\phi_x)\): once the perturbation is rewritten in terms of its descendants, the BRST contour variation and the prepared-state relation are expressed directly in terms of \(\cO_x^{[1]}\) and \(Q_B\cO_x\), without an additional \((2\pi)^{-1}\). The second equation of \eqref{eq:dQB} identifies the first-order shift of the special state \(F\) that appears in the mixed vertices of section \ref{sec:offcritical}.

The phrase ``induced special state'' should be read in the prepared-state sense: \(Q_B\cO_x\) is not identified with a state by a naive non-conformal state-operator correspondence on an arbitrary loop, but rather by the cap/hemisphere state-preparation geometry of figure \ref{fig:QB-action}, with the chosen local coordinate at the distinguished puncture supplying the circle on which the state is prepared. \eqref{eq:dQB} is therefore the first place where the tangent direction toward \(y_\lambda\) is allowed to differ from a conformal deformation. If \(Q_B\cO_x\neq 0\), then \(\delta_\lambda F\neq 0\) and the nearby theory is already off critical at order \(\lambda\). If \(Q_B\cO_x=0\), the special-state shift vanishes and one recovers the linearized Sen--Zwiebach situation; this means only that the deformation defines a conformal tangent direction modulo BRST-exact representatives, not that the tangent vector necessarily integrates to a finite conformal family.
In \eqref{eq:dQB} and below, unadorned \(Q_B\) denotes the BRST operator of the reference conformal theory \(x\) unless an explicit subscript is shown.
There is also a nontrivial local check of the trace sector. Work on a patch with flat Weyl frame and assume, for simplicity, that \(\phi_x\) is a scalar of fixed conformal dimensions \(h=\bar h=\Delta/2\). A perturbative computation of the stress-tensor variation \cite{Ahmadain:2024treelevel} then gives\footnote{Equation \eqref{eq:dThetaflat} is translated from the convention of \cite{Ahmadain:2024treelevel}, where the symbol \(\Theta\) denotes the full trace \(T^a{}_a\), while the present paper uses \(T^a{}_a=4\Theta\).}
\begin{equation}
\delta \Theta
=
\frac14(\Delta-2)\phi_x
+\frac12\sum_{n\ge 1}\frac{(-1)^n}{(n+1)!}
\left(\partial^n L_n\phi_x+\bar\partial^n \bar L_n\phi_x\right),
\label{eq:dThetaflat}
\end{equation}
This trace-sector relation admits a direct local comparison with the BRST expression \(Q_B(c\tilde c\,\phi_x)\). Combining \eqref{eq:dThetaflat} with the universal normalization \eqref{eq:F2WeylDeltac}, one obtains the corresponding linearized Weyl-response insertion
\begin{equation}
\Xi_{\phi_x}[v_\omega]
=
-4i\left[
\frac14(\Delta-2)\,\phi_x\,\delta\omega_v
+\frac12\sum_{n\ge1}\frac{1}{(n+1)!}
\left(L_n\phi_x\,\partial^n\delta\omega_v+\bar L_n\phi_x\,\bar\partial^n\delta\omega_v\right)
\right]d^2\sigma.
\label{eq:appXiPhix}
\end{equation}
Here the derivatives have been moved from \(\phi_x\) to \(\delta\omega_v\) by integration by parts, so the factors \((-1)^n\) in \eqref{eq:dThetaflat} cancel. On the BRST side, standard conformal-field-theory algebra gives
\begin{equation}
Q_B(c\tilde c\,\phi_x)
=
\frac12(\partial c+\bar\partial\tilde c)\,c\tilde c\,(\Delta-2)\phi_x
+c\tilde c\sum_{n\ge1}\frac{1}{(n+1)!}
\left(\partial^{n+1}c\,L_n\phi_x+\bar\partial^{n+1}\tilde c\,\bar L_n\phi_x\right).
\label{eq:appQBcctphi}
\end{equation}
To compare \eqref{eq:appQBcctphi} with \eqref{eq:appXiPhix}, one should view \eqref{eq:appQBcctphi} as the local insertion whose ordinary-puncture descendant is being evaluated on a pure Weyl tangent vector \(v_\omega\). The comparison has two ingredients. First, the undifferentiated factor \(c\tilde c\) supplies the ordinary-puncture two-form. Second, the ghost derivatives are converted into derivatives of the Weyl variation by the same local-coordinate calculus that underlies appendix \ref{app:antighost-descent}. Let \(\cB_{\mathrm{ord}}\) denote the ordinary-puncture component of \(\cB\). Then, on a patch with flat Weyl frame and with the puncture position held fixed, one has
\begin{equation}
\begin{aligned}
\frac12 \cB_{\mathrm{ord}}^{\,2}(c\tilde c)
&=
\frac12\,dz\wedge d\bar z
=
-i\,d^2\sigma,\\
\iota_{v_\omega}\cB_{\mathrm{ord}}(\partial c+\bar\partial\tilde c)
&=
2\,\delta\omega_v,\\
\iota_{v_\omega}\cB_{\mathrm{ord}}(\partial^{n+1}c)
&=
2\,\partial^n\delta\omega_v,\\
\iota_{v_\omega}\cB_{\mathrm{ord}}(\bar\partial^{n+1}\tilde c)
&=
2\,\bar\partial^n\delta\omega_v.
\end{aligned}
\label{eq:appOrdinaryGhostExchange}
\end{equation}
The first identity is the ordinary-puncture two-form associated with the standard \(c\tilde c\) insertion; the factor \(1/2\) is the usual antisymmetrization factor for the corresponding bundle two-form. The remaining identities are the pure-Weyl specialization of the same local-coordinate descent calculus used in appendix \ref{app:antighost-descent}: in a flat Weyl frame the puncture is fixed and only the Weyl scaling of the adapted local coordinate varies, so the ghost derivatives record the successive derivatives of \(\delta\omega_v\). We may use \eqref{eq:appOrdinaryGhostExchange} to compute the action of $\mathcal{B}_{\text{ord}}$ on \eqref{eq:appQBcctphi}, giving exactly \eqref{eq:appXiPhix} coefficient by coefficient. Thus \eqref{eq:dThetaflat} supplies a genuine local consistency check on the trace/Weyl sector of the first-order deformation.
In the present paper this is the only Weyl-response check we need. Since we are studying only the first-order deformation away from the conformal theory, we do not attempt here to analyze higher orders in Weyl variation or to formulate the corresponding trace/Weyl relation for a general varying metric background with the same level of detail.
At this first infinitesimal order, only the leading vertical component of the anomaly descendant contributes. Since \(\delta_\lambda F=O(\lambda)\), the bundle-degree completion involving the distinguished \(F^{[2]}\) insertion in \eqref{eq:kOmega} contributes only at higher order in \(\lambda\) and therefore does not alter the Sen--Zwiebach interpolation argument used below.

\begin{figure}[ht]
    \centering
    \begin{subfigure}[b]{0.48\textwidth}
        \centering
        \includegraphics[width=0.6\textwidth]{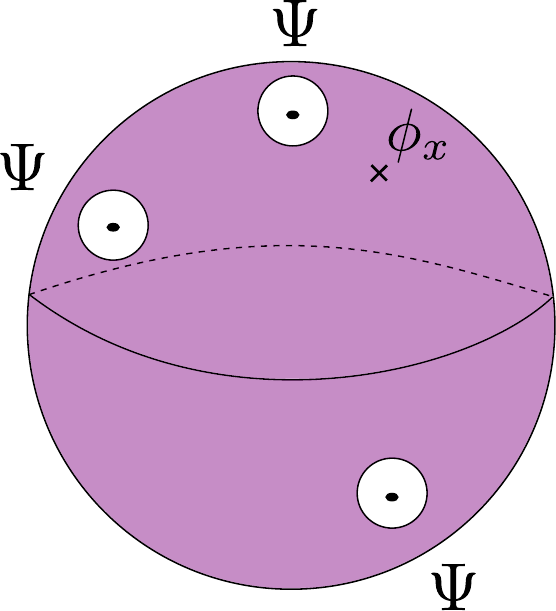}
        \caption{The asymmetric insertion of the local worldsheet field $\cO_x$ defining the deformed vertex.}
    \end{subfigure}
    \hfill
    \begin{subfigure}[b]{0.48\textwidth}
        \centering
        \includegraphics[width=0.6\textwidth]{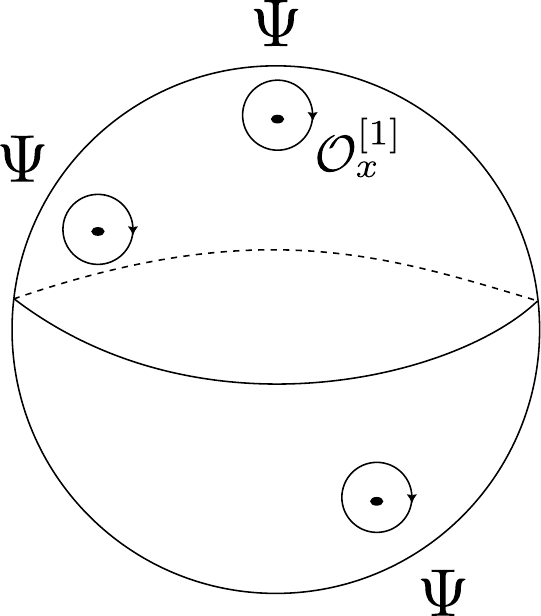}
        \caption{A boundary component obtained when the asymmetric puncture reaches a local-coordinate disk.}
    \end{subfigure}
    \caption{The worldsheet geometry underlying the first-order deformation by $\cO_x$. The same interpolation-space technology that appears in background independence also governs the off-critical vertices.}
    \label{fig:asymvertex}
\end{figure}

The corresponding asymmetric vertex is shown explicitly in figure \ref{fig:asymvertex}. The local insertion $\cO_x$ is integrated over the complement of the local-coordinate disks, and when it approaches one of those disks one lands on a boundary component described by the same interpolation-space geometry that already appeared at order $F$.

\subsection{First-Order BV Equivalence}

Write \(\Omega[\underline\Psi;\cO_x]\) for the genus-zero worldsheet form in the \emph{reference conformal theory \(x\)} with one distinguished movable insertion of \(\cO_x\) integrated over the complement of the ordinary local-coordinate disks; its integrated asymmetric vertex is denoted \(\{\Psi^n,\cO_x\}'\). Because both the interpolation chain and the distinguished puncture belong to the conformal theory \(x\), there is no off-critical bulk anomaly at this stage, and the form-level identity is the undeformed one:
\begin{equation}
\Omega[Q_B\underline\Psi;\cO_x]=-\delta\,\Omega[\underline\Psi;\cO_x].
\label{eq:criticalOxform}
\end{equation}
The first-order derivation is therefore hybrid in a precise sense. The interpolation geometry is the critical Sen--Zwiebach geometry of the reference theory \(x\), but the nearby theory may already be off critical because the special-state shift \(\delta_\lambda F=\lambda Q_B\cO_x\) need not vanish. Three explicit inputs are used: the tangent deformation \eqref{eq:Ox}; the contour-deformation identities \eqref{eq:dQB} with finite prepared state \(Q_B\cO_x\); and the critical interpolation-space relation \eqref{eq:Oxgamma1} below. Once those are granted, the remaining cancellation is an exact chain computation in the reference theory.
Let \(\Gamma^{(\cO)}_{0,n;1}\) denote the genus-zero Sen--Zwiebach interpolation chain in the reference conformal theory \(x\) whose distinguished puncture carries \(\cO_x\). Its boundary is
\begin{equation}
\partial \Gamma^{(\cO)}_{0,n;1}
=
\Gamma^{\prime(\cO)}_{0,n|1}
- \Gamma^{(\cO)}_{0,n|1}
- \sum_{l=1}^{n-2}\{\Gamma^{(\cO)}_{0,l+1;1},\Gamma_{0,n-l+1}\}. \reasoncheck
\label{eq:Oxgamma1}
\end{equation}
Here \(\Gamma^{\prime(\cO)}_{0,n|1}\) denotes the asymmetric chain in which \(\cO_x\) is integrated over the complement of the ordinary coordinate disks, while \(\Gamma^{(\cO)}_{0,n|1}\) denotes the chain obtained by moving the distinguished puncture onto one ordinary puncture and then interpreting the result using the ordinary state-operator map of the reference CFT \(x\). The identity \eqref{eq:Oxgamma1} is the genus-zero specialization of Sen--Zwiebach's interpolation-space equation for \(\mathcal U_{h,n|1}\), rewritten in the puncture notation used here. 

Using the canonical connection, the covariant derivative of the tree-level action takes the form
\begin{equation}
-D_\lambda(\wh\Gamma)S_0
=
\sum_{n\ge 2}\frac{1}{n!}\{\Psi^n,\cO_x\}'
+\sum_{n\ge 1}\frac{1}{n!}\{\Psi^n;Q_B\cO_x\}
. \reasoncheck
\label{eq:DSlambda}
\end{equation} \eqref{eq:DSlambda} records the two first-order sources of variation. The first sum is the ordinary Sen--Zwiebach term: the perturbing field is integrated over the complement of the ordinary local-coordinate disks. The second sum is the genuinely new off-critical correction coming from the first-order shift \(\delta_\lambda F=\lambda Q_B\cO_x\) in the special-state sector. Here \(\{\Psi^n,\cO_x\}'\) denotes the asymmetric vertex in which the local insertion \(\cO_x\) is integrated over the complement of the ordinary local-coordinate disks rather than placed at a symmetrized puncture. The term \(\{\Psi^n;Q_B\cO_x\}\) denotes the mixed vertex with one distinguished special puncture carrying the prepared state \(Q_B\cO_x\) obtained from \eqref{eq:dQB}.\footnote{This is the genus-zero analogue of the critical formula \( -D_\mu S = \sum \{\Psi^{\otimes n};{\cal O}_\mu\}_{h,n\bullet}\) in Sen--Zwiebach. As in their discussion, the \(n=2\) contribution comes from the deformation of the BRST charge appearing in the kinetic term.}
The coefficient \((2\pi)^{-1}\) in \eqref{eq:yperturbation} fixes the normalization of the worldsheet deformation itself. Formula \eqref{eq:DSlambda}, however, is written in terms of the corresponding tangent insertion \(\cO_x\) and its prepared-state descendant \(Q_B\cO_x\), so it carries the standard Sen--Zwiebach form once those tangent representatives have been chosen.

The BV generating functional for the first-order field redefinition is
\begin{equation}
\cU_x^{(1)}
=
\sum_{n\ge 1}\frac{1}{n!}\{\Psi^n;\cO_x\}. \bluecheck
\label{eq:U1}
\end{equation}
Its antibracket with $S_0$ generates the nonlinear part of the field redefinition in the BV sense. Because \(\cO_x\) belongs to the reference conformal theory \(x\), the ordinary CFT state-operator map of \(x\) identifies it with the state inserted at the distinguished puncture in \eqref{eq:U1}. No analogous canonical identification is being assumed for the nearby non-conformal theory \(y_\lambda\). The BRST variation of the distinguished \(\cO_x\) insertion is organized by the integrated form identity
\begin{equation}
\frac{1}{n!}\{Q_B\Psi,\Psi^{n-1};\cO_x\}
\;+\;
\frac{1}{n!}\{\Psi^n;Q_B\cO_x\}
=
-\int_{\partial \Gamma^{(\cO)}_{0,n;1}}\Omega[\underline\Psi;\cO_x]. \reasoncheck
\label{eq:OxQBchain}
\end{equation} \eqref{eq:OxQBchain} comes from integrating the critical form identity
\(\Omega[Q_B\underline\Psi;\cO_x]=-\delta\,\Omega[\underline\Psi;\cO_x]\)
over \(\Gamma^{(\cO)}_{0,n;1}\) and then shrinking the BRST contour onto the distinguished puncture. That shrinkage produces the prepared state \(Q_B\cO_x\). Expanding \(\{S_0,\cU_x^{(1)}\}\) gives the BRST-variation term on the left-hand side of \eqref{eq:OxQBchain} together with the ordinary sewing antibracket. Using \eqref{eq:Oxgamma1}, one obtains
\begin{equation}
-D_\lambda(\wh\Gamma)S_0+\{S_0,\cU_x^{(1)}\}
=
\sum_{n\ge2}\frac{1}{n!}
\int_{\partial \Gamma^{(\cO)}_{0,n;1}
-\Gamma^{\prime(\cO)}_{0,n|1}
+\Gamma^{(\cO)}_{0,n|1}
+\sum_{l=1}^{n-2}\{\Gamma^{(\cO)}_{0,l+1;1},\Gamma_{0,n-l+1}\}}
\Omega[\underline\Psi;\cO_x]
=0. \reasoncheck
\label{eq:OxfirstorderBV}
\end{equation} \eqref{eq:OxfirstorderBV} is therefore the actual first-order coefficient of the BV master equation written as a chain integral. The \(Q_B\cO_x\) terms have canceled already between \eqref{eq:DSlambda} and \eqref{eq:OxQBchain}. What remains is exactly the critical interpolation-space mechanism of Sen--Zwiebach, now applied to the distinguished \(\cO_x\) puncture. Thus the first-order deformation is controlled by critical interpolation geometry even though the deformed target theory may already be off critical through the induced special state \(Q_B\cO_x\).

We stop at first order in the background deformation. Higher orders require a careful handling of UV divergences endemic to higher orders in perturbation theory -- one may hope that a generalization of \eqref{eq:descent} removing the reliance on nice short-distance behavior may provide a natural UV regulator a la \cite{Sen:2019jpm}. A possible higher-order extrapolation of the first-order formulas is deferred to section \ref{sec:higher-order-bgi}, where it is recorded only as a tentative discussion point rather than a full proof.

\section{Conformal Matter Slightly Off Criticality}
\label{sec:deltac}

\noindent
We now turn from the general formalism to a concrete example in which every ingredient can be written down explicitly. The matter sector is still a CFT, but its central charge is shifted away from the critical value; the worldsheet theory is therefore conformal in the matter sector but off critical as a string background. In this setting one can identify the trace insertion \(\Theta\), the contour-dependent BRST charge \(Q_B[\gamma]\), the special state \(F\), and all of its descendants. The following section then applies the same antighost-dilaton state to organize the nearby linear-dilaton family.

\subsection{Conformal Matter Away from Criticality}

Consider a matter CFT with
\begin{equation}
c_m = 26+\Delta c_m. \bluecheck
\label{eq:cmdef}
\end{equation}
Here $\Delta c_m$ is the matter central-charge defect; $\Delta c_m=0$ is the critical case. A Weyl variation of the worldsheet metric inserts the trace anomaly
\begin{equation}
\Theta = -\frac{\Delta c_m}{48}\,R. \bluecheck
\label{eq:Theta}
\end{equation}
The symbol \(\Theta\) denotes the scalar trace insertion that measures Weyl non-invariance, not the tensor component \(T_{z\bar z}\) itself. In the complex-coordinate convention used throughout the paper, \(ds^2=2g_{z\bar z}\,dz\,d\bar z=e^{2\omega}|dz|^2\), so \(2g_{z\bar z}=e^{2\omega}\), \(\sqrt g=e^{2\omega}\), and \(\sqrt g\,R=-8\,\partial_z\bar\partial_z\omega\), as reviewed again in appendix \ref{app:antighost-descent}. In these conventions the Weyl anomaly is \(T^a{}_a=-(\Delta c_m/12)R\), while the corresponding \((1,1)\)-component satisfies
\begin{equation}
T_{z\bar z}=\sqrt g\,\Theta. \bluecheck
\label{eq:Thetazzbar}
\end{equation}
Equivalently, \(T^a{}_a=4\Theta\) in a locally flat frame. This fixes the coefficient \(1/48\) in \eqref{eq:Theta}. It is this same local insertion that appears in the curvature part of the BRST-anomaly descendant \(F^{[2]}\), and hence in the operator \(\mathfrak{k}\) through \eqref{eq:kOmega}. The corresponding BRST currents are
\begin{equation}
j_B
=
c\,T^{\mathrm m}
+ :bc\partial c:
+ \frac{3}{2}\partial^2 c
+ \tilde c\,T_{z\bar z},
\qquad
\bar j_B
=
\tilde c\,\bar T^{\mathrm m}
+ :\tilde b\tilde c\bar\partial \tilde c:
+ \frac{3}{2}\bar\partial^2 \tilde c
+ c\,T_{z\bar z}. \bluecheck
\label{eq:jB}
\end{equation}
where $T^{\mathrm m}$ and $\bar T^{\mathrm m}$ are the holomorphic and antiholomorphic matter stress tensors, and in the present conformal-matter example the trace component is the pure anomaly
\(
T_{z\bar z}=\sqrt g\,\Theta=-(\Delta c_m/48)\sqrt g\,R
\)
from \eqref{eq:Thetazzbar}. Writing the current in terms of \(T_{z\bar z}\) rather than a bare curvature term keeps the trace-anomaly origin visible and avoids suggesting spurious \(\partial R\) terms in the later divergence formula. Indeed, in \(\bar\partial j_B+\partial\bar j_B\) the terms in which the derivatives hit the stress tensor combine by stress-tensor conservation, so the divergence is controlled entirely by the same local trace component \(T_{z\bar z}\). The mixed OPE is especially simple at first order in $\Delta c_m$: \(T_{z\bar z}\) is a spectator coefficient, so the only singular contributions come from contracting the $c$ in the improvement term of $\bar j_B$ with the $b$ inside $:bc\partial c:$ in $j_B$, and similarly contracting the $\tilde c$ in the improvement term of $j_B$ with the $\tilde b$ inside $:\tilde b\tilde c\bar\partial\tilde c:$ in $\bar j_B$. Using $c(z)b(0)\sim z^{-1}$ and $\tilde c(\bar z)\tilde b(0)\sim \bar z^{-1}$ together with the holomorphic OPE check of the $j_B(z)j_B(0)$ term gives
\begin{equation}
j_B(z)j_B(0)
\sim
-\frac{\Delta c_m}{12z}\,c\partial^3 c(0),
\qquad
\bar j_B(\bar z)j_B(0)
\sim
\left(
\frac{c\partial c(0)}{z}
+\frac{\tilde c\bar\partial \tilde c(0)}{\bar z}
\right)T_{z\bar z}(0),
\reasoncheck
\label{eq:jjOPE}
\end{equation}
The antiholomorphic self-OPE is the left-right conjugate of the first term in \eqref{eq:jjOPE}, so the three channels needed for \(Q_B[\gamma]^2\) are fixed once \eqref{eq:jjOPE} is known. In this conformal-matter example one may rewrite the spectator coefficient as \(T_{z\bar z}=-(\Delta c_m/48)\sqrt g\,R\) if one wants the anomaly expressed explicitly in curvature language.
Using the standard contour representation
\begin{equation}
Q_B[\gamma]
=
\oint_\gamma \frac{dw}{2\pi i}\,j_B
-\oint_\gamma \frac{d\bar w}{2\pi i}\,\bar j_B, \bluecheck
\label{eq:QBcontour}
\end{equation}
and separating the holomorphic, antiholomorphic, and mixed OPE channels, together with the left-right conjugate of the first term in \eqref{eq:jjOPE}, one obtains
\begin{equation}
Q_B[\gamma]^2
=
-\frac{1}{2\pi i}\oint_\gamma
\left(
\frac{\Delta c_m}{12}\,c\partial^3 c\,dw
-\frac{\Delta c_m}{12}\,\tilde c\bar\partial^3\tilde c\,d\bar w
+\tilde c\bar\partial \tilde c\,T_{z\bar z}\,dw
-c\partial c\,T_{z\bar z}\,d\bar w
\right). \reasoncheck
\label{eq:QBsq}
\end{equation}
The defining relations \eqref{eq:descent} are then solved by
\begin{equation}
F
=
-\frac{\Delta c_m}{24}
(\partial c+\bar\partial \tilde c)
\bigl(c\partial^2 c - \tilde c\bar\partial^2 \tilde c\bigr). \reasoncheck
\label{eq:Fghostdilaton}
\end{equation}
This is the antighost dilaton familiar from the dilaton theorem \cite{Bergman:1994qq}---not to be confused with the ghost dilaton that describes a constant spacetime dilaton shift, although the two are closely related \cite{Bergman:1994qq,Rahman:1995ghost,Mazel:2024alu}. The special puncture is thus filled by the standard zero-momentum antighost-dilaton state.

\subsection{The Antighost-Dilaton Descendants}

It remains to verify that the antighost dilaton \eqref{eq:Fghostdilaton} realizes the full descendant family $F^{[k]}$ assumed in section \ref{sec:offcritical}: the one-form descendant \(F^{[1]}\), the two-form descendant \(F^{[2]}\), and the Weyl-response completion \eqref{eq:F2Weyl}. The contour-dependent charge \(Q_B[\gamma]\) is now explicit allowing us to fully fix $F^{[k]}$.

For later reference, the central-charge realization of the section-\ref{sec:offcritical} construction is
\begin{equation}
\begin{aligned}
Q_B[\gamma]
\text{ from \eqref{eq:jB} and \eqref{eq:QBcontour}}
&\Longrightarrow
F \text{ in \eqref{eq:Fghostdilaton}},
\\
\cB F
&\text{ gives the transgressed bundle one-form corresponding to }F^{[1]},
\\
F^{[2]}
&=
(\bar\partial j_B+\partial\bar j_B)\,dz\wedge d\bar z,
\\
\Xi[v_\omega]
&=
-4i\,\delta\omega_v\,\Theta\,d^2\sigma\,\sqrt g.
\end{aligned}
\label{eq:deltacdescentsummary}
\end{equation}
The computation of the local metric-adapted action of \(\cB\) on the antighost dilaton has three steps: evaluate the \(b,\tilde b\) contour OPEs; rewrite the resulting Taylor coefficients in terms of metric data using the Bergman--Zwiebach normalization; and act once more with \(\cB\) to recover the curvature descendant and the Weyl-response term.
To see directly how the metric enters its descendants, write
\begin{equation}
F= -\frac{\Delta c_m}{24}F_{\mathrm{agd}},
\qquad
F_{\mathrm{agd}}=XU,
\qquad
X:=\partial c+\bar\partial\tilde c,
\qquad
U:=c\partial^2 c-\tilde c\bar\partial^2\tilde c.
\end{equation}
Near a special puncture, let \(w\) be the metric-adapted local coordinate determined by \eqref{eq:specialmetriccoord}, let \(z=f(w)\) be the induced global map, and isolate the local contour part of \(\cB\) as
\begin{equation}
\cB
=
- \oint_{\partial D}\left(
\frac{dz}{2\pi i}\,\delta f(w)\,b(z)
-\frac{d\bar z}{2\pi i}\,\delta \bar f(\bar w)\,\tilde b(\bar z)
\right),
\qquad
\left(\frac{df}{dw}\right)^{-1}\sum_{m\ge 0}\delta f_m\,w^m
=:
\sum_{k\ge0}\alpha_k w^k. \bluecheck
\label{eq:Flocalalpha}
\end{equation}
The contour OPEs of \(b\) and \(\tilde b\) with the ghost factors in \(F_{\mathrm{agd}}\) then give the local transgression formula
\begin{equation}
\cB F_{\mathrm{agd}}
=
X\bigl(\alpha_0\partial^2 c-\bar\alpha_0\bar\partial^2\tilde c\bigr)
-(\alpha_1+\bar\alpha_1)U
-2X\bigl(\alpha_2 c-\bar\alpha_2\tilde c\bigr). \reasoncheck
\label{eq:Flocaldescent}
\end{equation}
This is where the metric-adapted geometry enters: the coefficients \(\alpha_k\) are not arbitrary local-coordinate data, but are determined by the Hermitian metric through the Bergman--Zwiebach normalization. In the fixed-metric regime used later in the genus-zero application, one has
\begin{equation}
\begin{split}
\delta\omega&=\partial_z\omega\,\delta z+\bar\partial_z\omega\,\delta\bar z,
\qquad
\alpha_0=e^{\omega}\delta z,
\qquad
\bar\alpha_0=e^{\omega}\delta\bar z,
\\
\alpha_1+\bar\alpha_1&=0,
\qquad
\alpha_2=-e^{-\omega}\partial_z\bar\partial_z\omega\,\delta\bar z,
\qquad
\bar\alpha_2=-e^{-\omega}\partial_z\bar\partial_z\omega\,\delta z. \reasoncheck
\end{split}
\label{eq:Fmetriccoeffs}
\end{equation}
Substituting these coefficients into \eqref{eq:Flocaldescent} gives the transgressed bundle one-form \(\cB F\), i.e. the pullback of the local worldsheet descendant \(F^{[1]}\) along the fixed-metric moving-puncture family:
\begin{equation}
\begin{split}
\cB F
=&\,
-\frac{\Delta c_m}{24}e^{\omega}(\partial c+\bar\partial\tilde c)
\bigl(\partial^2 c\,\delta z-\bar\partial^2\tilde c\,\delta\bar z\bigr)\\
&-\frac{\Delta c_m}{12}e^{-\omega}(\partial c+\bar\partial\tilde c)
\bigl(
c\,\partial_z\bar\partial_z\omega\,\delta\bar z
-\tilde c\,\partial_z\bar\partial_z\omega\,\delta z
\bigr)
+\delta(\cdots), \reasoncheck
\end{split}
\label{eq:F1ghostdilaton}
\end{equation}
The underlying local worldsheet one-form \(F^{[1]}\) is obtained from the same expression by replacing the bundle one-forms \(\delta z,\delta\bar z\) with the corresponding worldsheet differentials \(dz,d\bar z\). The omitted terms are exact worldsheet one-forms and therefore do not affect the contour integral entering \(Q_B[\gamma]^2\). Moreover, the same fixed-metric condition implies \(\cB X=-(\alpha_1+\bar\alpha_1)=0\), so a second application of \(\cB\) acts only on the one-form multiplying \(X\). This computes the ordered double bundle descendant. Since each contracted contour operator \(\cB(v)\) is odd, the associated bundle two-form satisfies
\begin{equation}
\iota_{v_2}\iota_{v_1}(\cB^2F)=2\,\cB(v_2)\cB(v_1)F. \reasoncheck
\label{eq:B2ordered}
\end{equation}
Therefore the worldsheet two-form descendant fixed by the \(Q_B\)-definition \eqref{eq:descent}, equivalently by the transgression relation \eqref{eq:Bdescent}, is represented by \(\frac12\cB^2F\), not by the ordered double action itself. The result is
\begin{equation}
F^{[2]}
=
-\frac{\Delta c_m}{48}
(\partial c+\bar\partial \tilde c)\,
\sqrt g\,R\,dz\wedge d\bar z. \reasoncheck
\label{eq:F2ghostdilaton}
\end{equation}
Appendix \ref{app:antighost-descent} contains the detailed local-coordinate algebra behind \eqref{eq:Flocaldescent}--\eqref{eq:F2ghostdilaton}, including the Taylor matching between the metric and the induced local coordinate.
The comparison with BRST current nonconservation is a comparison of two-forms, not of scalars. Rewriting \eqref{eq:QBcontour} as
\begin{equation}
Q_B[\gamma]
=
\frac{1}{2\pi i}\oint_\gamma (j_B\,dz-\bar j_B\,d\bar z), \reasoncheck
\label{eq:QBcontourForm}
\end{equation}
Stokes' theorem gives, for contours bounding an oriented region \(M\),
\begin{equation}
Q_B[\gamma_2]-Q_B[\gamma_1]
=
-\frac{1}{2\pi i}\int_M
(\bar\partial j_B+\partial \bar j_B)\,dz\wedge d\bar z. \reasoncheck
\label{eq:QBStokes}
\end{equation}
Comparing this with the defining descent relation in \eqref{eq:descent}, and using the trace-sector form of the anomaly discussed above, one identifies
\begin{equation}
F^{[2]}
=
(\bar\partial j_B+\partial \bar j_B)\,dz\wedge d\bar z
=
+(\partial c+\bar\partial \tilde c)\,T_{z\bar z}\,dz\wedge d\bar z. \reasoncheck
\label{eq:F2current}
\end{equation}
A word about normalizations: \eqref{eq:F2ghostdilaton} is a complex two-form, not a scalar. Using \(T_{z\bar z}=-(\Delta c_m/48)\sqrt g\,R\), the current-side result \eqref{eq:F2current} is equivalent to the curvature form \(F^{[2]}=-(\Delta c_m/48)(\partial c+\bar\partial \tilde c)\sqrt g\,R\,dz\wedge d\bar z\), so the comparison with Appendix \ref{app:antighost-descent} is immediate rather than requiring a second displayed equation.

The Weyl-response completion used in section \ref{sec:offcritical} is fixed by the standard Weyl-anomaly relation rather than by the fixed-metric second-descent calculation alone. What Appendix \ref{app:antighost-descent} does show directly is the local ghost-to-bundle exchange mechanism. From the middle term of \eqref{eq:appBFomega}, a pure Weyl variation at fixed puncture position produces the factor
\begin{equation}
(\cB X)(v_\omega)=2\,\delta\omega_v.
\end{equation}
This is the local mechanism by which the ghost factor \(X\) in the vertical descendant is exchanged for the bundle one-form \(\delta\omega_v\). Independently, the invariant measure convention of \eqref{eq:weylanomalyfunctional} gives the corresponding Weyl-response insertion
\begin{equation}
\Xi[v_\omega]
=
(2\pi i)\left(\frac{\Delta c_m}{24\pi}\,\delta\omega_v\,R\,d^2\sigma\,\sqrt g\right)
=
-4i\,\delta\omega_v\,\Theta\,d^2\sigma\,\sqrt g. \reasoncheck
\label{eq:F2WeylDeltacCheck}
\end{equation}
where the second equality uses \eqref{eq:Theta}. The prefactor \(+2\pi i\) is the net normalization already built into \eqref{eq:kOmega}: the standard extra-puncture factor together with the additional minus from shrinking the BRST contour on the outside of the ordinary local-coordinate disks. This is the anomaly-normalized Weyl-response insertion required by the universal axiom \eqref{eq:F2WeylDeltac}. The local complex-two-form representative \eqref{eq:F2ghostdilaton} is still the object that enters the contour-change formula \eqref{eq:QBStokes}; only the Weyl-response insertion is rewritten in invariant measure notation. With the coefficient in \eqref{eq:F2ghostdilaton}, the same local exchange law \((\cB X)(v_\omega)=2\,\delta\omega_v\) gives
\begin{equation}
(\cB F^{[2]})(v_\omega)
=
+(\cB X)(v_\omega)\,T_{z\bar z}\,dz\wedge d\bar z
=
-4i\,\delta\omega_v\,\Theta\,d^2\sigma\,\sqrt g
=
\Xi[v_\omega],
\end{equation}
where we used \(T_{z\bar z}=\sqrt g\,\Theta\) and \(dz\wedge d\bar z=-2i\,d^2\sigma\).
\diffnote{The sign now matches the universal Weyl-response insertion once the contour-orientation sign in \eqref{eq:kOmega} is included. In particular, the explicit central-charge realization and the anomaly-normalized insertion \(\Xi[v_\omega]\) agree.}

Physically, the antighost dilaton is the same obstruction appearing in recent conformal-perturbative SFT analyses: a shift of matter central charge creates an antighost-dilaton source in the string-field equation, and a linear-dilaton sector supplies the compensating deformation that trivializes that source in BRST cohomology \cite{Mazel:2024alu}. The mixed-vertex language encodes this mechanism directly through the special punctures and their descendants.

\diffnote{The coefficient in \eqref{eq:F2ghostdilaton} is obtained by converting the Appendix-A ordered double action of the odd contour operators into the corresponding antisymmetric bundle two-form. This restores agreement with the older formula recorded in \texttt{off-critical SFT.tex} after translating conventions.}

\diffnote{The convention \eqref{eq:cmdef} is fixed from the outset and then used uniformly. This corrects the unlabeled opening convention of section 4 in \texttt{main.tex} and is what reconciles the later formula \texttt{eqn:epsilon-0-solution} (4.25), the unlabeled first-order linear-dilaton relation in subsection 4.4.1, and the unlabeled exact central-charge relation near the end of subsection 4.4.2.}

\section{Nearly Marginal Linear-Dilaton Sector}
\label{sec:deltac-linear}

\noindent
The central-charge example of the previous section identified the antighost dilaton as the special state; we now ask what the string-field equations of motion look like in its presence. The simplest testing ground is a one-parameter linear-dilaton family, where the mixed-vertex formalism can be pushed beyond the abstract descendant family and into explicit differential equations.

\subsection{Linearized Setup}

To study explicit string-field solutions, isolate one matter coordinate $Y$ with linear-dilaton stress tensor
\begin{equation}
T_Y
=
-\frac{1}{\alpha'}(\partial Y)^2
+\frac{\beta}{\sqrt{2\alpha'}}\,\partial^2 Y,
\qquad
c_\beta = 1+3\beta^2. \bluecheck
\label{eq:TY}
\end{equation}
where $\beta$ is the slope of the reference linear-dilaton background and $c_\beta$ is the contribution of the $Y$ sector to the matter central charge. Let the remaining matter sector be a spectator CFT with whatever central charge is needed to maintain \eqref{eq:cmdef}. We use the nearly marginal basis
\begin{equation}
\cV_D := c\partial^2 c - \tilde c\bar\partial^2\tilde c,
\qquad
\cV_G := c\tilde c\,\partial Y\,\bar\partial Y,
\qquad
\cV_K := (\partial c+\bar\partial \tilde c)(c\partial Y-\tilde c\bar\partial Y).
\bluecheck
\label{eq:VDVGVK}
\end{equation}
The three basis states have distinct roles: $\cV_D$ is the ghost-dilaton direction, $\cV_G$ is the metric fluctuation in the $Y$ direction, and $\cV_K$ is a pure-gauge descendant that mixes with them under BRST variation. This basis is the one used in the string-field treatment of nearly marginal sigma-model deformations and linear-dilaton shifts \cite{Bergman:1994qq,Mazel:2024alu,Frenkel:2025wko,Kim:2026stringloops}.
Write the string field as
\begin{equation}
\Psi = \sum_{n=0}^\infty \eps^{n+1}\Psi_n,
\bluecheck
\label{eq:epsansion}
\end{equation}
where $\eps$ is a bookkeeping parameter for the deformation amplitude and the coefficient fields $\Psi_n$ are independent of $\eps$,
and at leading order use the ansatz
\begin{equation}
\Psi_0
=
\mathcal A_0(Y)\,\cV_D
+\mathcal B_0(Y)\,\cV_G
+\mathcal K_0(Y)\,\cV_K.
\bluecheck
\label{eq:Psi0general}
\end{equation}
Here and below, primes on $\mathcal A(Y)$, $\mathcal B(Y)$, and $\mathcal K(Y)$ denote ordinary derivatives with respect to the single matter coordinate $Y$.

Three parameters should be kept distinct: $\beta$, the slope of the \emph{reference} linear-dilaton background; $\Delta c_m$, the matter central-charge defect of that reference background; and $\eps$, the amplitude of the string-field deformation toward a nearby critical background. The logic has three steps. First, BRST variation gives the differential system \eqref{eq:linearizedsystem}. Second, that system rules out a linearized flat-space solution. Third, for $\beta\neq 0$ the same string field is reinterpreted as a shift of the linear-dilaton slope.

Acting with the BRST charge in the linear-dilaton background and matching the coefficients of the three independent basis states gives the linearized system
\begin{equation}
\begin{split}
&-\mathcal A_0'(Y)-\frac{\alpha'}{4}\mathcal B_0'(Y)+\mathcal K_0(Y)+\frac{\sqrt{\alpha'}\beta}{2\sqrt2}\mathcal B_0(Y)=0,\\
&-\frac{\alpha'}{4}\mathcal A_0''(Y)+\frac{\alpha'}{4}\mathcal K_0'(Y)+\frac{\sqrt{\alpha'}\beta}{2\sqrt2}\bigl(\mathcal A_0'(Y)+\mathcal K_0(Y)\bigr)=\frac{\Delta c_m}{24},\\
&-\frac{\alpha'}{4}\mathcal B_0''(Y)=0. \reasoncheck
\end{split}
\label{eq:linearizedsystem}
\end{equation}
These three equations organize both the flat-space and linear-dilaton analyses below. They arise from the worldsheet contractions of the BRST current with the three basis states; in flat space, before including the linear-dilaton improvement term, one finds
\begin{equation}
\begin{split}
Q_B\bigl(\mathcal A(Y)\cV_D\bigr)
&=
-\mathcal A'(Y)\,c\tilde c\bigl(\partial^2 c\,\bar\partial Y+\bar\partial^2\tilde c\,\partial Y\bigr)
-\frac{\alpha'}{4}\mathcal A''(Y)(\partial c+\bar\partial\tilde c)\cV_D,\\
Q_B\bigl(\mathcal B(Y)\cV_G\bigr)
&=
-\frac{\alpha'}{4}\mathcal B''(Y)(\partial c+\bar\partial\tilde c)\cV_G
-\frac{\alpha'}{4}\mathcal B'(Y)c\tilde c\bigl(\partial^2 c\,\bar\partial Y+\bar\partial^2\tilde c\,\partial Y\bigr),\\
Q_B\bigl(\mathcal K(Y)\cV_K\bigr)
&=
\mathcal K(Y)c\tilde c\bigl(\partial^2 c\,\bar\partial Y+\bar\partial^2\tilde c\,\partial Y\bigr)
+\frac{\alpha'}{4}\mathcal K'(Y)(\partial c+\bar\partial\tilde c)\cV_D. \reasoncheck
\end{split}
\label{eq:QBbasis}
\end{equation}
The linear-dilaton improvement adds the terms proportional to $\beta$ in \eqref{eq:linearizedsystem}. The local-coordinate transformations needed for the explicit vertex computations are collected in appendix \ref{app:nlsm-local-coords}, while the descent of the antighost dilaton is derived in appendix \ref{app:antighost-descent}.

\subsection{Flat Space: No Linearized Solution}
\label{subsec:flat-space}

Set $\beta=0$. Then the system \eqref{eq:linearizedsystem} becomes
\begin{equation}
\begin{split}
&-\mathcal A_0'(Y)-\frac{\alpha'}{4}\mathcal B_0'(Y)+\mathcal K_0(Y)=0,\\
&-\frac{\alpha'}{4}\mathcal A_0''(Y)+\frac{\alpha'}{4}\mathcal K_0'(Y)=\frac{\Delta c_m}{24},\\
&-\frac{\alpha'}{4}\mathcal B_0''(Y)=0. \reasoncheck
\end{split}
\label{eq:flatsystem}
\end{equation}
These equations are inconsistent unless $\Delta c_m=0$: differentiating the first and using the third contradicts the second. The linearized equation of motion
\begin{equation}
Q_B\Psi_0 = -F \reasoncheck
\label{eq:flatlinearized}
\end{equation}
therefore has no solution in flat space. This is not a failure of the formalism but a diagnosis: flat space is the wrong background. The obstruction is first-order in $\Delta c_m$, while the matter central charge varies only quadratically with the linear-dilaton slope, so the natural expansion parameter is $\eps \sim \sqrt{|\Delta c_m|}$ rather than $\Delta c_m$ itself.

At order $\eps$ one accordingly solves the homogeneous equation $Q_B\Psi_0=0$, and a representative is
\begin{equation}
\Psi_0
=
\frac{\eps_0}{\sqrt{2\alpha'}}
\bigl(
Y\,\cV_D + \cV_K
\bigr). \calccheck
\label{eq:Psi0flat}
\end{equation}
This state represents an infinitesimal shift of the linear-dilaton slope away from the flat reference background, in agreement with the recent conformal-perturbative string-field analysis of \cite{Mazel:2024alu}.

At the next order, the equation of motion is
\begin{equation}
Q_B\Psi_1 + \frac{1}{2}[\Psi_0^2] + F = 0. \reasoncheck
\label{eq:flatsecond}
\end{equation}
where $[\Psi_0^2]$ denotes the ordinary CSFT cubic product generated by the three-string vertex.
Decomposing \eqref{eq:flatsecond} along the basis $\cV_D,\cV_G,\cV_K$ produces a coupled system for $\mathcal A_1,\mathcal B_1,\mathcal K_1$. Appendix \ref{app:flat-compatibility} evaluates the ordinary quadratic bracket \([\Psi_0^2]\) directly from the flat-space three-string vertices and BPZ pairings of the same nearly marginal sector \cite{Mazel:2024alu,Mazel:2025diffeo}, rewrites the result in the present basis, and derives the compatibility condition \eqref{eq:flatrelation}.
Mutual consistency of the resulting differential equations then requires
\begin{equation}
\Delta c_m = -12 \eps_0^2. \reasoncheck
\label{eq:flatrelation}
\end{equation}
Thus a real solution exists only for $\Delta c_m<0$, namely when the reference matter theory is subcritical and can relax toward a linear-dilaton background. \eqref{eq:flatrelation} is also the $\beta\to 0$ limit of the exact central-charge relation \eqref{eq:exactdelta} derived in subsection \ref{subsec:linear-dilaton-bookkeeping}. That exact relation should be read as a consistency condition in the neighboring linear-dilaton family, not as a substitute for the cubic-vertex derivation of \eqref{eq:flatrelation}; it reproduces the same final formula without determining the individual coefficients in $[\Psi_0^2]$.

\diffnote{The flat-space subsection of \texttt{main.tex} already contained the physically correct relation \eqref{eq:flatrelation}, namely \texttt{eqn:epsilon-0-solution} (4.25). The present section uses one central-charge convention throughout so that this relation agrees with the rest of the central-charge discussion.}

\diffnote{The flat-space ``Second Order Solution'' block in \texttt{main.tex} already contained the correct obstruction relation \eqref{eq:flatrelation}, but not a direct derivation of the source coefficients in the \(\cW_1,\cW_2,\cW_3\) basis. Appendix \ref{app:flat-compatibility} now evaluates the ordinary quadratic bracket directly from the flat-space cubic products and BPZ pairings. This corrects the intermediate coefficients while leaving the final relation \eqref{eq:flatrelation} unchanged.}

\subsection{Linear Dilaton and the Exact Central-Charge Relation}
\label{subsec:linear-dilaton-bookkeeping}

Now take $\beta\neq 0$, so the reference background already has a linear dilaton. The same string-field state \eqref{eq:Psi0flat} shifts the slope:
\begin{equation}
\beta \longrightarrow \beta - 2\eps. \bluecheck
\label{eq:betashift}
\end{equation}
The induced change in matter central charge is therefore
\begin{equation}
\Delta c_{\beta}
=
3\bigl[(\beta-2\eps)^2-\beta^2\bigr]
=
-12\beta\eps + 12\eps^2. \calccheck
\label{eq:Deltac}
\end{equation}
Reaching a critical background requires $\Delta c_{\beta} = -\Delta c_m$, which gives the exact relation
\begin{equation}
\Delta c_m = 12\beta\eps - 12\eps^2. \calccheck
\label{eq:exactdelta}
\end{equation}
To first order this gives \(\Delta c_m = 12\beta\eps + O(\eps^2)\), which is precisely the linearized solution of \eqref{eq:linearizedsystem}. Expanding the exact solution for small $\Delta c_m$ gives
\begin{equation}
\eps
=
\frac{\beta-\sqrt{\beta^2-\Delta c_m/3}}{2}
=
\frac{\Delta c_m}{12\beta}
+\frac{\Delta c_m^2}{144\beta^3}
+O(\Delta c_m^3). \calccheck
\label{eq:epsseries}
\end{equation}
We choose the branch for which $\eps\to 0$ as $\Delta c_m\to 0$. The square root remains real for $\Delta c_m\le 3\beta^2$, which is precisely the regime in which the nearby critical background stays within this one-parameter linear-dilaton family.

\eqref{eq:exactdelta} is the exact central-charge relation within the nearby one-parameter linear-dilaton family. Its $\beta\to 0$ limit reproduces \eqref{eq:flatrelation}, while its first-order expansion reproduces the first-order relation extracted from \eqref{eq:linearizedsystem}. The explicit genus-zero mixed-vertex computation in the nearby linear-dilaton family should reproduce the next term in \eqref{eq:epsseries}; subsection \ref{subsec:lowpoint-second-order} does so in one explicit low-point representative, and the remaining task is to obtain the same result by a direct intrinsic evaluation of the mixed bracket.

\diffnote{The sign choice in \eqref{eq:exactdelta} is the key repair to the linear-dilaton discussion. It reconciles the unlabeled first-order relation in subsection 4.4.1 of \texttt{main.tex} with the unlabeled exact central-charge formula near the end of subsection 4.4.2 and the flat-space result \texttt{eqn:epsilon-0-solution} (4.25).}

\subsection{Low-Point Scheme for the Second-Order Check}
\label{subsec:lowpoint-second-order}

We fix a low-point representative of the local-coordinate and metric data in order to compare the mixed bracket \([\Psi_0;F]\) with the kinetic correction built from \(F^{[2]}\). Three choices enter: a Hermitian metric on the sphere, a contour representative for the kinetic term, and local-coordinate representatives of \(\Gamma_{0,3}\), \(\Gamma_{0,2;1}\), and \(\Gamma_{0,1;2}\). We describe these in turn. Take the sphere to have global coordinate \(z\) and regard it as two unit disks glued along the seam \(|z|=1\), with metric
\begin{equation}
g_{z\bar z}
=
\begin{cases}
1, & |z|\le 1,\\[3pt]
|z|^{-4}, & |z|\ge 1.
\end{cases}. \reasoncheck
\label{eq:gluedmetric}
\end{equation}
This metric is continuous across the seam, while its derivatives jump there. The exterior region is flat because the inversion coordinate \(\zeta=1/z\) pulls the metric back to \(|d\zeta|^2\).

In a matter theory with central-charge defect \(\Delta c_m\), changing the Weyl frame of the worldsheet partition function is governed by the local anomaly functional. With the convention \(ds^2=e^{2\omega}|dz|^2\), we write the Weyl shift as \(g'_{ab}=e^{2\omega}g_{ab}\). The partition functions are then related by
\begin{equation}
Z[g']
=
Z[g]\,
\exp\!\left(
\frac{\Delta c_m}{24\pi}
\int d^2\sigma\,\sqrt g\,
\bigl[(\nabla\omega)^2+R\,\omega\bigr]
\right). \bluecheck
\label{eq:weylanomalyfunctional}
\end{equation}
\diffnote{Equation \eqref{eq:weylanomalyfunctional} disagrees with \texttt{main.pdf} equation \texttt{eqn:CFT-Z-transformation} (4.12). The latter writes \(Z[e^{\omega}g]\) with coefficient \(\Delta c_m/6\) and measure \(d^2z\), whereas the present paper uses the book-compatible Weyl-anomaly formula in invariant measure \(d^2\sigma\sqrt g\) with coefficient \(\Delta c_m/(24\pi)\), translated to the local convention \(g'_{ab}=e^{2\omega}g_{ab}\).}
Thus a change of Weyl frame can contribute at the same perturbative order as the insertion of additional special punctures. For the kinetic term we take the BRST contour to have two connected components, one on either side of the seam. In this glued-disk representative the contour is the analogue of the central geodesic of the cylinder, and dragging it toward the puncture sweeps out exactly one hemisphere. Figure \ref{fig:QB-action} should be read with this contour choice in mind.

For the basic three-string vertex we choose the flat local coordinates
\begin{equation}
f_1(w)=w,
\qquad
f_2(w)=\frac{1}{w},
\qquad
f_3(w)=1+w. \bluecheck
\label{eq:Gamma03coords}
\end{equation}
The first two maps place punctures at \(z=0\) and \(z=\infty\) in the flat patches of the two hemispheres, while the third places a puncture at \(z=1\) with local coordinate disk tangent to the seam from one side. The puncture at \(z=1\) is understood as lying infinitesimally to that chosen side of the seam, so that its local coordinate disk sits entirely in a flat patch of the glued metric. As usual, the cubic string product is obtained by symmetrizing over the ordinary punctures; the displayed maps are one convenient representative of that symmetrized geometry.

\begin{figure}[H]
\centering
\includegraphics[width=0.8\textwidth]{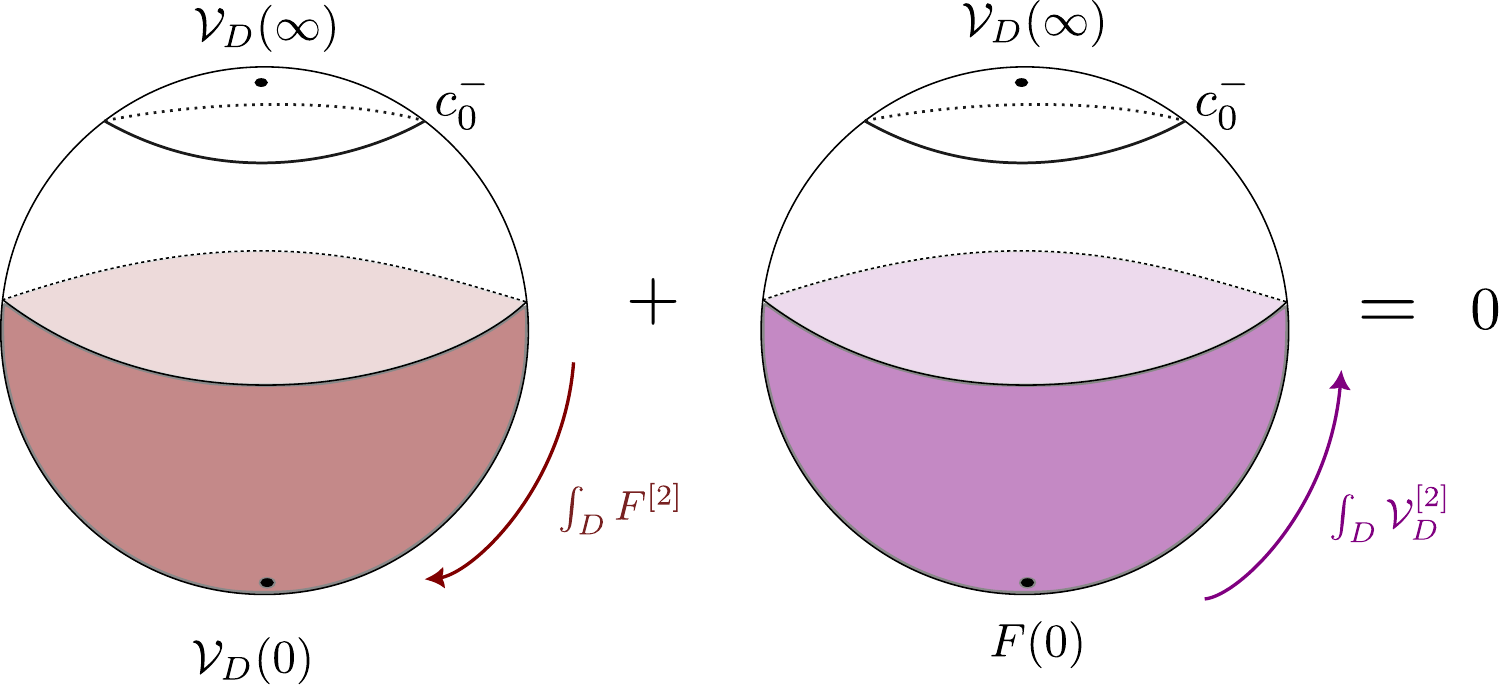}
\caption{First step in the low-point organization of the mixed bracket $[\Psi_0;F]$ in the linear-dilaton sector. Interpolating along $\Gamma_{0,2;1}$ symmetrizes the dilaton insertion and the special puncture. The two orientations of the angular average contribute with opposite signs.}
\label{fig:psi-F-evaluation-1-streamlined}
\end{figure}

\begin{figure}[H]
\centering
\includegraphics[width=\textwidth]{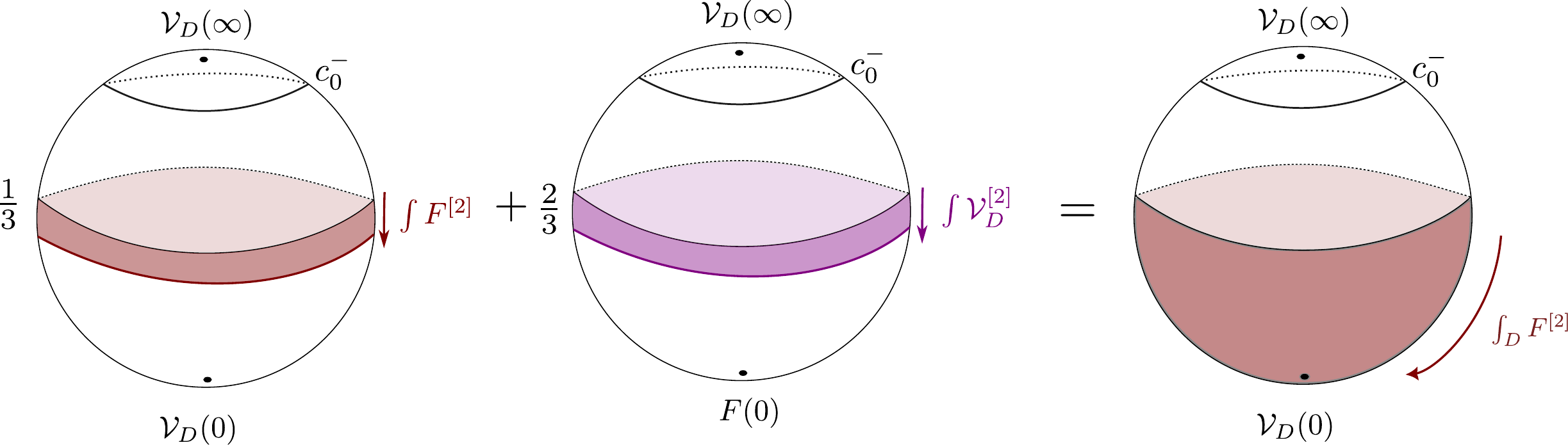}
\caption{Second step in the low-point organization of $[\Psi_0;F]$. After symmetrization, the insertion is transported from the midpoint of the sphere to the third puncture of the chosen representative of $\Gamma_{0,3}$.}
\label{fig:psi-F-evaluation-2-streamlined}
\end{figure}

The one-parameter chain \(\Gamma_{0,2;1}\) is then represented by moving the distinguished special puncture through the ordinary puncture in this glued-disk geometry; the corresponding picture is shown in figure \ref{fig:Gamma-021-viz-streamlined}. In this representative the punctures are literally moved through one another along a one-dimensional path on the sphere. Because the antighost dilaton has regular OPEs with the low-point fields used in the nearly marginal sector, this interpolation requires no additional local counterterms at the order considered here. The corresponding \(\Gamma_{0,1;2}\) is obtained from the same local geometry by reinterpreting one ordinary puncture as the distinguished special puncture. These are the low-point conventions used again in figures \ref{fig:psi-F-evaluation-1-streamlined} and \ref{fig:psi-F-evaluation-2-streamlined}.

The choices above define one explicit low-point representative for organizing the equation of motion, not a preferred action-defining string-field frame. Different choices of string vertices or low-point representatives are related by field redefinition, and the background-independence framework allows the calculation to be carried out in any such representative. The coefficient-level comparison in \eqref{eq:linear-dilaton-second-order}--\eqref{eq:tentativeepsshift} is written in this representative.

\diffnote{Compared with the old ``Vertices and Local Coordinates'' block and the formula \texttt{eqn:CFT-Z-transformation} in \texttt{main.tex}, the low-point discussion here separates ingredients with different status: the Weyl-anomaly formula is corrected, the geometric organization of the mixed bracket is stated explicitly, and the calculation is organized in one convenient low-point representative rather than presented as a preferred action-level frame.}

At the next order in the nearly marginal sector, the equation of motion becomes
\begin{equation}
Q_B\Psi_1
-\frac{1}{2\pi i}\int_D F^{[2]}\,\Psi_0
+\frac{1}{2}[\Psi_0^2]
+[\Psi_0;F]
+[;F^2]
=0. \reasoncheck
\label{eq:linear-dilaton-second-order}
\end{equation}
Here \(D\) denotes the hemisphere of the glued sphere bounded by the seam contour and containing the third puncture of \eqref{eq:Gamma03coords}; dragging the BRST contour across the complement of the local-coordinate disks in the low-point representative fixed by \eqref{eq:gluedmetric} and \eqref{eq:Gamma03coords} produces precisely the integral over that region. The notation \([;F^2]\) denotes the mixed-vertex state with no ordinary punctures and two special punctures. For the projected nearly marginal equation \eqref{eq:linear-dilaton-second-order}, one does not need the full state \([;F^2]\), only its projection onto the span of \(\cV_D,\cV_G,\cV_K\). Since the special state \(F\) in \eqref{eq:Fghostdilaton} is purely ghost and contains no \(Y\)-field, that projection cannot carry a nontrivial function of \(Y\); it must therefore take the form
\begin{equation}
[;F^2]_{\rm nm}
=
a_D\,\cV_D
+a_G\,\cV_G
+a_K\,\cV_K,
\label{eq:F2nmform}
\end{equation}
with coefficients depending on the background parameters but not on \(Y\).

To determine these coefficients, one pairs \([;F^2]_{\rm nm}\) with test insertions obtained by multiplying the basis insertions by Schwartz functions of \(Y\). Testing against \(f(Y)\cV_G\) and \(f(Y)\cV_K\) gives zero for every \(f\in\mathcal S(\mathbb R)\): each contributing correlator contains a single \(Y\)-oscillator coming from \(\cV_G\) or \(\cV_K\), while neither copy of \(F\) contains any \(Y\)-field into which it can Wick contract. Hence \(a_G=a_K=0\).

For the remaining ghost-dilaton coefficient one must use the background-charge-dressed test insertion \(f(Y)e^{\beta Y}\cV_D\). Using the linearized equation of motion \(Q_B\Psi_0=-F\), one may rewrite the quadratic bracket \(\{\,\cdot\,;F^2\}\) as \(\{\,\cdot\,;(Q_B\Psi_0)^2\}\); the sign of \(\Psi_0\) is irrelevant because the bracket is quadratic in \(Q_B\Psi_0\). At the order considered here, one may evaluate the resulting matrix element with the undeformed form identity \eqref{eq:QBcritical}. The reason is that the off-critical correction \eqref{eq:QBplusk} would insert one additional distinguished copy of \(F^{[2]}\), and therefore one more power of the defect sector, so it starts only at order \(F^3\) rather than in the present order-\(F^2\) computation. Let \(\Omega[f(Y)e^{\beta Y}\cV_D;\Psi_0,Q_B\Psi_0]\) denote the antisymmetrized form obtained by replacing the two special insertions by \(\Psi_0\) and \(Q_B\Psi_0\), namely \(\Omega[f(Y)e^{\beta Y}\cV_D;\Psi_0,Q_B\Psi_0]:=\left.\frac{d}{d\lambda}\right|_{\lambda=0}\Omega[f(Y)e^{\beta Y}\cV_D;(\Psi_0+\lambda Q_B\Psi_0)^{\otimes 2}]\). Then
\begin{equation}
\bra{f(Y)e^{\beta Y}\cV_D}c_0^-\ket{[;F^2]}
=
\{f(Y)e^{\beta Y}\cV_D;(Q_B\Psi_0)^2\}
=
-\int_{\partial\Gamma_{0,1;2}}\Omega[f(Y)e^{\beta Y}\cV_D;\Psi_0,Q_B\Psi_0].
\label{eq:F2VDboundary}
\end{equation}
The boundary correlator on the right-hand side is evaluated in the same low-point representative \eqref{eq:Psi0flat}. The \(\cV_K\) component of \(\Psi_0\) drops out immediately, because it carries \(\partial Y\) or \(\bar\partial Y\) while \(Q_B\Psi_0=-F\) and the dressed test insertion contain no \(Y\)-oscillator. Thus only the \(Y\cV_D\) term in \eqref{eq:Psi0flat} contributes.

The factor \(e^{\beta Y}\) in the test insertion precisely compensates the linear-dilaton background charge, so the remaining zero-mode dependence of the boundary term is proportional to the first moment \(\int dY_0\,Y_0\,f(Y_0)\). In the standard function-valued/NLSM treatment of the linear-dilaton zero mode, this vanishes for every even Schwartz function \(f\). On the other hand, if \(a_D\neq 0\) in \eqref{eq:F2nmform}, then the left-hand side of \eqref{eq:F2VDboundary} is proportional to \(a_D\int dY_0\,f(Y_0)\): the BPZ pairing \(\langle e^{\beta Y}\cV_D\,|\,c_0^-\,|\,\cV_D\rangle\) is nonvanishing because the factor \(e^{\beta Y}\) saturates the background charge. Choosing an even \(f\) with \(\int dY_0\,f(Y_0)\neq 0\) therefore forces \(a_D=0\).

Hence \([;F^2]_{\rm nm}=0\), and \([;F^2]\) does not contribute to \eqref{eq:linear-dilaton-second-order} in the nearly marginal sector of this low-point scheme.
In this representative the bundle-degree completion of the distinguished \(F^{[2]}\) insertion in \eqref{eq:kOmega} does not contribute, because the low-point interpolation keeps the metric fixed. Only the vertical fixed-metric component of \(F^{[2]}\) enters the projected equation of motion at this order.

The ordinary cubic bracket \([\Psi_0^2]\) is the same cubic bracket that appears in the flat-vertex analysis of the \(\beta=0\) sector discussed in subsection \ref{subsec:flat-space}. The genuinely new off-critical input is the comparison between the mixed vertex \([\Psi_0;F]\) and the kinetic correction built from the descendant \eqref{eq:F2ghostdilaton}. Because the OPE of \(F^{[2]}\) with \(\Psi_0\) is regular, only the dilaton component \(Y\cV_D\) of \eqref{eq:Psi0flat} contributes to that kinetic term.

The mixed bracket is organized by two geometric steps. First, one interpolates along \(\Gamma_{0,2;1}\) so that the dilaton insertion and the special puncture are symmetrized. Averaging the resulting one-form insertion over the angular direction converts it to the corresponding two-form descendant, and the two orientations contribute with opposite signs. This is the step shown in figure \ref{fig:psi-F-evaluation-1-streamlined}.

The second move transports the resulting insertion from the midpoint of the sphere to the third puncture of the chosen low-point representative of $\Gamma_{0,3}$. Since the curvature is localized at the seam between the two hemispheres, this transport picks out one hemisphere's contribution in the same Gauss--Bonnet normalization that appears in the kinetic term. This is the step shown in figure \ref{fig:psi-F-evaluation-2-streamlined}.

Taken together, these two steps identify the mixed bracket with the same hemisphere integral that appears in the kinetic correction, namely \((2\pi i)^{-1}\int_D F^{[2]}\,\Psi_0\). In the first step, the two symmetrized contributions come with opposite orientations and combine into the angularly averaged two-form descendant. In the second step, that descendant sweeps exactly one copy of the hemisphere \(D\) in the chosen representative of \(\Gamma_{0,3}\). The mixed bracket therefore reproduces precisely the same hemisphere integral as the kinetic correction:
\begin{equation}
-\frac{1}{2\pi i}\int_D F^{[2]}\,\Psi_0 + [\Psi_0;F] = 0. \reasoncheck
\label{eq:tentativecancellation}
\end{equation}
Combining this cancellation with the \(\beta=0\) second-order equations discussed above gives
\begin{equation}
\begin{split}
&-\mathcal A_1'(Y)-\frac{\alpha'}{4}\mathcal B_1'(Y)+\mathcal K_1(Y)+\frac{\sqrt{\alpha'}\beta}{2\sqrt2}\mathcal B_1(Y)+\frac{5}{12}\left(\frac{\Delta c_m^2}{72\beta^2\alpha'}\right)Y=0,\\
&-\frac{\alpha'}{4}\mathcal A_1''(Y)+\frac{\alpha'}{4}\mathcal K_1'(Y)+\frac{\sqrt{\alpha'}\beta}{2\sqrt2}\bigl(\mathcal A_1'(Y)+\mathcal K_1(Y)\bigr)-\frac{\alpha'}{48}\left(\frac{\Delta c_m^2}{72\beta^2\alpha'}\right)=0,\\
&-\frac{\alpha'}{4}\mathcal B_1''(Y)+\frac{\sqrt{\alpha'}\beta}{2\sqrt2}\mathcal B_1'(Y)-\frac{1}{2}\left(\frac{\Delta c_m^2}{72\beta^2\alpha'}\right)=0. \reasoncheck
\end{split}
\label{eq:tentativesecondordersystem}
\end{equation}
The three inhomogeneous terms are exactly the flat-space coefficients from the \(\beta=0\) cubic computation, rewritten using the first-order relation \(\Delta c_m=12\beta\eps+O(\eps^2)\), namely \(\eps_0=\Delta c_m/(12\beta)\), while the terms proportional to \(\beta\) are the same linear-dilaton BRST contributions that already appear in \eqref{eq:linearizedsystem}. The individual coefficients in these inhomogeneous terms depend on the chosen representative, but the compatibility condition extracted from the existence of a solution is the invariant quantity.
Differentiating the first equation in \eqref{eq:tentativesecondordersystem} and using the other two to eliminate \(\mathcal A_1''\) and \(\mathcal B_1''\) gives
\begin{equation}
-\frac{2\beta}{\sqrt{2\alpha'}}\bigl(\mathcal A_1'(Y)+\mathcal K_1(Y)\bigr)
+\frac{\Delta c_m^2}{72\beta^2\alpha'}
=0. \calccheck
\label{eq:tentativecompatibility}
\end{equation}
so that the constant term would imply a second-order shift
\begin{equation}
\Delta\eps=\frac{\Delta c_m^2}{144\beta^3}. \calccheck
\label{eq:tentativeepsshift}
\end{equation}
This is exactly the \(O(\Delta c_m^2)\) term in the series \eqref{eq:epsseries}. Thus the low-point computation in this explicit representative reproduces the second-order term in the nearby linear-dilaton family.

\section{Discussion and Outlook}
\label{sec:discussion}

The main lesson of this paper is that Zwiebach's off-critical construction gives a concrete description of nearby worldsheet backgrounds when exact BRST nilpotency fails. The defect is carried by a single fixed state \(F\), and the resulting deformation of the BV geometry is organized by mixed vertices with special punctures whose local coordinates are determined by the worldsheet metric. Rewritten in this language, the off-critical formalism dovetails naturally with Sen--Zwiebach background independence \cite{Sen:1990hh,Sen:1993mh,Sen:1993kb,Zwiebach:1996jc,Zwiebach:1996ph}: the same interpolation-space geometry that relates neighboring conformal backgrounds continues to operate off the conformal locus, supplemented by an additional layer of worldsheet data recording the failure of conformal invariance.

\subsection{Relation to Weyl-Orbit Averaging}

The construction of this paper does not require a prior notion of averaging over Weyl frames, and for that reason sections \ref{sec:offcritical} and \ref{sec:bgindep} were formulated directly in terms of punctured spheres equipped with a Hermitian metric. Nevertheless, the Weyl-orbit picture provides a useful heuristic guide to the formalism. If \(\omega_1\) and \(\omega_2\) are two Weyl frames on the same punctured sphere, then the trace insertion \(\Theta\) (related to the worldsheet stress tensor by \(T^a{}_a=4\Theta\)) generates their relative change through
\begin{equation}
Z[\omega_2]
=
\left\langle
\exp\!\left(
\frac{2}{\pi}\int d^2\sigma\,\sqrt g\,(\omega_2-\omega_1)\,\Theta
\right)
\right\rangle_{\omega_1}. \reasoncheck
\label{eq:weylshift}
\end{equation}
Equivalently, the exponent can be written as \((1/2\pi)\int d^2\sigma\,\sqrt g\,(\omega_2-\omega_1)\,T^a{}_a\).
Writing \(\delta\omega:=\omega_2-\omega_1\) for motion along the Weyl orbit, we can define a Weyl-invariant quantity $\tilde{Z}$ by formally averaging over the Weyl orbit:
\begin{equation}
\widetilde Z
:=
\frac{1}{\mathcal V}\int\mathcal D\delta\omega\,
\left\langle
\exp\!\left(
\frac{2}{\pi}\int d^2\sigma\,\sqrt g\,\delta\omega\,\Theta
\right)
\right\rangle_{\omega_1}. \reasoncheck
\label{eq:weylaverage}
\end{equation}
This expression is only a heuristic starting point: one must still gauge-fix the noncompact Diff$\rtimes$Weyl orbit and control the functional measure. But its expansion in the number of $\Theta$ insertions (equivalent to expanding in the number of $F^{[k]}$ insertions) already makes visible why the off-critical vertices carry one extra real direction per special puncture. Each additional integrated \(\Theta\) insertion carries its worldsheet position together with one local scale direction, so the first correction to an ordinary correlator is naturally three-dimensional rather than two-dimensional at each special insertion.

If \(C_i:=\left\langle \int d^2z\,\cO_i(z)\right\rangle\), then the first-order correction from averaging over the Weyl orbit is formally
\begin{equation}
\widetilde C_i
=
C_i+\frac{2}{\pi\mathcal V}\int\mathcal D\delta\omega\int d^2z_1\,d^2\sigma_2\,\sqrt g\,
\delta\omega(z_2)\,\langle \cO_i(z_1)\Theta(z_2)\rangle_{\omega_1}
+O(\Theta^2).
\reasoncheck
\label{eq:weylaverageobservable}
\end{equation}
At this stage the extra direction is still described as a local scale variable. Formally gauge fixing the two insertion positions together with the local scale at the \(\Theta\) insertion produces
\begin{equation}
\cO_i(z_1)\,d^2z_1 \mapsto c\tilde c\,\cO_i(z_1),
\qquad
\frac{2}{\pi}\,\Theta(z_2)\,d^2\sigma_2\,\sqrt g\,d[\delta\omega(z_2)]
\mapsto
(\partial c+\bar\partial\tilde c)\,c\tilde c\,\Theta(z_2).
\reasoncheck
\label{eq:weylghostfixing}
\end{equation}
Fixing the remaining relative twist angle between the two punctures then gives
\begin{equation}
\widetilde C_i
=\,
C_i+
\bra{c\tilde c\,\cO_i}c_0^-
\ket{(\partial c+\bar\partial\tilde c)\,c\tilde c\,\Theta}
+O(\Theta^2).
\reasoncheck
\label{eq:weylaverageobservableghost}
\end{equation}
Considering integrated higher-point correlation functions similarly leads to expressions resembling the moduli space integrals encountered in the construction of the off-shell action in \ref{sec:offcritical}.

The construction of this paper is the geometric refinement of this heuristic picture. In sections \ref{sec:offcritical} and \ref{sec:bgindep} we do not introduce an independent Weyl parameter at a special puncture; instead the Hermitian metric determines the normalized special local coordinate through \eqref{eq:specialmetriccoord}, and the extra interpolation direction is encoded by the metric dependence of that coordinate. The special puncture thereby packages the local failure of Weyl invariance in a form compatible with the contour formalism for \(\cB\).

\subsection{Toward Higher-Order Background Independence}
\label{sec:higher-order-bgi}

Section \ref{sec:bgindep} establishes background independence only at first order in the deformation parameter. The natural question is whether the same pattern persists to higher orders. The most direct extrapolation would introduce mixed vertices carrying one distinguished \(\cO_x\) insertion together with \(l\) pre-existing special punctures, and from them build the candidate BV generator
\begin{equation}
\cU_x
=
\sum_{n\ge 1}\sum_{l\ge 0}\frac{1}{n!\,l!}\{\Psi^n;F^l,\cO_x\}. \bluecheck
\label{eq:Uall}
\end{equation}
Here the comma before \(\cO_x\) indicates that this insertion is distinguished from the \(l\) fully antisymmetrized copies of \(F\), and \(\{\Psi^n;F^l,\cO_x\}\) is meant to denote the corresponding mixed vertex with one distinguished \(\cO_x\) puncture.

We emphasize that \eqref{eq:Uall} is an ansatz, not a theorem. It is suggested by the first-order formulas and by the all-orders mixed-vertex recursion of section \ref{sec:offcritical}, but we have not derived it. To turn it into a theorem one would need, at minimum, an explicit construction of the mixed interpolation chains with one distinguished \(\cO_x\) insertion, the analogues of the first-order chain and BRST identities \eqref{eq:Oxgamma1} and \eqref{eq:OxQBchain}, finiteness of the prepared state \(Q_B\cO_x\) in the presence of existing special punctures, and control of any additional contact terms compatible with the recursion \eqref{eq:gammageneral}. More conceptually, one would like a general framework for state spaces and their connections beyond the conformal locus. Even for conformal theories this geometry is subtle and was clarified by the work of Ranganathan, Sonoda, and Zwiebach on connections over families of CFTs \cite{Ranganathan:1992nb,Ranganathan:1993vj}. Until the corresponding non-conformal structure is understood, the higher-order extension is best viewed as a conjectural program.

The nearly marginal linear-dilaton sector provides a natural testing ground. There \(\cO_x\) is the standard ghost-number-two insertion that shifts the dilaton slope, and \(Q_B\cO_x\) produces the antighost-dilaton state \eqref{eq:Fghostdilaton}. The antisymmetrized self-collision of \(F\) is harmless in this example---the antighost dilaton is built entirely from \(c\)-type ghost fields, whose OPE is regular---so the analytic questions that remain are unusually concrete, even if they do not yet justify \eqref{eq:Uall} in full.

\subsection{Future Directions}

The off-critical formalism is a natural candidate framework for deformations that change the central charge, for noncompact theories with continuous families of nearly marginal operators, and for situations in which ordinary conformal perturbation theory is strongly scheme-dependent or singular \cite{Sen:2019jpm,Mazel:2024alu}. The string vertices play the role of off-shell renormalization data. For NLSMs in particular, the choice of string-field frame controls how geometric spacetime fields are represented inside the string field \cite{Mazel:2024alu,Mazel:2025diffeo}. The same mechanism already appears in curved examples such as the \(S^3\) solution with \(H\)-flux \cite{Mazel:2025diffeo} and in perturbative string-field constructions of Ramond--Ramond backgrounds \cite{Cho:2018nfn}, suggesting that the central-charge deformation studied here is the simplest member of a much wider class. 

Immediate further work is to find the necessary off-shell corrections $\{\Psi^{n};F^m\}_{g}$ for all genera (and open string topologies) following the same interpolation logic. Such a formulation, together with direct control of contact terms in the off-critical conformal perturbation theory, should eventually be compared with Sen's off-shell amplitude, Wilsonian effective-action, shifted-vacuum, and 1PI frameworks \cite{Sen:2015offshell,Sen:2016qap,Pius:2014gza,Pius:2014iaa,Sen:2014dqa} and with the recent NLSM-inspired off-shell string actions and earlier sigma-model effective-action approaches \cite{Ahmadain:2022offshell,Ahmadain:2024treelevel,Fradkin:1985qj,Tseytlin:1988tv, Frenkel:2025wko}; this would bring the program closer to a genuinely background-independent formulation of closed string field theory. In a similar vein, deformations of the worldsheet action away from $c=0$ CFTs may provide a systematic realization of the mechanism describing loop-corrected actions suggested by Fischler and Susskind \cite{Fischler:1986ci}. 

More broadly, an extension of the SFT action following the manifestly background independent logic of \cite{Ahmadain:2024treelevel} may extend CSFT past the perturbative regime, providing new insights into the fundamental structure of the string theory target space.

\section*{Acknowledgements}

We thank Manki Kim, Justin Kulp, and Jacob McNamara for invaluable and insightful discussions. We thank Minjae Cho, Ben Mazel, Jaroslav Scheinpflug, and Ashoke Sen for insightful comments on an early draft. We thank the participants of the 2026 String Field Theory and Flux Compactification workshop in Benasque for many valuable discussions. We thank OpenAI Codex (GPT 5.4) and Anthropic Claude Code (Opus 4.6) for assistance with drafting, editing, consistency checks, and manuscript organization. Some of the equations in this manuscript were also checked using the string-code repository \href{https://github.com/jscheinpflug/string-code}{\texttt{string-code}} (GitHub project maintained by Jaroslav Scheinpflug, with contributions from Joao Gomide, Alex Michel, Xi Yin) and against XY's string notes \href{https://github.com/xiyin137/stringbook}{\texttt{stringbook}}. The authors are responsible for all remaining errors. XY's work is supported by DOE grant DE-SC0007870. AA is supported by STFC Consolidated Grant No. ST/X000648/1.

\appendix

\section{Descent of the Antighost Dilaton}
\label{app:antighost-descent}

This appendix computes the action of the operator-valued one-form $\cB$ on the antighost dilaton $F_{\mathrm{agd}}$ in a given local coordinate system as used in \eqref{eq:deltacdescentsummary}. The conceptual chain
\begin{equation}
Q_B[\gamma]\;\Longrightarrow\;F\;\Longrightarrow\;\cB F\;\Longrightarrow\;F^{[2]}\;\Longrightarrow\;(\cB F^{[2]})(v_\omega)
\label{eq:appendixchain}
\end{equation}
was already stated and interpreted in section \ref{sec:deltac}; what remains here is the Taylor/OPE bookkeeping that produces the coefficients in \eqref{eq:Flocaldescent}, \eqref{eq:F1ghostdilaton}, and \eqref{eq:F2ghostdilaton}. We work with the ghost-sector local insertion
\begin{equation}
F_{\mathrm{agd}}
:=
(\partial c+\bar\partial \tilde c)
\bigl(c\partial^2 c - \tilde c\bar\partial^2 \tilde c\bigr),
\bluecheck
\label{eq:Fagddef}
\end{equation}
So the physical special state used in the main text is \(F = -\frac{\Delta c_m}{24}\,F_{\mathrm{agd}}\).
We compute the transgressed descendants \(\cB F_{\mathrm{agd}}\) and \(\cB^2 F_{\mathrm{agd}}\) in a local coordinate system induced by the worldsheet metric. The operator-valued one-form \(\cB\) must be distinguished from the contracted odd contour operator \(\cB(v)=\iota_v\cB\). The notation \(\cB F_{\mathrm{agd}}\) means the bundle one-form whose value on \(v\) is obtained by letting \(\cB(v)\) act on the local insertion \(F_{\mathrm{agd}}\) through its OPE; acting once more gives an ordered bundle two-form. In the fixed-metric families used here, the transgression from local worldsheet forms to bundle forms is implemented locally by replacing \(dz,d\bar z\) with \(\delta z,\delta\bar z\). The derivation has three steps: first compute the local OPE with the \(b\)-ghost contour, then convert the resulting Taylor coefficients into metric data, and finally act once more with \(\cB\) to recover the ordered double descendant whose associated antisymmetric bundle two-form \(\frac12\cB^2F\) determines the local worldsheet descendant \(F^{[2]}\).

Let the puncture sit at $w=0$, let $z=f(w)=\sum_{m\ge 0}f_m w^m$ be the local coordinate map, and write the Grassmann-even one-form $\cB$ as
\begin{equation}
\cB
=
- \oint_{\partial D}\left(
\frac{dz}{2\pi i}\,\delta f(w)\,b(z)
-\frac{d\bar z}{2\pi i}\,\delta \bar f(\bar w)\,\tilde b(\bar z)
\right),
\qquad
\delta f(w)=\sum_{m\ge 0}\delta f_m\,w^m.
\bluecheck
\label{eq:appBlocal}
\end{equation}
Using $dz=(df/dw)\,dw$ and the primary transformation of the $b$-ghost,
\begin{equation}
b(z)=\left(\frac{df}{dw}\right)^{-2}b(w),
\qquad
\tilde b(\bar z)=\left(\frac{d\bar f}{d\bar w}\right)^{-2}\tilde b(\bar w). \bluecheck
\label{eq:appbghosttransform}
\end{equation}
Define
\begin{equation}
\left(\frac{df}{dw}\right)^{-1}\sum_{m\ge 0}\delta f_m\,w^m
=:
\sum_{k\ge 0}\alpha_k w^k.
\bluecheck
\label{eq:appalpha}
\end{equation}
The bundle exterior derivative \(\delta\) acts on the family of local-coordinate and metric data, whereas \(\partial_z\) and \(\bar\partial_z\) act on the local worldsheet coordinates at the puncture. Write \(X:=\partial c+\bar\partial\tilde c\) and \(U:=c\partial^2 c-\tilde c\bar\partial^2\tilde c\), so that \(F_{\mathrm{agd}}=XU\).
The action of $\cB$ is then computed by the graded Leibniz rule together with the contour OPEs of $b$ and $\tilde b$ with the ghost factors sitting at the puncture. Only poles through order $w^{-3}$ and $\bar w^{-3}$ contribute, so only $\alpha_0,\alpha_1,\alpha_2$ can appear. Equivalently, the entire computation is local and depends only on the first three Taylor coefficients of the local coordinate map.
The contour OPEs
\begin{equation}
b(w)\,\partial^n c(0)\sim (-1)^n n!\,w^{-n-1},
\qquad
\tilde b(\bar w)\,\bar\partial^n\tilde c(0)\sim (-1)^n n!\,\bar w^{-n-1}. \bluecheck
\label{eq:appghostopes}
\end{equation}
then imply
\begin{equation}
\begin{split}
\cB F_{\mathrm{agd}}
=&\,
(\partial c+\bar\partial\tilde c)
\bigl(\alpha_0\partial^2 c-\bar\alpha_0\bar\partial^2\tilde c\bigr)\\
&-(\alpha_1+\bar\alpha_1)\bigl(c\partial^2 c-\tilde c\bar\partial^2\tilde c\bigr)
-2(\partial c+\bar\partial\tilde c)\bigl(\alpha_2 c-\bar\alpha_2\tilde c\bigr). \calccheck
\end{split}
\label{eq:appBFalpha}
\end{equation}
\diffnote{Equation \eqref{eq:appBFalpha} agrees in its coefficient pattern with the older formulas \texttt{eqn:BF-2} and \texttt{eqn:BF-3} in \texttt{main.pdf}, after translating to the present Weyl-factor convention \(ds=e^{\omega}|dz|\).}

To relate the coefficients \(\alpha_k\) to geometric data, introduce the metric factor \(ds=\rho^z |dz|=e^{\omega(z,\bar z)}|dz|\) and impose the Bergman--Zwiebach normalization already stated abstractly in \eqref{eq:specialmetriccoord}. In local coordinates this means requiring that the induced metric in the adapted coordinate be flat to the required order:
\begin{equation}
\partial_w^n\rho^w|_{w=0}=\partial_{\bar w}^n\rho^w|_{w=0}=0,
\qquad
\rho^w=\left|\frac{df}{dw}\right|\rho^z. \bluecheck
\label{eq:apprhoframe}
\end{equation}
To the order needed for \eqref{eq:appBFalpha}, one may expand
\begin{equation}
\left|\frac{df}{dw}\right|
=
|f_1|
\left(
1+\frac{f_2}{f_1}w+\frac{\bar f_2}{\bar f_1}\bar w
+\Bigl(\frac{3f_3}{2f_1}-\frac{f_2^2}{2f_1^2}\Bigr)w^2
+\Bigl(\frac{3\bar f_3}{2\bar f_1}-\frac{\bar f_2^2}{2\bar f_1^2}\Bigr)\bar w^2
+\frac{|f_2|^2}{|f_1|^2}w\bar w
\right). \calccheck
\label{eq:appmetricexpand1}
\end{equation}
and
\begin{equation}
\begin{aligned}
\rho^z(f(w),\bar f(\bar w))
=\,&
\rho^z
+(f_1w+f_2w^2)\partial_z\rho^z
+(\bar f_1\bar w+\bar f_2\bar w^2)\partial_{\bar z}\rho^z\\
&+\frac{1}{2}f_1^2w^2\partial_z^2\rho^z
+\frac{1}{2}\bar f_1^2\bar w^2\partial_{\bar z}^2\rho^z
+|f_1|^2w\bar w\,\partial_z\partial_{\bar z}\rho^z. \calccheck
\end{aligned}
\label{eq:appmetricexpand2}
\end{equation}
Matching coefficients in $\rho^w=|df/dw|\,\rho^z$ gives
\begin{equation}
\rho^z = |f_1|^{-1},
\qquad
\frac{f_2}{f_1^2}=-\partial_z\omega,
\qquad
\frac{3f_3}{f_1^3}-\frac{4f_2^2}{f_1^4}=-\partial_z^2\omega, \calccheck
\label{eq:appfcoeffs}
\end{equation}
and, taking $f_1$ real,
\begin{equation}
\alpha_0=f_1^{-1}\delta z,
\qquad
\alpha_1=f_1^{-1}\delta f_1-2\frac{f_2}{f_1^2}\delta z,
\qquad
\alpha_2=f_1^{-1}\delta f_2-2\frac{f_2}{f_1^2}\delta f_1+\left(4\frac{f_2^2}{f_1^3}-3\frac{f_3}{f_1^2}\right)\delta z. \calccheck
\label{eq:appalphas}
\end{equation}
In particular, the first equality in \eqref{eq:appfcoeffs} gives $f_1=e^{-\omega}$ once we choose $f_1$ to be real. The remaining equalities then convert the Taylor coefficients of the local coordinate map into derivatives of the Weyl factor at the puncture.
Substituting \eqref{eq:appfcoeffs} and \eqref{eq:appalphas} into \eqref{eq:appBFalpha} yields the compact geometric formula
\begin{equation}
\begin{split}
\cB F_{\mathrm{agd}}
=&\,
e^{\omega}(\partial c+\bar\partial\tilde c)
\bigl(\partial^2 c\,\delta z-\bar\partial^2\tilde c\,\delta\bar z\bigr)\\
&+2\bigl(\delta\omega-\partial_z\omega\,\delta z-\bar\partial_z\omega\,\delta\bar z\bigr)
\bigl(c\partial^2 c-\tilde c\bar\partial^2\tilde c\bigr)\\
&+2(\partial c+\bar\partial\tilde c)
\bigl(
ce^{-\omega} \delta(\partial_z\omega)-\tilde c e^{-\omega} \delta(\bar\partial_z\omega)
-ce^{-\omega} \partial_z^2\omega\,\delta z+\tilde c e^{-\omega}\bar\partial_z^2\omega\,\delta\bar z
\bigr). \calccheck
\end{split}
\label{eq:appBFomega}
\end{equation}
\diffnote{Equation \eqref{eq:appBFomega} is the present-convention version of the older appendix formula in \texttt{main.pdf}, with \(ds=e^{\omega}|dz|\) replacing \(ds=e^{-\omega}|dz|\). This formula is independently checked by \texttt{checks/verify\_antighost\_descent\_geometric.py}.}

The genus-zero application uses the case in which the Hermitian metric is held fixed while the puncture moves, so that
\begin{equation}
\delta\omega=\partial_z\omega\,\delta z+\bar\partial_z\omega\,\delta\bar z. \bluecheck
\label{eq:appdomega}
\end{equation}
In that regime the middle line of \eqref{eq:appBFomega} cancels, and one obtains after a short simplification
\begin{equation}
\begin{split}
\cB F_{\mathrm{agd}}
=&\,
e^{\omega}(\partial c+\bar\partial\tilde c)
\bigl(\partial^2 c\,\delta z-\bar\partial^2\tilde c\,\delta\bar z\bigr)\\
&+2e^{-\omega}(\partial c+\bar\partial\tilde c)
\bigl(
c\,\partial_z\bar\partial_z\omega\,\delta\bar z
-\tilde c\,\partial_z\bar\partial_z\omega\,\delta z
\bigr)
+ \delta(\cdots). \calccheck
\end{split}
\label{eq:appBFfixed}
\end{equation}
\diffnote{Equation \eqref{eq:appBFfixed} is the fixed-metric specialization of \eqref{eq:appBFomega}. In the present convention it reproduces the same coefficient pattern as the older formula in \texttt{off-critical SFT.tex}. This specialization is independently machine-checked.}
where the omitted terms are total derivatives along the contour and therefore do not affect the integrated descendant.

Isolate the insertion-valued one-form multiplying $X=\partial c+\bar\partial\tilde c$ in \eqref{eq:appBFfixed}:
\begin{equation}
J
:=
e^{\omega}
\bigl(\partial^2 c\,\delta z-\bar\partial^2\tilde c\,\delta\bar z\bigr)
+2e^{-\omega}
\bigl(
c\,\partial_z\bar\partial_z\omega\,\delta\bar z
-\tilde c\,\partial_z\bar\partial_z\omega\,\delta z
\bigr). \reasoncheck
\label{eq:appJdef}
\end{equation}
Then $\cB F_{\mathrm{agd}}=XJ+\delta(\cdots)$. In the fixed-metric regime one has
\begin{equation}
\cB X
=
-(\alpha_1+\bar\alpha_1)
=
\bigl(2\delta\omega-2\partial_z\omega\,\delta z-2\bar\partial_z\omega\,\delta\bar z\bigr)
=0. \reasoncheck
\label{eq:appBX}
\end{equation}
Thus the second descent acts only on $J$. The elementary descendants needed for this step are
\begin{equation}
\cB c=-\alpha_0,
\qquad
\cB\tilde c=-\bar\alpha_0,
\qquad
\cB\partial^2 c=-2\alpha_2,
\qquad
\cB\bar\partial^2\tilde c=-2\bar\alpha_2. \calccheck
\label{eq:appBelementary}
\end{equation}
In the fixed-metric regime,
\begin{equation}
\alpha_0=e^{\omega}\delta z,
\qquad
\bar\alpha_0=e^{\omega}\delta\bar z,
\qquad
\alpha_2=-e^{-\omega}\partial_z\bar\partial_z\omega\,\delta\bar z,
\qquad
\bar\alpha_2=-e^{-\omega}\partial_z\bar\partial_z\omega\,\delta z. \reasoncheck
\label{eq:appalphafixed}
\end{equation}
Here $X=\partial c+\bar\partial\tilde c$ is a Grassmann-odd zero-form. Therefore, for every tangent vector $v$, the odd contour operator $\cB(v)$ satisfies
\begin{equation}
\cB(v)(XJ)=(\cB(v)X)J-X\,\cB(v)J. \reasoncheck
\label{eq:BvLeibniz}
\end{equation}
Repackaging this statement into one-form notation gives
\begin{equation}
\cB^2 F_{\mathrm{agd}}
=
-X\,\cB J. \reasoncheck
\label{eq:appB2FagdStep}
\end{equation}
Applying the four elementary descendants to the four terms in $J$, one finds that each contributes the same curvature coefficient and therefore
\begin{equation}
\cB J
=
+8\,\partial_z\bar\partial_z\omega\,\delta z\wedge \delta\bar z. \calccheck
\label{eq:appBJ}
\end{equation}
For a metric \(ds^2=e^{2\omega}|dz|^2\), one has \(\sqrt g=\sqrt{\det g_{ab}}=e^{2\omega}\) in real worldsheet coordinates and the scalar curvature obeys \(\sqrt g\,R=-8\,\partial_z\bar\partial_z\omega\).
Using this identity in the previous line gives
\begin{equation}
\cB^2 F_{\mathrm{agd}}
=
+(\partial c+\bar\partial\tilde c)\,\sqrt g\,R\,\delta z\wedge \delta\bar z. \calccheck
\label{eq:appB2Fagdfinal}
\end{equation}
Restoring the overall factor in the physical special state $F=-(\Delta c_m/24)F_{\mathrm{agd}}$ then gives
\begin{equation}
\cB^2F
=
-\frac{\Delta c_m}{24}
(\partial c+\bar\partial\tilde c)\,\sqrt g\,R\,\delta z\wedge \delta\bar z. \calccheck
\label{eq:appB2Ffinal}
\end{equation}
This is the explicit realization, in the fixed-metric moving-puncture family, of the ordered bundle two-form obtained by acting twice with the odd contour operator \(\cB(v)\). Because the corresponding antisymmetric bundle two-form satisfies \(\iota_{v_2}\iota_{v_1}(\cB^2F)=2\,\cB(v_2)\cB(v_1)F\), the second relation in \eqref{eq:Bdescent} identifies the local descendant \(F^{[2]}\) with the half-sized two-form \(\frac12\cB^2F\). Replacing the bundle area form \(\delta z\wedge\delta\bar z\) by the corresponding worldsheet area form \(dz\wedge d\bar z\) then gives
\begin{equation}
F^{[2]}
=
-\frac{\Delta c_m}{48}
(\partial c+\bar\partial\tilde c)\,\sqrt g\,R\,dz\wedge d\bar z
\reasoncheck
\label{eq:appF2final}
\end{equation}
which is the formula quoted in the main text.

\diffnote{The downstream second-descent formulas written here follow directly from the local contour algebra encoded in \eqref{eq:appBFalpha}, together with the fact that the antisymmetric bundle two-form corresponding to the ordered double action is \(\frac12\cB^2\). After translating conventions, this restores agreement with the older note \texttt{off-critical SFT.tex}; with the contour-orientation sign in \eqref{eq:kOmega}, the appendix formula \eqref{eq:appF2final} also agrees with the anomaly-normalized Weyl-response insertion used in the main text.}

\diffnote{Compared with the ``Action of $\mathcal{B}$ on Antighost Dilaton'' appendix of \texttt{main.tex}, the derivation behind \texttt{eqn:BF-2} (A.19) and \texttt{eqn:BF-3} (A.22) is written so that the product structure of the antighost dilaton, the metric expansion, and the final curvature identity are explicit. The longer $z$-frame rewriting leading to \texttt{eqn:BF-z} (A.24) and the fully expanded formula for $\mathcal B^2F$ that follows it are not reproduced here because, after specialization to the fixed-metric regime used later in the paper, they add no genus-zero input beyond the curvature insertion already displayed.}

\section{Local-Coordinate Dependence of the Nearly Marginal Basis}
\label{app:nlsm-local-coords}

This appendix computes the local-coordinate transformations needed in the nearly marginal sector of section \ref{sec:deltac-linear}. Let
\begin{equation}
z=f(w)=f_0+f_1 w+f_2 w^2+\cdots,
\qquad
\eta:=\frac{f_2}{f_1^2}. \bluecheck
\label{eq:appfexpansion}
\end{equation}
We work in the nonlinear-sigma-model (NLSM) approximation in which logarithmic pieces that would generate branch cuts are dropped, since they cancel from the gauge-invariant quantities of interest \cite{Bergman:1994qq,Frenkel:2025wko,Kim:2026stringloops}. In this approximation all three basis states in \eqref{eq:VDVGVK} mix under a general local-coordinate change through the same combination $\eta$.
Recall from \eqref{eq:VDVGVK} that $\cV_D$ is the ghost-dilaton direction, $\cV_G$ the graviton-type direction, and $\cV_K$ the vector-type descendant.

The basic transformations are the usual ones for a scalar field and the reparametrization ghosts:
\begin{equation}
c(w)=\left(\frac{df}{dw}\right)^{-1}c(z),
\qquad
\tilde c(\bar w)=\left(\frac{d\bar f}{d\bar w}\right)^{-1}\tilde c(\bar z),
\qquad
\partial_w=\frac{df}{dw}\,\partial_z,
\qquad
\bar\partial_w=\frac{d\bar f}{d\bar w}\,\bar\partial_z. \bluecheck
\label{eq:appghostcoordtransform}
\end{equation}
Expanding at $w=0$ gives
\begin{equation}
\frac{f''(0)}{f'(0)^2}=2\eta,
\qquad
\frac{\bar f''(0)}{\bar f'(0)^2}=2\bar\eta. \bluecheck
\label{eq:appeta}
\end{equation}
Those two numbers are the only local-coordinate data that survive in the quadratic approximation used in \eqref{eq:appvector}, meaning that we keep terms through second order in derivatives of the profile functions and through the corresponding order in the local-coordinate expansion.

For the graviton-type insertion one may write
\begin{equation}
G(Y)\partial_w Y\bar\partial_w Y
=
-\frac{1}{\sqrt{2\pi}}\int dk\,\frac{\tilde G(k)}{k^2}\,\partial_w\bar\partial_w e^{ikY},
\bluecheck
\label{eq:appGidentity}
\end{equation}
where $\tilde G(k)$ is the Fourier transform of the profile $G(Y)$. This formula serves only to reduce the transformation law of the graviton-type insertion to that of exponentials $e^{ikY}$. After transforming to the $z$ coordinate and expanding to quadratic order in derivatives of $G$, one finds
\begin{equation}
\begin{split}
c\tilde c\,G(Y)\partial_wY\bar\partial_wY
\mapsto\,&
c\tilde c\,G(Y)\partial_zY\bar\partial_zY\\
&-\frac{\alpha'}{2}c\tilde c
\left(
\bar\partial_zY\,\eta+\partial_zY\,\bar\eta
\right)\partial_Y G(Y)
+\frac{\alpha'^2}{4}c\tilde c\,|\eta|^2\,\partial_Y^2G(Y).
\end{split}
\calccheck
\label{eq:appgraviton}
\end{equation}
The Fourier representation makes the $Y$-derivative structure transparent: each derivative with respect to $Y$ is generated by a factor of $ik$. After the coordinate transformation, the Jacobian produces the two structures proportional to $\partial_YG$ and $\partial_Y^2G$, and both are controlled by $\eta$ and $\bar\eta$.

The ghost-dilaton part is simpler because only the ghost transformation is needed:
\begin{equation}
c\,\partial_w^2 c
\mapsto
c\,\partial_z^2 c - 2\eta\,c\,\partial_z c,
\qquad
\tilde c\,\bar\partial_w^2\tilde c
\mapsto
\tilde c\,\bar\partial_z^2\tilde c - 2\bar\eta\,\tilde c\,\bar\partial_z\tilde c. \calccheck
\label{eq:appdilaton}
\end{equation}
Consequently,
\begin{equation}
\cV_D
\mapsto
\cV_D - 2\eta\,c\partial c + 2\bar\eta\,\tilde c\bar\partial\tilde c. \reasoncheck
\label{eq:appVD}
\end{equation}
This formula makes explicit why the ghost-dilaton direction mixes with the pure-ghost descendants under a generic change of local coordinates.

For the antighost dilaton one also needs the first derivative of the ghost:
\begin{equation}
\partial_w c \mapsto \partial_z c-2\eta\,c,
\qquad
\bar\partial_w\tilde c \mapsto \bar\partial_z\tilde c-2\bar\eta\,\tilde c, \calccheck
\label{eq:appghosttransformsimple}
\end{equation}
which gives
\begin{equation}
\begin{split}
(\partial_w c+\bar\partial_w\tilde c)
\bigl(c\partial_w^2 c-\tilde c\bar\partial_w^2\tilde c\bigr)
\mapsto\,&
(\partial_z c+\bar\partial_z\tilde c-2\eta c-2\bar\eta\tilde c)\\
&\times
\bigl(c\partial_z^2 c-2\eta c\partial_z c-\tilde c\bar\partial_z^2\tilde c+2\bar\eta \tilde c\bar\partial_z\tilde c\bigr).
\end{split}
\calccheck
\label{eq:appantighosttransform}
\end{equation}
This partially factorized form makes clear that the same combination $\eta=f_2/f_1^2$ governs both the deformation of $\partial c+\bar\partial\tilde c$ and the deformation of $c\partial^2 c-\tilde c\bar\partial^2\tilde c$. That is what makes the nearly marginal sector manageable in low-point string vertices.

Finally, for the vector-type insertion one obtains
\begin{equation}
\begin{split}
(\partial_w c+\bar\partial_w\tilde c)
\bigl(c\partial_wY-\tilde c\bar\partial_wY\bigr)\mathcal K(Y)
\mapsto\,&
(\partial_z c-2\eta c+\bar\partial_z\tilde c-2\bar\eta\tilde c)\\
&\times\left[
\mathcal K(Y)\bigl(c\partial_zY-\tilde c\bar\partial_zY\bigr)
-\frac{\alpha'}{2}\mathcal K'(Y)\bigl(\eta c-\bar\eta\tilde c\bigr)
\right].
\end{split}
\calccheck
\label{eq:appvector}
\end{equation}

Equations \eqref{eq:appgraviton}, \eqref{eq:appVD}, and \eqref{eq:appvector} are the local-coordinate inputs used when evaluating the cubic products in the nearly marginal sector. They explain why the basis \eqref{eq:VDVGVK} closes under BRST variation and under the low-point products relevant for the flat-space and linear-dilaton computations.

\diffnote{Compared with the second appendix of \texttt{main.tex}, this appendix centers the local-coordinate technology on the basis \eqref{eq:VDVGVK} and makes explicit the transformations underlying \texttt{eqn:graviton-transformation} (B.3), \texttt{eqn:dilaton-transformation} (B.4), \texttt{eqn:antighost-dilaton-transformation} (B.5), and \texttt{eqn:vector-transformation} (B.6), including the factor-of-two convention in the definition of $\eta=f_2/f_1^2$. The longer Fourier manipulations and logarithmic branch-cut terms present in \texttt{main.tex} are omitted here because, in the NLSM approximation stated at the start of this appendix, they cancel from the gauge-invariant low-point data retained in the black text.}

\section{Deriving the Linearized Nearly Marginal Equations}
\label{app:linearized-nearly-marginal}

This appendix derives the first order equations of motion used in section \ref{sec:deltac-linear}. The reduction consists of tracking which of the three structures $\cW_1,\cW_2,\cW_3$ each BRST variation lands in and then matching coefficients. Separating that algebra from the low-point vertex calculations of subsection \ref{subsec:lowpoint-second-order} makes clear which statements are kinematic and which depend on explicit string products.

Define three independent ghost-matter structures,
\begin{equation}
\cW_1:=c\tilde c\bigl(\partial^2 c\,\bar\partial Y+\bar\partial^2\tilde c\,\partial Y\bigr),
\qquad
\cW_2:=(\partial c+\bar\partial\tilde c)\,\cV_D,
\qquad
\cW_3:=(\partial c+\bar\partial\tilde c)\,\cV_G. \bluecheck
\label{eq:appWbasis}
\end{equation}
These three structures are linearly independent in the restricted nearly marginal sector considered here, so the BRST equation is equivalent to matching their coefficients one by one.
Recall from \eqref{eq:VDVGVK} that the basis vectors $\cV_D,\cV_G,\cV_K$ denote, respectively, the ghost-dilaton, graviton-type, and vector-type directions in the single-coordinate sector.
Using \eqref{eq:QBbasis}, a general linear combination
\begin{equation}
\Psi_0=\mathcal A_0(Y)\,\cV_D+\mathcal B_0(Y)\,\cV_G+\mathcal K_0(Y)\,\cV_K \bluecheck
\label{eq:apppsi0ansatz}
\end{equation}
obeys, in flat space,
\begin{equation}
\begin{aligned}
Q_B\Psi_0
=\,&
\left(
-\mathcal A_0'(Y)-\frac{\alpha'}{4}\mathcal B_0'(Y)+\mathcal K_0(Y)
\right)\cW_1\\
&+
\left(
-\frac{\alpha'}{4}\mathcal A_0''(Y)+\frac{\alpha'}{4}\mathcal K_0'(Y)
\right)\cW_2
-\frac{\alpha'}{4}\mathcal B_0''(Y)\,\cW_3. \reasoncheck
\end{aligned}
\label{eq:appQBpsi0flat}
\end{equation}
The special state
\begin{equation}
F=-\frac{\Delta c_m}{24}\,(\partial c+\bar\partial\tilde c)\,\cV_D. \reasoncheck
\label{eq:appFsector}
\end{equation}
has support only along the $\cW_2$ direction. Hence the linearized equation $Q_B\Psi_0=-F$ is equivalent to the three scalar equations
\begin{equation}
\begin{split}
&-\mathcal A_0'(Y)-\frac{\alpha'}{4}\mathcal B_0'(Y)+\mathcal K_0(Y)=0,\\
&-\frac{\alpha'}{4}\mathcal A_0''(Y)+\frac{\alpha'}{4}\mathcal K_0'(Y)=\frac{\Delta c_m}{24},\\
&-\frac{\alpha'}{4}\mathcal B_0''(Y)=0. \reasoncheck
\end{split}
\label{eq:appflatlinearizedsystem}
\end{equation}
This is exactly the flat-space system in \eqref{eq:flatsystem}.

The inconsistency for $\Delta c_m\neq 0$ is immediate once one writes the equations in this basis. Differentiating the first equation and using the third gives
\begin{equation}
-\frac{\alpha'}{4}\mathcal A_0''(Y)+\frac{\alpha'}{4}\mathcal K_0'(Y)=0, \reasoncheck
\label{eq:appflatconsistencyeq}
\end{equation}
which contradicts the second equation unless $\Delta c_m=0$. The absence of a flat-space solution at order $\Delta c_m$ therefore follows before any cubic product is evaluated.

In a linear-dilaton background with slope $\beta$, the BRST current contains the improvement term proportional to $\partial^2Y$. Its effect on the nearly marginal sector is to shift only the coefficients of $\cW_1$ and $\cW_2$, leaving the $\cW_3$ equation unchanged. The result is
\begin{equation}
\begin{aligned}
Q_B\Psi_0
=\,&
\left(
-\mathcal A_0'(Y)-\frac{\alpha'}{4}\mathcal B_0'(Y)+\mathcal K_0(Y)
+\frac{\sqrt{\alpha'}\beta}{2\sqrt2}\mathcal B_0(Y)
\right)\cW_1\\
&+
\left(
-\frac{\alpha'}{4}\mathcal A_0''(Y)+\frac{\alpha'}{4}\mathcal K_0'(Y)
+\frac{\sqrt{\alpha'}\beta}{2\sqrt2}\bigl(\mathcal A_0'(Y)+\mathcal K_0(Y)\bigr)
\right)\cW_2
-\frac{\alpha'}{4}\mathcal B_0''(Y)\,\cW_3, \reasoncheck
\end{aligned}
\label{eq:applinearizedQB}
\end{equation}
which reproduces \eqref{eq:linearizedsystem} after setting $Q_B\Psi_0=-F$.

The special solution \eqref{eq:Psi0flat} is obtained by taking
\begin{equation}
\mathcal A_0(Y)=\frac{\eps_0}{\sqrt{2\alpha'}}\,Y,
\qquad
\mathcal B_0(Y)=0,
\qquad
\mathcal K_0(Y)=\frac{\eps_0}{\sqrt{2\alpha'}}. \calccheck
\label{eq:appeps0profile}
\end{equation}
The first and third equations are then automatic. The second reduces to \(\beta\eps_0/2=\Delta c_m/24\), hence \(\Delta c_m=12\beta\eps_0\). This is the first-order relation quoted above.

\diffnote{In \texttt{main.tex}, the algebra leading to the differential systems of section 4 was spread across unnumbered BRST computations in the flat-space and linear-dilaton subsections. Here it is isolated so that the structural input to \eqref{eq:flatsystem} and \eqref{eq:linearizedsystem} is explicit.}

\section{Second Order Equations of Motion}
\label{app:flat-compatibility}

In this appendix we report our setup and results for the string products necessary for the second order equations of motion. We therefore evaluate the ordinary quadratic bracket of \(\Psi_0\) directly in the same nearly marginal flat-space sector, rewrite it in the \(\cW_1,\cW_2,\cW_3\) basis, and then extract the resulting compatibility condition. The BRST variation of
\begin{equation}
\Psi_1=\mathcal A_1(Y)\,\cV_D+\mathcal B_1(Y)\,\cV_G+\mathcal K_1(Y)\,\cV_K \bluecheck
\label{eq:apppsi1ansatz}
\end{equation}
has the same form as \eqref{eq:appQBpsi0flat}, with the index \(0\) replaced by \(1\):
\begin{equation}
\begin{aligned}
Q_B\Psi_1
=\,&
\left(
-\mathcal A_1'(Y)-\frac{\alpha'}{4}\mathcal B_1'(Y)+\mathcal K_1(Y)
\right)\cW_1\\
&+
\left(
-\frac{\alpha'}{4}\mathcal A_1''(Y)+\frac{\alpha'}{4}\mathcal K_1'(Y)
\right)\cW_2
-\frac{\alpha'}{4}\mathcal B_1''(Y)\,\cW_3. \reasoncheck
\end{aligned}
\label{eq:appQBpsi1flat}
\end{equation}

The ordinary cubic bracket quoted below is evaluated in the same local-coordinate convention as the Mathematica files used to check \eqref{eq:appflatbracket}: the three ordinary punctures carry local maps
\begin{equation}
f_1(w)=w,\qquad f_2(w)=1+w,\qquad f_3(w)=1/w.
\label{eq:appflatcoords}
\end{equation}
This differs from the low-point representative \eqref{eq:Gamma03coords} by a permutation of the ordinary puncture labels; since the ordinary cubic bracket is fully symmetrized, the resulting bracket is unchanged. In the scripts this is implemented by fixing the global puncture positions to \(z_1=0\), \(z_2=1\), \(z_3=e^{i\theta}/y\), where \(y\to 0\) parametrizes the approach to the sewing degeneration. One then sums over the relevant puncture orderings, extracts the coefficient of \(y^0\), and finally takes the angular average in \(\theta\).

To convert the raw correlator outputs into bracket components, one pairs with the BPZ-conjugate test states \(\cV_G^C\), \(\cV_D^C\), and \(\cV_K^C\). The corresponding normalizations are
\begin{equation}
\bra{\cV_G^C}c_0^-\ket{\cV_G}=\frac{\alpha'^2}{4},
\qquad
\bra{\cV_D^C}c_0^-\ket{\cV_D}=-8,
\qquad
\bra{\cV_K^C}c_0^-\ket{\cV_K}=2\alpha'. \calccheck
\label{eq:appflatbpz}
\end{equation}
Here the conjugate directions are
\begin{equation}
\begin{aligned}
\cV_G^C&=(\partial c+\bar\partial\tilde c)\,\cV_G=\cW_3,\\
\cV_D^C&=(\partial c+\bar\partial\tilde c)\,\cV_D=\cW_2,\\
\cV_K^C&=c\tilde c\bigl(\partial^2 c\,\bar\partial Y+\bar\partial^2\tilde c\,\partial Y\bigr)=\cW_1.
\end{aligned}
\end{equation}
The notebook therefore first returns coefficients against \(\cV_G^C\), \(\cV_D^C\), and \(\cV_K^C\); dividing by the BPZ factors in \eqref{eq:appflatbpz} and rewriting the result in the equivalent \(\cW_3,\cW_2,\cW_1\) basis yields the bracket components below.

Direct evaluation of the ordinary flat-space three-string vertices and the corresponding BPZ pairings gives the bracket components
\begin{equation}
\begin{split}
\frac{1}{2}[D(Y)\cV_D,D(Y)\cV_D]
=&\;
\frac{1}{6}\Bigl(
-16(\partial_YD)^2
+32\partial_Y(D\partial_YD)
-8\partial_Y^2(D^2)
-4D\partial_Y^2D
\Bigr)\cW_3\\
&+\frac{1}{6}\Bigl(
14D\partial_YD-4\partial_Y(D^2)
\Bigr)\cW_1,
\end{split}
\label{eq:appDDbracket}
\end{equation}
\begin{equation}
\begin{split}
[D(Y)\cV_D,K(Y)\cV_K]
=&\;
\frac{1}{6}\Bigl(
-16K\partial_YD
+8\partial_Y(DK)
-4D\partial_YK
\Bigr)\cW_3\\
&-\frac{\alpha'}{6}\Bigl(
\frac72 K\partial_YD
-\frac72 \partial_Y(DK)
+4D\partial_YK
\Bigr)\cW_2
+\frac{1}{6}(4DK)\cW_1,
\end{split}
\label{eq:appDKbracket}
\end{equation}
\begin{equation}
\frac{1}{2}[K(Y)\cV_K,K(Y)\cV_K]
=
-\frac{1}{6}(4K^2)\cW_3
-\frac{1}{6}\frac{\alpha'}{2}K^2\cW_2.
\label{eq:appKKbracket}
\end{equation}
Now substitute the flat-space profile \eqref{eq:appeps0profile}. Since \(\mathcal B_0=0\), only the functions
\[
D(Y)=\frac{\eps_0}{\sqrt{2\alpha'}}\,Y,
\qquad
K(Y)=\frac{\eps_0}{\sqrt{2\alpha'}}
\]
appear. Summing \eqref{eq:appDDbracket}, \eqref{eq:appDKbracket}, and \eqref{eq:appKKbracket} then gives
\begin{equation}
\frac{1}{2}[\Psi_0^2]
=
+\frac{5}{6\alpha'}\eps_0^2 Y\,\cW_1
-\frac{1}{24}\eps_0^2\,\cW_2
-\frac{1}{\alpha'}\eps_0^2\,\cW_3. \reasoncheck
\label{eq:appflatbracket}
\end{equation}
Combining \eqref{eq:appQBpsi1flat}, \eqref{eq:appflatbracket}, and \(F=-(\Delta c_m/24)\cW_2\) gives the corresponding inhomogeneous flat-space system
\begin{equation}
\begin{split}
&-\mathcal A_1'(Y)-\frac{\alpha'}{4}\mathcal B_1'(Y)+\mathcal K_1(Y)+\frac{10}{12\alpha'}\eps_0^2Y=0,\\
&-\frac{\alpha'}{4}\mathcal A_1''(Y)+\frac{\alpha'}{4}\mathcal K_1'(Y)-\frac{\Delta c_m}{24}-\frac{1}{24}\eps_0^2=0,\\
&-\frac{\alpha'}{4}\mathcal B_1''(Y)-\frac{12}{12\alpha'}\eps_0^2=0. \reasoncheck
\end{split}
\label{eq:appflatsourced}
\end{equation}
Now differentiate the first line of \eqref{eq:appflatsourced} and use the third line to eliminate \(\mathcal B_1''\). This gives
\begin{equation}
-\frac{\alpha'}{4}\mathcal A_1''(Y)+\frac{\alpha'}{4}\mathcal K_1'(Y)+\frac{11}{24}\eps_0^2=0. \calccheck
\label{eq:appflatcompatibility1}
\end{equation}
Comparing \eqref{eq:appflatcompatibility1} with the second line of \eqref{eq:appflatsourced} immediately yields
\begin{equation}
\Delta c_m=-12\eps_0^2. \calccheck
\label{eq:appflatcompatibility2}
\end{equation}
This is the relation quoted in \eqref{eq:flatrelation}.

\diffnote{Compared with the ``Second Order Solution'' block in \texttt{main.tex}, appendix D now records the direct ordinary cubic/BPZ evaluation of \([\Psi_0^2]\) in the \(\cW_1,\cW_2,\cW_3\) basis. The intermediate source coefficients differ from the ones written in \texttt{main.tex}, but the final compatibility relation \eqref{eq:flatrelation} is unchanged.}

\bibliographystyle{JHEP}
\bibliography{refs}

\end{document}